\documentclass[aps,pra,letterpaper,reprint,%
showpacs,nofootinbib,groupedaddress,%
longbibliography,floatfix]{revtex4-1}

\usepackage{amsmath}

\usepackage{txfonts}
\usepackage{mathrsfs}

\usepackage{microtype}

\usepackage{xcolor}
\usepackage{hyperref}
\hypersetup{hypertexnames=false,% required for autonum
	bookmarksopen=true,%
	bookmarksnumbered=true,%
	colorlinks=true,%
    	linkcolor=blue,% color of internal links
    	citecolor=blue,% color of links to bibliography
    	filecolor=blue,% color of file links
    	urlcolor=blue,% color of external links
           linkbordercolor=blue,% color of internal links % if colorlinks=false, borders are drawn
    	citebordercolor=blue,% color of links to bibliography
    	filebordercolor=blue,% color of file links
    	urlbordercolor=blue,% color of external links
	pdfborder={0 0 0.5},
	pdfstartview=FitH,%
	pdfnewwindow=true% links in new window
}

\usepackage{graphicx}

\newcommand{\bra}[1]{\langle#1\rvert} % Bra
\newcommand{\ket}[1]{\lvert#1\rangle} % Ket
\newcommand{\braket}[2]{ \langle #1 | #2 \rangle} %Inner Product
\newcommand{\braopket}[3]{\langle {#1} | {#2} | {#3}\rangle} % Matrix Element
\makeatletter
\newcommand{\vast}{\bBigg@{2}}
\newcommand{\Vast}{\bBigg@{4}}
\makeatother

\newcommand{\me}{\mathrm{e}}
\newcommand{\mi}{\mathrm{i}}

\newcommand{\dif}{\mathrm{d}}
\usepackage{bm} 
\newcommand{\abs}[1]{\lvert#1\rvert}

\DeclareMathOperator{\Real}{\mathrm{Re}}

\DeclareMathOperator{\res}{\mathrm{Res}}
\newcommand{\da}{{\downarrow}} %downarraw in {} so as to prevent the arrow from being used as a binary operator
\newcommand{\ua}{{\uparrow}} %uparrow in {}

\begin{document}
\title{Effects of Modal Dispersion on Few Photon -- Qubit Scattering in One-Dimensional Waveguides}
\author{ \c{S}\"ukr\"u Ekin Kocaba\c{s}}
\email{ekocabas@ku.edu.tr}
\affiliation{Department of Electrical  \& Electronics Engineering, Ko\c{c} University, Rumeli Feneri Yolu, TR34450 Sar{\i}yer, \.Istanbul, Turkey}
\homepage[]{http://home.ku.edu.tr/~ekocabas/}
\date{March, 2016 --- Published version}

\begin{abstract}
We study one- and two-photon scattering from a qubit embedded in a one-dimensional waveguide in the presence of modal dispersion. We use a resolvent based analysis and utilize techniques borrowed from the Lee model studies. Modal dispersion leads to atom-photon bound states which necessitate the use of multichannel scattering theory. We present multichannel scattering matrix elements in terms of the solution of a Fredholm integral equation of the second kind. Through the use of the Lippmann-Schwinger equation, we derive an infinite series of Feynman diagrams that represent the solution to the integral equation. We use the Feynman diagrams as vertex correction terms to come up with closed form formulas that successfully predict the trapping rate of a photon in the atom-photon bound state. We verify our formalism through Krylov-subspace based numerical studies with pulsed excitations. Our results provide the tools to calculate the complex correlations between scattered photons in a dispersive environment.

\end{abstract}
%\pacs{03.65.Nk, 32.50.+d, 42.50.Ct, 42.50.Pq}
% 03.65.Nk 	Scattering theory 
% 32.50.+d 	Fluorescence, phosphorescence (including quenching)
% 42.50.Ct 	Quantum description of interaction of light and matter; related experiments 
% 42.50.P
\maketitle

\section{Introduction}
Recent experimental advances in coupling the radiation from quantum emitters into waveguides \cite{OShea2013,Makhonin2014,Goban2015,Kress2015,Sollner2015} motivates us to analyze light-matter interactions in waveguiding structures. In a waveguide geometry, the propagating modes slow down due to modal dispersion, which leads to an increased photon-qubit coupling in plasmonic and photonic crystal waveguides \cite{Lodahl2015}. Additionally, tight confinement of the modes boosts coupling as well \cite{Tame2013}. Furthermore, effects of dispersion are also important for single-photon and squeezed light sources that rely on phase matching \cite{Walmsley2015}. Dispersive effects, particularly around the band edges, also provide the means to simulate many-body systems using atoms trapped near photonic waveguides \cite{Douglas2015}. Many quantum information processing and quantum communication proposals require the generation of highly correlated photonic states that encode information in them. Controlled interaction of photons and qubits is a key element of these proposals \cite{Chang2014}. 

Initial studies on qubit-photon dynamics in a dispersive medium focused on single photon events near a photonic bandgap edge \cite{John1991,John1994,Kofman1994,Gaveau1995,Lambropoulos2000,Vats2002} and near the cut-off frequency of propagating waveguide modes \cite{Kofman1996}. In these systems, large changes in the density of photonic states lead to effects such as non-exponential decay of excited qubits and enhancements in photon emission spectrum.

Through the use of the Bethe ansatz technique, analysis of two-photon scattering in waveguides with embedded qubits became possible for a linearized modal dispersion \cite{Shen2007}. Similar results were later obtained via LSZ reduction \cite{Shi2009}, input-output method \cite{Fan2010} and diagrammatic series summation \cite{Pletyukhov2012} as well. Gaussian pulse scattering \cite{Chen2011a,Chumak2014} and interaction of photons with more than one qubit in a waveguide with a linearized dispersion relation were subsequently presented \cite{Gonzalez-Ballestero2013,Laakso2014,Fang2015a,Shi2015}.

Meanwhile, Krylov-subspace \cite{Longo2009,Longo2010,Longo2011}, DMRG \cite{Prior2013}, MPS \cite{Sanchez-Burillo2014}, master equation \cite{Caneva2015} based numerical methods were developed. Recently, numerical analysis of arbitrary dispersion of waveguide modes in a multi-photon setting became possible \cite{Nysteen2015}.

Investigation of photon-qubit scattering in dispersive waveguides remained mostly at the single photon level. Tight-binding cosine dispersion \cite{Shi2009,Longhi2007,Zhou2008,Roy2011a,Lombardo2014,Wang2014,Biondi2014}, a quadratic dispersion \cite{Huang2013b} and rectangular waveguide dispersion \cite{Li2014} were modeled, extensions were made for multi-mode waveguides. 

One of the interesting effects that develops as a result of dispersion is the existence of atom-photon bound states \cite{Lombardo2014,Vega2014,Calajo2015,Shi2015b}. Trapping photons in such a bound state was numerically shown in \cite{Longo2010} by investigating the time evolution of a two-photon wavepacket. Two-photon scattering matrix elements, involving the bound states, for a tight-binding system were presented in \cite{Shi2009a} through the use of the scattering eigenstates by solving the time-independent Schr\"odinger equation. The solution in \cite{Shi2009a} utilizes techniques developed for the Lee model \cite{Lee1954,Kallen1955,Glaser1956–1957}. The Lee model involves two neutral fermions and a relativistic boson field. Although it has been described as a `guinea pig' \cite{Wick1955} on which mathematical tricks can be deployed at will, the Lee model played an important role in the birth of the renormalization group idea which led to the solution of the Kondo problem \cite{Peskin2014}.

We will follow the lead in \cite{Shi2009,Shi2009a} and look further into the links between the Lee model and the multi-mode Jaynes-Cummings Hamiltonian that describes photon-qubit scattering. We will use a resolvent (i.e.\ Green's function) based analysis method, different than the Schr\"odinger equation based one used in \cite{Shi2009a}. Our analysis will require us to utilize multi-channel scattering formalism to come up with formulas for predicting elements of the scattering matrix for the case of two-photons interacting with a qubit in a dispersive environment. In our quest to obtain the $S$-matrix elements, we will derive the rules for Feynman diagrams and use the diagrams to understand the scattering pathways. We will also provide the analysis with which one can correlate the results from Krylov-subspace based numerical simulations with pulsed excitations, and the $S$-matrix elements so as to obtain an independent verification of the formulas we derive.

In Sec.\ \ref{sec:definition} we will make certain definitions used throughout the manuscript. In Sec.\ \ref{sec:onephoton} we will derive one-photon scattering results and introduce the machinery of Feynman diagrams. In Sec.\ \ref{sec:twophoton} we will extend the results to the case of two-photons. We will introduce the concept of a `channel' and derive the scattering matrix elements in between every possible channel in Sec.\ \ref{sec:multichannel}. In Sec.\ \ref{sec:results} we test our formalism by comparing it against pulse based numerical scattering simulations. We discuss our results in Sec.\ \ref{sec:discuss} and conclude in Sec.\ \ref{sec:conclusion}.

Very recently, analysis of one- and two-photon scattering in a waveguide with an arbitrary dispersion relationship is investigated in \cite{Schneider2016} through the use of path-integrals and Feynman-diagrams. We will be referring to \cite{Schneider2016} in the upcoming sections.

\section{Definitions} \label{sec:definition}
We are interested in the coupling of a two-level atom, also referred to as a qubit, to a single-mode waveguide. The Hamiltonian for the system is of a multi-mode Jaynes-Cummings type \cite{*[{}] [{ Sec. 24.}] Shore1993}. A real-space, discrete version of the Hamiltonian can be obtained by treating the waveguide as composed of a linear array of cavities connected to  each other in a 1D lattice with a tight-binding coupling between neighboring cavities \cite{Zhou2008,Longo2009,Lombardo2014}. The discrete version of the Hamiltonian, $H_\text{d}$, is given by $(\hbar=1)$
\begin{align}
H_\text{d}= -J\sum_{x=1}^N \Bigl(\hat{a}^\dag_{x+1} \hat{a}_x + \hat{a}^\dag_x \hat{a}_{x+1}\Bigr) + \frac{\Omega}{2} \sigma_z + g' (\sigma^+ \hat{a}_0 + \hat{a}^\dag_0 \sigma^-). \label{eq:Hd}
\end{align}
Here, $\hat{a}^\dag_x$ ($\hat{a}_x$) are the creation (annihilation) operators for photons at lattice site $x$, $J$ is the coupling constant between neighboring cavities, $\Omega$ is the level separation for the qubit, $\sigma_i$ are the Pauli spin matrices, $g'$ is the constant coupling constant between the qubit at $x=0$ and the photons, $\sigma^\pm$ are the raising/lowering operators for the qubit, $N$ is the lattice size. We take the separation among coupled cavities as $a=1$. $H_\text{d}$ can be written in terms of free propagating photon states with a wave vector $k$,
\begin{align}
\ket{k} &= \frac{1}{\sqrt{N}} \sum_x \me^{\mi k x} \ket{x}, \nonumber \\
\intertext{to get}
H_\text{d}&=\sum_k \omega_k \hat{a}^\dag_k \hat{a}_k + \frac{\Omega}{2} \sigma_z + \frac{g'}{\sqrt{N}} \sum_k (\sigma^+ \hat{a}_k + \hat{a}^\dag_k \sigma^-), \nonumber
\intertext{where}
\label{eq:wk}\omega_k & = -2J \cos k.
\end{align}
We take the large $N$ limit with diminishing $\Delta_k=2\pi/N$, to transition to a continuous set of $k$ values by changing the sums to integrals via $\Delta_k \sum_k \rightarrow \int\dif k$, and the discrete operators to continuous ones with $\hat{a}_k/\sqrt{\Delta_k}\rightarrow a_k$ to get \cite{Fan2010}
\begin{align}
\label{eq:Hamiltonian}H = \underbrace{\int_{-\pi}^\pi \dif k \omega_k a^\dag_k a_k + \frac{\Omega}{2} \sigma_z}_{H_0} +\underbrace {g \int_{-\pi}^\pi\dif k (\sigma^+ a_k + a^\dag_k \sigma^-)}_V,
\end{align}
in which the continuous coupling parameter $g$ is related to the discrete one as 
\begin{align}
\label{eq:gp}g=\frac{g'}{\sqrt{2\pi}}.
\end{align}
We separate the Hamiltonian into a non-interacting part, $H_0$, and a coupling part, $V$, as well. 

We define the resolvent of the Hamiltonian as \cite{Cohen-Tannoudji1992}
\begin{align*}
G(z) = \frac{1}{z-H}.
\end{align*} The resolvent is closely related to the time evolution operator via
\begin{align*}
U(t>0)=\lim_{\eta\rightarrow 0^+}\frac{1}{2\pi\mi}\int_{+\infty}^{-\infty}\me^{-\mi x t} G(x+\mi \eta).
\end{align*}
The time evolution operator $U(t>0)$ is effectively the inverse Laplace transform of $\mi G(\mi s)$ such that
\begin{align*}
U(t>0)=\mathscr{L}^{-1}\{\mi G(\mi s)\}(t),
\end{align*}
which can be seen from the Bromwich integral representation of $\mathscr{L}^{-1}$. The resolvent operator also satisfies the Lippmann-Schwinger equation for $G(z)$\footnote{See Chap. 8 of \cite{Taylor2006}.}
\begin{align}\label{eq:LS_G1}
G(z) & = G_0(z)+G_0(z) V G(z), \\ 
	 & = G_0(z) + G(z) V G_0(z),\label{eq:LS_G2} \intertext{where} 
\label{eq:G0def}G_0(z) & = \frac{1}{z-H_0}. 
\end{align}

\section{One-Photon Scattering, Bound States, Feynman Diagram Representation} \label{sec:onephoton}

We will first start by calculations in the one-photon sector. The notation we use is borrowed from \cite{Maxon1965} in which the Lee model is solved in its lowest excitation sector. There are four relevant matrix elements of $G(z)$\footnote{These four functions are analogous to the four $\hat{\tau}$ functions in \cite{Maxon1965}.}
\begin{align*}
G_1(z) & \equiv \braopket{{\uparrow}}{G(z)}{{\uparrow}}, \\
G_2(z;k) & \equiv \braopket{k{\downarrow}}{G(z)}{{\uparrow}}, \\
G_3(z;k) & \equiv \braopket{{\uparrow}}{G(z)}{k{\downarrow}}, \\
G_4(z;p,k) & \equiv \braopket{p{\downarrow}}{G(z)}{k{\downarrow}}.
\end{align*}
Here $\ket{\ua}$ and $\ket{\da}$ represent the excited and ground states of the qubit. Through the use of \eqref{eq:LS_G1}-\eqref{eq:LS_G2} we get coupled equations among the matrix elements of $G(z)$. To do so observe that 
\begin{align}
\begin{split}\label{eq:G0use}
\braopket{p{\downarrow}}{G_0(z)}{k{\downarrow}} &= \frac{\delta(k-p)}{z-\omega_k+\Omega/2}, \\
\braopket{{\uparrow}}{G_0(z)}{{\uparrow}} &= \frac{1}{z-\Omega/2},
\end{split}
\end{align}
since $\ket{k\da}$ and $\ket{\ua}$ are eigenstates of $H_0$.
Additionally, the identity operator in the one-photon sector, $\openone_\text{1P}$, is given by
\begin{align}
\label{eq:Identity1}\openone_\text{1P}=\ket{{\uparrow}}\bra{{\uparrow}} + \int_{-\pi}^\pi \dif k \ket{k}\bra{k}.
\end{align}
By sandwiching \eqref{eq:Identity1} in between $V$ and $G$ in \eqref{eq:LS_G1} and \eqref{eq:LS_G2}, using \eqref{eq:G0use} one can derive 
\begin{align*}
G_1(z) & = \frac{1}{z-\Omega/2} \left( 1+ g \int_{-\pi}^\pi \dif k G_2(z;k) \right), \\
G_2(z;k) & = \frac{g G_1(z)}{z+\Omega/2 - \omega_k} = G_3(z;k), \\
G_4(z;p,k) & = \frac{\delta(k-p)+g G_3(z;k)}{z+\Omega/2-\omega_p}.
\end{align*}
Substitution of $G_2(z)$ into the expression for $G_1(z)$ gives us
\begin{align*}
G_1(z) = \left( z - \Omega/2-g^2 \int_{-\pi}^\pi \dif k \frac{1}{z+\Omega/2 - \omega_k}  \right)^{-1}.
\end{align*}
We call the integral in the expression above as 
\begin{align}
\label{eq:Iz}I(z) \equiv  \int_{-\pi}^\pi \dif k \frac{1}{z - \omega_k},
\end{align}
and with this definition the solutions to other matrix elements become
\begin{align}
\label{eq:G1}G_1(z) & = \frac{1}{z - \Omega/2-g^2 I(z+\Omega/2)}, \\
G_2(z;k) & = \frac{g}{(z+\Omega/2-\omega_k)}G_1(z), \nonumber \\
G_3(z;k) & = G_2(z;k), \nonumber \\
\label{eq:G4}G_4(z;p,k) & = \frac{\delta(k-p)}{(z+\Omega/2-\omega_k)} \\ & \quad + \frac{g^2 G_1(z)}{(z+\Omega/2-\omega_k)(z+\Omega/2-\omega_p)}.\nonumber
\end{align}
Properties of $I(z)$ are summarized in Appendix \ref{App:Iz}.

\begin{figure}
\includegraphics{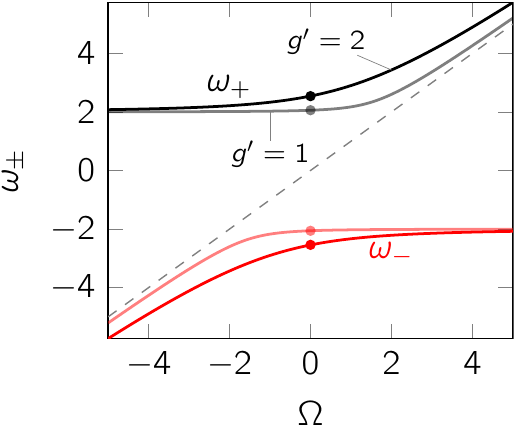}
\caption{(Color online) $\omega_\pm$ as a function of $\Omega$ for $g'=1$ and $g'=2$. $J=1$ taken. The values at $\Omega=0$, are highlighted with filled circles.} \label{fig:BoundStateEnergy}
\end{figure}

\begin{figure}
\includegraphics{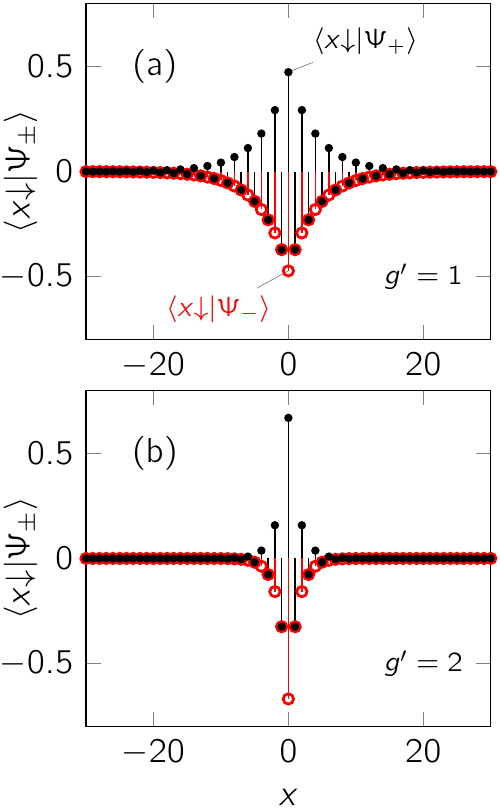}
\caption{(Color online) Photonic part of the bound state as a function of $x$ for (a) $g'=1$ and (b) $g'=2$, $\Omega=0$, $J=1$ taken. Black ($\bullet$) and red ($\circ$) curves refer to the $+$ and $-$ bound states, respectively. Description of the bound states is available in Appendix \protect\ref{App:bound}.} \label{fig:BoundStateSpace}
\end{figure}

The poles of $G_1(z)$ are related to the bound states of the system. There are two poles with energies $\omega_\pm$ outside the $[-2J,2J]$ band. Properties of these two bound states for the $\Omega=0$ case were investigated in \cite{Lombardo2014}. They are also summarized in Appendix \ref{App:bound} for completeness. When we work with bound states, we will also take $\Omega=0$ due to the simplicity of the closed form formulas for $\omega_\pm$ in that particular case. In Fig \ref{fig:BoundStateEnergy} we plot $\omega_\pm$ as a function of $\Omega$ for different $g'$ values. The special case for $\Omega=0$ is highlighted in the figure with filled circles. At large $\abs{\Omega}$ values one of the poles get  close to $\Omega$ whereas the other one converges to a value near $\pm 2J$, but still remaining outside of the band of allowed states, i.e. $\abs{\omega_\pm} > 2J$ for all $\Omega$. In Fig \ref{fig:BoundStateSpace} we present the photonic part of the bound state for the $\Omega=0$ case for two values of $g'$. As can be seen from the figures, a higher $g'$ value leads to a tighter confinement of the photons around the qubit.

\begin{figure}
\includegraphics{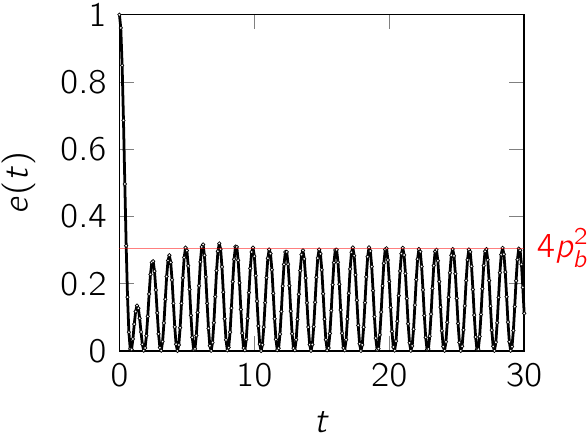}
\caption{(Color online) $e(t)$ for $g'=2$ with $\Omega=0$ and $J=1$. } \label{fig:BoundStateTime}
\end{figure}

The spontaneous emission from an initially excited atom at $t=0$ is of the form of an exponential decay in a non-dispersive waveguide where the photon energy and the $k$-vector are linearly proportional \cite{Rephaeli2010}. However, in a structured photon medium with dispersion, the picture changes \cite{John1994}. We can calculate the time evolution, $e(t)$, of an initially excited atom (with $\Omega=0$) as, 
\begin{align}\label{eq:time_evolve}\begin{split}
e(t) & = \braopket{\ua}{U(t>0)}{\ua} = \mathscr{L}^{-1}\{\mi G_1(\mi s)\}(t) \\
     & = \mathscr{L}^{-1}\left\{ \frac{1}{s+{g'^2}/{\sqrt{4 J^2+s^2}}} \right\}(t) \\
     & = \mathscr{L}^{-1}\left\{ \frac{s(s^2+4J^2)}{s^4+4J^2s^2-g'^4} - \frac{g'^2\sqrt{s^2+4J^2}}{s^4+4J^2s^2-g'^4} \right\}(t).\end{split} 
\end{align}
The first term of the last line can easily be inverted by partial fraction expansion. The second term can be written as a convolution integral in time-domain. Details are left to Appendix \ref{App:spontaneous}. In Fig \ref{fig:BoundStateTime}\footnote{This figure is the same as Fig 2(e) in \cite{Lombardo2014} and verifies our formalism in the one-photon sector. See equation (27) in \cite{Lombardo2014} for the derivation of the $4p_b^2$ limiting value.} we plot $e(t)$ for $g'=2$. As seen in the figure, the evolution of atomic excitation is far from being an exponential decay and shows constant oscillations at steady state. These oscillations are a signature of the excitation of the atom-photon bound state in which the atomic excitation probability is trapped at the fixed value $4p_b^2$ with a corresponding photonic cloud around it as in Fig \ref{fig:BoundStateSpace}.

Now that we have an understanding of the link between time evolution and the resolvent, in principle we can start from an asymptotic state in the infinite past and use the time evolution operator to arrive at an asymptotic state in the infinite future to get the scattering matrix element in between those two states.\footnote{Such a calculation is available on pp. 222-225 of \cite{Cohen-Tannoudji1992}.} However, a more concise way would be to make use of the relationship between the $S$-matrix and the $T$ operator given as
\begin{align}
\label{eq:S1photon}\braopket{p\da}{S}{k\da}&=\braket{p}{k}\\
&-2\pi\mi \, \delta(\omega_p-\omega_k)\lim_{\eta\rightarrow 0^+}\braopket{p\da}{T(\omega_p -\Omega/2+\mi\eta)}{k\da}, \nonumber
\end{align}
where $T$ satisfies the Lippmann-Schwinger equation
\begin{align}
\label{eq:Tz}T(z) = V + V G_0(z) T(z).
\end{align}
$T(z)$ and $G(z)$ are intimately related via 
\begin{align}
T(z) &= V + V G(z) V, \nonumber \\
\label{eq:GtoT}G(z) &= G_0(z) + G_0(z) T(z) G_0(z),
\end{align}
hence the knowledge of one is sufficient to obtain the other.

\begin{figure}
\includegraphics{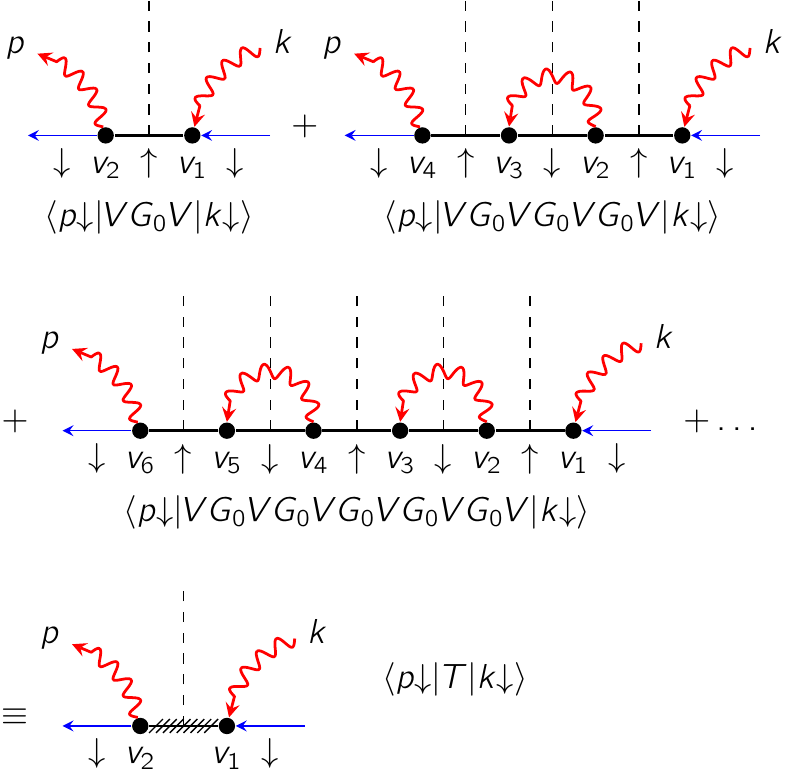}
\caption{(Color online) Feynman graphs for the first three non-zero elements of $\braopket{p\da}{T}{k\da}$ are labeled with the corresponding matrix element underneath each graph. Formal summation of all graphs lead to $\braopket{p\da}{T}{k\da}$ represented by the last graph.} \label{fig:OnePhotonFeynman}
\end{figure}

Due to the close relationship between $S$ and $T$ matrix elements, it is of use to get a series representation of $T(z)$. To do so, we use \eqref{eq:Tz}. We first ignore the  $T(z)$ term on the right hand side and set $T(z)\approx V$, the inhomogeneous term. We then use this expression as an approximation to $T(z)$ on the right hand side to get $T(z)\approx V + V G_0 V$. We continue the iteration process to get an infinite series representation of $T(z)$ as
\begin{align}
\label{eq:Tz_expansion}\begin{split}T(z) = V + V G_0(z) V + V G_0(z) V G_0(z) V \\ + V G_0(z) V G_0(z) V G_0(z) V + \ldots \end{split}
\end{align}
To get the $S$-matrix elements, we need to put \eqref{eq:Tz_expansion} in between the bra and the ket associated with the initial and final asymptotic states. For one-photon scattering those states are denoted by $\bra{p\da}$ and $\ket{k\da}$. Terms with an odd number of $V$ operators in \eqref{eq:Tz_expansion} result in a zero value for this particular choice of initial and final states. We therefore choose terms with only an even power of $V$ in the expansion. To calculate those matrix elements, we need to contract a progressively larger number of creation and annihilation operators. Luckily, in the one-photon sector, there is only a single way to form the contraction. We can denote those non-zero contractions in terms of Feynman graphs as shown in Fig \ref{fig:OnePhotonFeynman}. The graphs go from right to left to make the connection with the underlying matrix element in the bra-ket notation more clear. Our graphical representation is a slightly modified version of previously published results in impurity scattering.\footnote{See p. 227 of \cite{Cohen-Tannoudji1992} and p. 17, Fig 1.6(ii) of \cite{Hewson1993}. We merge the two representations to draw Fig \ref{fig:OnePhotonFeynman}. In \cite{Schneider2016} Feynman   diagrams are drawn in an alternative manner, for the infinite series representation obtained for $G(z)$.}

Horizontal straight lines denote the state of the qubit, labeled underneath the line. Wavy lines represent the photons. Each intersection of a photon line with a qubit line represents an interaction term, marked with a filled circle at the vertex. The number of vertices correspond to the order of the expansion. When a photon line with a label $m$ terminates at a vertex with label $v$, the interaction is of the type $g \int \dif v \sigma^+ a_v$ which represents the absorption of a photon and the excitation of the qubit, whereas, when a photon line exits a vertex, there is an interaction with $g \int \dif v a^\dag_v \sigma^-$ that shows the emission of a photon with label $v$ from the qubit and a lowering of the qubit state. These terms come from the definition of $V$ in \eqref{eq:Hamiltonian}.

There are external vertices to the right and to the left of the diagrams. These vertices represent the incoming and outgoing state of the photons, respectively. External vertices only have labels such as $k,p$. Initial and final states of the qubit are differentiated by blue lined arrows going towards the first internal vertex, and going away from the last internal vertex. 

Vertical dashed lines, in between the internal vertices, represent the effect of the $G_0$ operators in the matrix element. We look at the intersection of photon lines and the qubit line with the dashed line to get the state, and thus the energy of the system at the dashed line position. We associate a factor $(z-\sum \omega_n \pm \Omega/2)^{-1}$ with each dashed line: $\sum \omega_n$ for the total energy of all photons intersecting with the dashed line, $\pm \Omega/2$ for the state of the qubit. This factor comes directly from the definition of $G_0$ in \eqref{eq:G0def}. 

For a given Feynman graph, the rules for constructing the associated matrix element value are given below.
{\begin{enumerate}
\item Photons carry the label of the internal or external vertex they originate from.
\item Associate a factor of $g$ for each internal vertex.
\item Associate a $\delta(m-v)$ term for photons with a label $m$ terminating at an internal or an external vertex $v$.
\item Draw dashed lines in between internal vertices and for each dashed line, associate a $(z-\sum \omega_n \pm \Omega/2)^{-1}$ term as described above.
\item Integrate over all internal vertex variables.
\end{enumerate}
The application of these rules to the diagrams in Fig \ref{fig:OnePhotonFeynman} leads to
\begin{widetext}
\begin{align*}
& \iint_{-\pi}^\pi \dif v_1 \dif v_2 g^2 \frac{\delta(k-v_1)\delta(p-v_2)}{z-\Omega/2} 
+ \iiiint_{-\pi}^\pi \dif v_1 \dif v_2 \dif v_3 \dif v_4 g^4 \frac{\delta(k-v_1)\delta(v_2-v_3) \delta(p-v_4)}{(z-\Omega/2)(z+\Omega/2-\omega_{v_2})(z-\Omega/2)} \\
& + \idotsint_{-\pi}^\pi  \dif v_1 \dots \dif v_6 g^6 \frac{\delta(k-v_1)\delta(v_2-v_3) \delta(v_4-v_5) \delta(p-v_6)}{(z-\Omega/2)(z+\Omega/2-\omega_{v_2})(z-\Omega/2)(z+\Omega/2-\omega_{v_4})(z-\Omega/2)} + \dots \\
& = \frac{g^2}{z-\Omega/2}\left[ 1 + \frac{g^2 I(z+\Omega/2)}{z-\Omega/2}+ \left( \frac{g^2 I(z+\Omega/2)}{z-\Omega/2}\right)^2 + \dots  \right] \\
& = \frac{g^2}{z-\Omega/2-g^2 I(z+\Omega/2)} \equiv \braopket{p\da}{T(z)}{k\da}.
\end{align*}
\end{widetext}
In the last two lines, we integrate out the Dirac delta functions, and employ the definition of $I(z)$ from \eqref{eq:Iz}. We see that all the internal photon contributions, in the shape of `bubbles,' can be summed to come up with a closed form expression for $\braopket{p\da}{T}{k\da}$. We graphically denote the summation of all internal bubble diagrams with a hatched qubit line as shown on the last line in Fig \ref{fig:OnePhotonFeynman}. Since the hatched line contribution includes both excited and ground states of the qubit, no particular qubit label is put underneath it. The term associated with the summed $G_0$ propagator---a dashed line intersecting the hatched line---with $\ket{\da}$ qubit states on both sides is given by $[z-\Omega/2-\sum \omega_n-g^2 I(z+\Omega/2-\sum \omega_n)]^{-1}$. In the one-photon case, $\sum \omega_n$ terms  do not show up, however, those terms will be important when there are two or more photons. The summed $G_0$ propagator can be used directly when deriving matrix elements and will be utilized in the two-photon sector.

Our derivation of the matrix element $\braopket{p\da}{T}{k\da}$ was through the use of \eqref{eq:Tz_expansion}. On the other hand, we could have gotten the same result by noting that we have a closed form formula for $G_4$ in \eqref{eq:G4}. It is easy to transition from $G_4$ to $\braopket{p\da}{T}{k\da}$ via \eqref{eq:GtoT}. Even though it would have been faster to obtain $\braopket{p\da}{T}{k\da}$ in this alternative way, the use of \eqref{eq:Tz_expansion} allowed us to derive the Feynman diagram representations which will be invaluable in two-photon scattering analysis in the following sections.

At this moment, we have all the information necessary to calculate the one-photon scattering matrix. By using the properties of $I(z)$ from Appendix \ref{App:Iz} we can write the on-shell,\footnote{Note that we calculated $\braopket{p\da}{T(z)}{k\da}$ for an arbitrary $z$ value. To transition to the relevant $S$-matrix element in \eqref{eq:S1photon} we need the value at $z=\omega_k-\Omega/2$, the energy value of the incoming state $\ket{k\da}$, which leads to the terminology ``on the energy shell'' or abbreviated as ``on-shell.'' Through the use of the Dirac delta function in \eqref{eq:S1photon}, incoming and outgoing energies are set equal to each other.} that is $z=\omega_k-\Omega/2$, amplitude of the $T$-matrix element as
\begin{align*}
\lim_{\eta\rightarrow 0^+}&\braopket{p\da}{T(\omega_k- \Omega/2+\mi \eta)}{k\da} \\
&\qquad = \frac{-\mi g^2 2J \abs{\sin k}}{g'^2+\mi (2J \cos k +\Omega )2J \abs{\sin k}},
\end{align*}
where $g'$ and $g$ are related via \eqref{eq:gp}. In \eqref{eq:S1photon}, $\braket{p}{k}$ term introduces a $\delta(k-p)$ factor. We convert  $\delta(\omega_p-\omega_k)$ by the properties of Dirac delta function and \eqref{eq:wk} as
\begin{align}
\label{eq:dirac_delta}\delta(\omega_p-\omega_k)& =\frac{\delta(k-p)}{2J \abs{\sin k}} + \frac{\delta(k+p)}{2J \abs{\sin k}}, \\
\intertext{to get}
\braopket{p\da}{S}{k\da} & = t_k \delta(k-p) + r_k \delta(k+p),\nonumber \\
\intertext{where}
\label{eq:reflection} r_k & = \frac{-g'^2}{g'^2+\mi (2J \cos k +\Omega)2J \abs{\sin k}}, \\
\label{eq:transmission} t_k & = 1+r_k.
\end{align}
These equations agree with those in \cite{Zhou2008}.

Our formalism can also be used to derive results for single photon scattering in a waveguide with a linear dispersion relationship \cite{Shen2007}. When the dispersion is linear, $I(z)$ becomes a constant \cite{Pletyukhov2012}. The interested reader can look at \cite{Schneider2016} where the correspondence between linear and arbitrary dispersion analysis is made in detail.

\section{Two-Photon Scattering}\label{sec:twophoton}
Now that we have analyzed one-photon scattering, we will move into the two-photon domain. Following \cite{Maxon1966} we define\footnote{The four functions $G_{5}$ through $G_8$ correspond to $\hat{\tau}^5$ through $\hat{\tau}^8$ in \cite{Maxon1966}.} the matrix elements in the two-photon sector as
\begin{align*}
G_5(z;p,k) & \equiv \braopket{p\ua}{G(z)}{k\ua}, \\
G_6(z;p_1,p_2,k) & \equiv \braopket{p_1p_2\da}{G(z)}{k\ua}, \\
G_7(z;p,k_1,k_2) & \equiv \braopket{p\ua}{G(z)}{k_1k_2\da}, \\
G_8(z;p_1,p_2,k_1,k_2) & \equiv \braopket{p_1p_2\da}{G(z)}{k_1k_2\da}.
\end{align*}
We also need the identity operator in the two-photon subspace given by
\begin{align}
\label{eq:Identity2}\openone_\text{2P}= \frac{1}{2!}\iint_{-\pi}^\pi \dif k_1 \dif k_2 \ket{k_1k_2\da}\bra{k_1k_2\da} + \int_{-\pi}^\pi \dif k \ket{k\ua}\bra{k\ua}.\:\:\:
\end{align}
The term $\frac{1}{2!}$ arises due to the orthogonality condition $\braket{p_1p_2}{k_1k_2}=\delta(p_1-k_1)\delta(p_2-k_2)+\delta(p_1-k_2)\delta(p_2-k_1)$ and the need to have a unity operator when acting on an arbitrary two photon state.

Through the use of \eqref{eq:LS_G1} and sandwiching \eqref{eq:Identity2} in between $V$ and $G$ we get
\begin{align}
\label{eq:G_5}\begin{split}G_5(z;p,k) &= \frac{\delta(p-k)}{z-\Omega/2-\omega_p} \\
&\quad +\frac{g}{z-\Omega/2-\omega_p}\int_{-\pi}^\pi\dif p_i G_6(z;p,p_i,k).
\end{split}
\end{align}
Similarly we get
\begin{align}
\label{eq:G_6}G_6(z;p_1,p_2,k) &= g\frac{G_5(z;p_1,k) + G_5(z;p_2,k)}{z-\omega_{p_1}-\omega_{p_2}+\Omega/2}, \\
\label{eq:G_7}G_7(z;k,p_1,p_2) & = g\frac{G_5(z;k,p_1) + G_5(z;k,p_2)}{z-\omega_{p_1}-\omega_{p_2}+\Omega/2}, \\
\label{eq:G_8}G_8(z;p_1,p_2,k_1,k_2) & = g\frac{G_7(z;p_1,k_1,k_2) + G_7(z;p_2,k_1,k_2)}{z-\omega_{p_1}-\omega_{p_2}+\Omega/2}.
\end{align}
As seen above, the knowledge of $G_5$ is sufficient to obtain the remaining matrix elements. We substitute \eqref{eq:G_6} into \eqref{eq:G_5} and get
\begin{align*}
\bigl( z-\Omega/2 - \omega_p -g^2 I(z+\Omega/2-\omega_p) \bigr) G_5(z;p,k) \\
\quad = \delta(k-p) + g^2 \int_{-\pi}^\pi \dif p_i \frac{G_5(z;p_i,k)}{z+\Omega/2-\omega_p-\omega_{p_i}}.
\end{align*}
The equation above can be written via $z\rightarrow z-\Omega/2$ as
\begin{align*}
\bigl( z-\Omega - \omega_p -g^2 I(z-\omega_p) \bigr) G_5(z-\Omega/2;p,k) \\
\quad = \delta(k-p) + g^2 \int_{-\pi}^\pi \dif p_i \frac{G_5(z-\Omega/2;p_i,k)}{z-\omega_p-\omega_{p_i}}.
\end{align*}
Let 
\begin{align}
\label{eq:Hz}H(z;p)=z-\Omega - \omega_p -g^2 I(z-\omega_p).
\end{align}
Now we will define
\begin{align}
\label{eq:U_def}U(z;p,k) = \frac{H(z;k)\left[H(z;p)  G_5(z-\Omega/2;p,k)-\delta(k-p)\right]}{g^2}.
\end{align}
As a result we get the following integral equation\footnote{This equation is analogous to equation (13) in \cite{Maxon1966}.}
\begin{align}
\label{eq:U_sing_inteq} U(z+\mi \eta;p,k)&=\frac{1}{z-\omega_{p}-\omega_{k}+ \mi \eta} \\
&+ g^2 \int_{-\pi}^\pi \dif p_i \frac{U(z;p_i,k)}{H(z+\mi\eta;p_i)(z-\omega_{p}-\omega_{p_i}+\mi \eta)}.\nonumber
\end{align}
The equation above is an integral equation in the variable $p$ for fixed $z$ and $k$. We will see in the following sections that the limit $\eta \rightarrow 0^+$ is to be taken, for a variety of $z$ values both on and off the energy shell. The solution of $U(z;p,k)$ gives the entire picture in the two-photon sector. However, in order to calculate the scattering matrix elements we will first need to understand multichannel scattering theory.

\section{Multichannel Scattering}\label{sec:multichannel}

Unlike one-photon scattering, when two photons are present the system has enough energy to excite the atom-photon bound states with energies $\omega_\pm$. Processes---with no analogues in the one-photon sector---such as two photons coming in, one photon getting trapped in a bound state, the other photon scattering away, become possible. In the two-photon sector, there are two different sets of stable states that can be the incoming or outgoing asymptotic states in a scattering matrix description: a) Two free photons with the qubit in its ground state, b) One free photon with the other photon forming a bound state with the qubit. We refer to these two sets as two different `channels' \cite{Newton1982,Taylor2006}. From now on, the first set with two free photons will be called as channel 0, the second set as channel 1. We will use results from three-particle scattering analysis \cite{Lovelace1964,Alt1967,Osborn1971}. We will see that off-shell values of the two-photon matrix elements will be required to describe multichannel scattering \cite{Lovelace1964}.

The multichannel $S$-matrix is given by transitions among different possible scattering states. Channel 0 states are denoted by $\ket{k_1k_2\da}$ with two free photons ($k_1$ and $k_2$) and qubit in the ground state ($\da$). Channel 1 states are $\ket{k\Psi_\pm}$ with one free photon ($k$) and a photon-qubit bound state (${\Psi_\pm}$) with energy $\omega_\pm$. In this part and the following two subsections, we will be taking $\Omega=0$ as in \cite{Lombardo2014}, in order to have a simpler algebra. For $\Omega=0$ we have $\omega_+ = -\omega_-$. There are three possible sets of transitions between the two channels; from channel 0 to 0, 0 to 1 and 1 to 1. We will call these free to free, free to bound and bound to bound scattering matrix elements. The multichannel scattering matrix element is given by 
\begin{align}
\label{eq:S_multi}\braopket{\phi_f}{S}{\phi_i} & = \braket{\phi_f}{\phi_i} \\
& \quad - 2\pi\mi \delta(E_i-E_f) \lim_{\eta\rightarrow0^+}\braopket{\phi_f}{U_{fi}(E_i+\mi\eta)}{\phi_i},\nonumber
\end{align}
where $\ket{\phi_j}$ is an arbitrary asymptotic state in channel $j\in\{0,1\}$ with energy $E_j$ and $U_{fi}(z)$ are channel dependent transition operators. The relationship between the resolvent, $G(z)$, and $U_{fi}(z)$ is given by\footnote{See equation (III.23) in \cite{Osborn1971} and (2.11) in \cite{Alt1967}, and note the sign in our definition of $G(z)$.}
\begin{align}
\label{eq:G_to_U}G(z) = G_f(z) \delta_{f i}+G_f(z)U_{f i}(z)G_i(z),
\end{align}
where $G_j(z) = (z-H_j)^{-1}$. $G_0(z)$ is as previously defined in \eqref{eq:G0def}. $G_1(z)$ has $H_1=H_0+V'$ where $V'$ has a similar form to $V$ but redefined to act only on the bound state part of the wave function of channel 1, and not on the free photon part. Let us analyze the result of the operation $H_1\ket{k\Psi_\pm}$. Expanding $\ket{\Psi_\pm}$ via  \eqref{eq:bound_p} we get
\begin{align*}
&H_1\ket{k\Psi_\pm} = (H_0+V')\left( \sqrt{p_b}\ket{k\ua} + \sqrt{p_b} g \int_{-\pi}^\pi \frac{\dif p \ket{kp\da}}{\omega_\pm - \omega_p} \right) \\
 &= \omega_k \sqrt{p_b}\ket{k\ua} \\
 &\quad + \sqrt{p_b} g \int_{-\pi}^\pi \dif p \frac{(\omega_k+\omega_p)\ket{kp\da}}{\omega_\pm - \omega_p} + \ket{k} \otimes V'\ket{\Psi_\pm} \\
 &= \omega_k \ket{k\Psi_\pm} + \ket{k} \otimes H_1 \ket{\Psi_\pm} = (\omega_k+\omega_\pm) \ket{k\Psi_\pm},
\end{align*}
where we used the fact that the bound state is an eigenstate of $H_1$. As a result, we arrive at
\begin{align}
\label{eq:G_1}G_1(z)^{-1}\ket{k\Psi_\pm} = (z-\omega_k-\omega_\pm)\ket{k\Psi_\pm}.
\end{align}
We are now ready to calculate the multichannel scattering matrix elements.

\subsection{Bound to Bound Scattering}
In bound to bound scattering, the free photon scatters off of the atom-photon bound state while the bound state remains intact during the scattering process. We will be interested in calculating the reflection and transmission coefficients for the free photon, similar to the one-photon scattering case. We note that, during bound to bound scattering, the bound state cannot change its energy from $\omega_+$ to $\omega_-$ or vice versa. In such a case it would be impossible to satisfy energy conservation $\omega_k + \omega_\pm = \omega_p + \omega_\mp$ since  $\abs{\omega_\pm-\omega_\mp} = 2 \abs{\omega_\pm} > 4J$ whereas $\abs{\omega_{p,k}}\le 2J$. Therefore, the general form of the bound to bound scattering matrix element can be obtained from \eqref{eq:S_multi} as
\begin{align*}
\braopket{p\Psi_\pm}{S}{k\Psi_\pm} &= \delta(p-k)-2\pi \mi \delta(\omega_p-\omega_k)\\
& \quad \times \lim_{\eta\rightarrow 0^+}\braopket{p\Psi_\pm}{U_{11}(\omega_k+\omega_\pm+\mi \eta)}{k\Psi_\pm}. 
\end{align*}
We will drop $\lim_{\eta\rightarrow0^+}$ expression in the remaining parts of this section to simplify the notation. In order to calculate the $U_{11}$ matrix element we begin by rewriting \eqref{eq:G_to_U} as
\begin{align*}
U_{11}(z) = G_1^{-1}(z) G(z) G_1^{-1}(z)-G_1^{-1}(z),
\end{align*}
and hence
\begin{align}
\label{eq:U11}\begin{split}\braopket{p\Psi_\pm}{U_{11}(z)}{k\Psi_\pm} = (z-\omega_p-\omega_\pm)(z-\omega_k-\omega_\pm)\\
\times \braopket{p\Psi_\pm}{G(z)}{k\Psi_\pm} - \delta(p-k) (z-\omega_p-\omega_\pm),\end{split}
\end{align}
by the use of \eqref{eq:G_1}. Additionally, since we are interested in the on energy shell limit $z\rightarrow \omega_p+\omega_\pm=\omega_k+\omega_\pm$ the Dirac delta term above will drop down as it will be multiplied by a zero factor. Similarly, we will only be interested in those terms in $\braopket{p\Psi_\pm}{G(z)}{k\Psi_\pm}$ that have a double pole at $z=\omega_p+\omega_\pm$, the remaining terms will automatically go to zero due to the presence of the $(z-\omega_p-\omega_\pm)(z-\omega_k-\omega_\pm)$ multiplier.

We expand $\bra{p\Psi_\pm}$ and $\ket{k\Psi_\pm}$ by the use of \eqref{eq:bound_p} for $\Omega=0$ and get 
\begin{align}
&\braopket{p\Psi_\pm}{G(z)}{k\Psi_\pm} = \underbrace{p_b \braopket{p\ua}{G(z)}{k\ua}}_{\text{Term 1}}\nonumber \\
&+\underbrace{p_b g \int_{-\pi}^\pi \dif k' \frac{\braopket{p\ua}{G(z)}{kk'\da}}{\omega_\pm+2J \cos k' }}_{\text{Term 2}} 
+ \underbrace{p_b g \int_{-\pi}^\pi \dif p' \frac{\braopket{pp'\da}{G(z)}{k\ua}}{\omega_\pm+2J \cos p' }}_{\text{Term 3}} \nonumber \\
&+ \underbrace{p_b g^2 \iint_{-\pi}^\pi \dif p' \dif k' \frac{\braopket{pp'\da}{G(z)}{kk'\da}}{(\omega_\pm+2J \cos p' )(\omega_\pm+2J \cos k' )}}_{\text{Term 4}}.\label{eq:Gz_terms}
\end{align}
We will now investigate the four terms one by one. The first term can be written via \eqref{eq:U_def} as
\begin{align}
\label{eq:Gz_pk_to_U}\braopket{p\ua}{G(z-\Omega/2)}{k\ua} = \frac{g^2 U(z;p,k)}{H(z;k)H(z;p)}+\frac{\delta(k-p)}{H(z;p)}.
\end{align}
Note that
\begin{align}
\label{eq:H_residue}\lim_{z\rightarrow \omega_p+\omega_\pm}\frac{z-\omega_p-\omega_\pm}{H(z;p)}=p_b,
\end{align}
therefore the contribution of the first term to the on-shell bound to bound transmission matrix element of \eqref{eq:U11} by the use of \eqref{eq:Gz_terms}, \eqref{eq:Gz_pk_to_U} and \eqref{eq:H_residue} becomes (remember that $\Omega=0$)
\begin{align*}
\text{Term 1}= \quad p_b^3 g^2 U(z;p,k).
\end{align*}
For the second term we use \eqref{eq:G_7} to write it in terms of $G_5$. We then follow the same path as in the  calculation of the first term to get 
\begin{align*}
\text{Term 2}= \quad p_b^3 g^4 \left( \int_{-\pi}^\pi \frac{\dif k'}{(\omega_\pm+2 J \cos k' )^2} \right)U(z;p,k).
\end{align*}
Term 3 has the same value as term 2. For the last term we make use of \eqref{eq:G_8} and get
\begin{align*}
\text{Term 4}= \quad p_b^3 g^6 \left( \int_{-\pi}^\pi \frac{\dif k'}{(\omega_\pm+2 J \cos k' )^2} \right)^2 U(z;p,k).
\end{align*}
We add the four terms and simplify the resulting expression through the use of the normalization condition of the bound state
\begin{align}
\label{eq:Psi_norm}p_b + p_b g^2 \int_{-\pi}^\pi \frac{\dif k'}{(\omega_\pm+2 J \cos k' )^2} = 1,
\end{align}
to arrive at the concise form 
\begin{align*}
\lim_{z\rightarrow\omega_p+\omega_\pm}\braopket{p\Psi_\pm}{U_{11}(z)}{k\Psi_\pm} = p_b g^2 U(\omega_p+\omega_\pm;p,k).
\end{align*}
We can now write down the scattering matrix element through the use of \eqref{eq:dirac_delta} as
\begin{align}
\label{eq:b_to_b}\braopket{p\Psi_\pm}{S}{k\Psi_\pm} &= t_{k\pm}^\odot \delta(p-k) + r_{k\pm}^\odot \delta(p+k)  \quad\text{where}\quad \\
r_{k\pm}^\odot &= \frac{-2\pi\mi p_b g^2 U(\omega_k+\omega_\pm;k,k)}{2 J \abs{\sin k}}, \nonumber \\
t_{k\pm}^\odot &= 1+r_{k\pm}^\odot. \nonumber
\end{align}
Here, $r_{k\pm}^\odot$ and $t_{k\pm}^\odot$ refer to the reflection and transmission coefficients of the free photon with a wave vector $k$, scattering off of the bound qubit-photon system with energy $\omega_\pm$.

\subsection{Free to Bound Scattering}
In free to bound scattering, one of the two free incoming photons gets trapped and forms a bound state with the qubit. This process is described by the matrix element $\braopket{p\Psi_\pm}{S}{k_1 k_2 \da}$. We use \eqref{eq:S_multi} to get
\begin{align*}
\braopket{p\Psi_\pm}{S}{k_1 k_2 \da} & = -2\pi\mi \delta(\omega_{k_1}+\omega_{k_2}-\omega_{p}-\omega_{\pm})\\
& \quad \braopket{p\Psi_\pm}{U_{10}(\omega_{k_1}+\omega_{k_2})}{k_1 k_2 \da}.
\end{align*}
The relationship between $U_{10}$ and $G$ is written via \eqref{eq:G_to_U} as
\begin{align*}
U_{10}(z) = G_1^{-1}(z) G(z) G_0^{-1}(z).
\end{align*}
By using \eqref{eq:G_1} and observing that $\ket{k_1 k_2 \da}$ is an eigenstate of $H_0$, we get
\begin{align}
\braopket{p\Psi_\pm}{U_{10}(z)}{k_1 k_2 \da} & = (z-\omega_p-\omega_\pm)(z-\omega_{k_1}-\omega_{k_2})\nonumber \\
\label{eq:U10}& \qquad \times \braopket{p\Psi_\pm}{G(z)}{k_1 k_2 \da}.
\end{align}
We expand $\ket{\Psi_\pm}$ via \eqref{eq:bound_p} as in the previous subsection and use \eqref{eq:G_6}, \eqref{eq:G_7} with \eqref{eq:Gz_pk_to_U} to rewrite \eqref{eq:U10} in terms of the $U$ function. Then, by keeping terms that remain finite when $z$ takes its on-shell value $z_\text{os}=\omega_p+\omega_\pm$ and using the normalization condition \eqref{eq:Psi_norm} we arrive at the final result
\begin{align}
\label{eq:free_to_b_S}\braopket{p\Psi_\pm}{S}{k_1 k_2 \da} & = -2\pi\mi \delta(\omega_{k_1}+\omega_{k_2}-\omega_{p}-\omega_{\pm})\\
&\quad \times g^3 \sqrt{p_b} \left[ \frac{U(z_\text{os};p,k_1)}{H(z_\text{os};k_1)} + \frac{U(z_\text{os};p,k_2)}{H(z_\text{os};k_2)} \right].\nonumber
\end{align}

\subsection{Free to Free Scattering}
In free to free scattering, bound states will not explicitly be used and thus we will allow for finite $\Omega\neq 0$, in contrast to the previous two subsections. The matrix element we are after is
\begin{align*}
&\braopket{p_1 p_2 \da}{S}{k_1 k_2 \da}=\braket{p_1 p_2\da}{k_1 k_2 \da}\\
&\quad - 2\pi \mi \delta(\omega_{p_1}+\omega_{p_2}-\omega_{k_1}-\omega_{k_2})\braopket{p_1 p_2 \da}{U_{00}(z_\text{os})}{k_1 k_2 \da},
\end{align*}
where this time the on shell energy value is given by $z_\text{os}=\omega_{p_1}+\omega_{p_2}-\Omega/2=\omega_{k_1}+\omega_{k_2}-\Omega/2$. A comparison of \eqref{eq:GtoT} and \eqref{eq:G_to_U} reveals that $U_{00}(z) = T(z)$. Through the use of \eqref{eq:G_to_U} we get
\begin{align*}
& \mspace{-36mu} \braopket{p_1 p_2 \da}{U_{00}(z)}{k_1 k_2 \da} \\ 
& = (z+\Omega/2-\omega_{p_1}-\omega_{p_2})(z+\Omega/2-\omega_{k_1}-\omega_{k_2})\\
&\qquad \times  \braopket{p_1 p_2 \da}{G(z)}{k_1 k_2 \da}\\
& \quad - (z+\Omega/2-\omega_{k_1}-\omega_{k_2}) \braket{p_1 p_2 \da}{k_1 k_2 \da}.
\end{align*}
The last line on the right hand size goes to zero as $z\rightarrow z_\text{os}$. We can write the scattering matrix element by the help of \eqref{eq:Gz_pk_to_U}, \eqref{eq:G_7} and \eqref{eq:G_8}  as
\begin{align}
\label{eq:free_to_free_a}&\braopket{p_1 p_2 \da}{S}{k_1 k_2 \da} = \braket{p_1 p_2\da}{k_1 k_2 \da} \\
&\quad - 2\pi \mi \delta(\omega_{p_1}+\omega_{p_2}-\omega_{k_1}-\omega_{k_2}) \nonumber \\
&\qquad \times \sum_{m,n} \left(g^2\frac{\delta(p_m-k_n)}{H(z_\text{pos};p_m)} + g^4\frac{U(z_\text{pos};p_m,k_n)}{H(z_\text{pos};p_m)H(z_\text{pos};k_n)}\right)\nonumber
\end{align}
where we defined the photon on-shell energy as $z_\text{pos} = z_\text{os}+\Omega/2$ for notational brevity. Furthermore $m,n \in \{1,2\}$ are photon indices, hence the summation above has four different pairings of $(p_m,k_n)$. We will wait until after the end of the next subsection to put \eqref{eq:free_to_free_a} into a more familiar form in terms of single photon reflection and transmission coefficients. As a prerequisite, let us first analyze $U(z;p,k)$ perturbatively.

\begin{figure}
\includegraphics{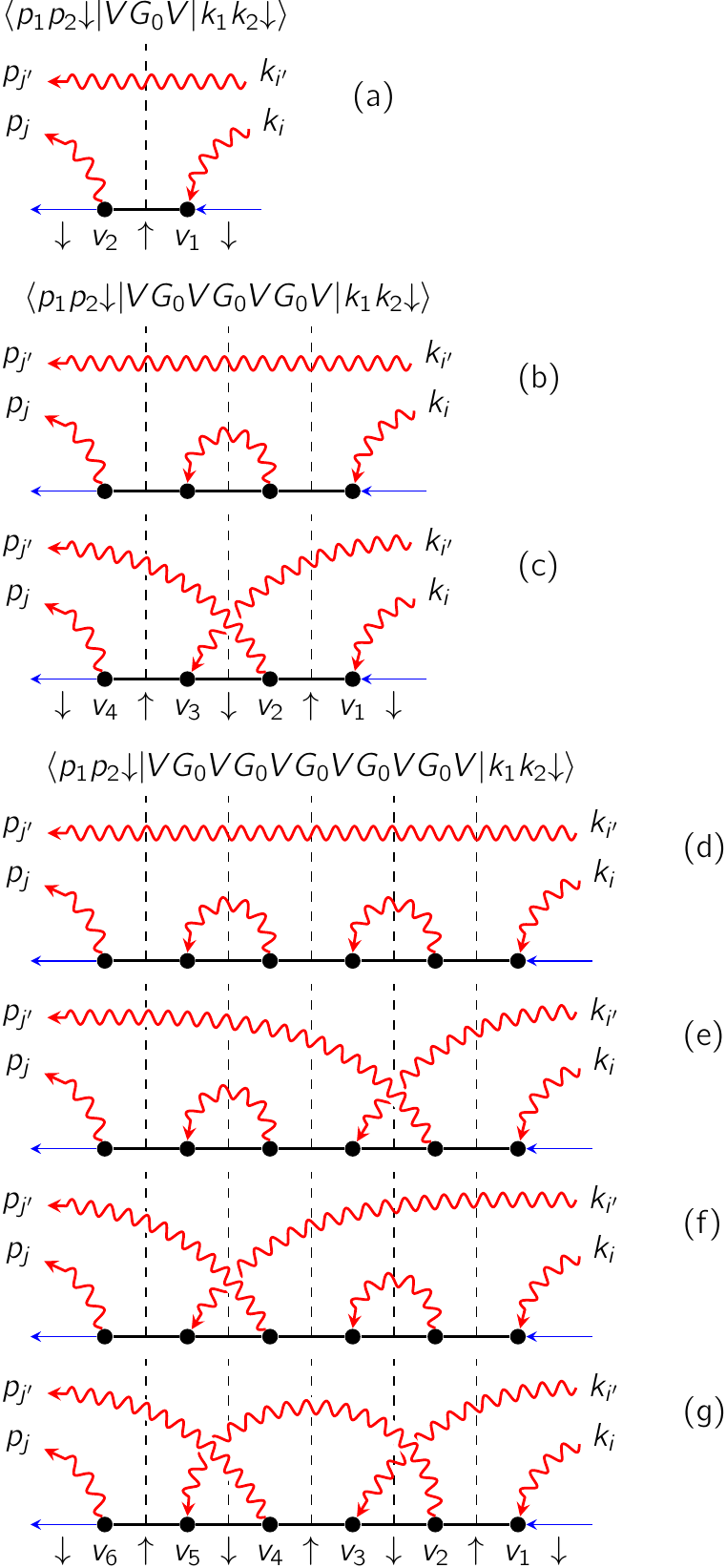}
\caption{(Color online) Feynman graphs for two-photon scattering. } \label{fig:TwoPhotonFeynman}
\end{figure}

\subsection{Perturbative Analysis of Free to Free Scattering}
In the case of one-photon scattering, there was only a single way to contract the creation and annihilation operators for all terms in \eqref{eq:Tz_expansion}. The situation gets more complicated when there are multiple photons. Similar to Fig \ref{fig:OnePhotonFeynman}, in Fig \ref{fig:TwoPhotonFeynman} we illustrate the Feynman graphs for the first three non-zero terms in \eqref{eq:Tz_expansion} for the case of two free photons in the incoming and outgoing states. Each plot refers to four similar looking graphs due to the way the incoming and outgoing photons can be labeled, that is $\{i,j\} \in \{1,2\}$ with $i'\neq i$, $j'\neq j$. The number of plots doubles as the order of the expansion $n$ in $\braopket{p_1p_2\da}{V(G_0V)^{2n+1}}{k_1k_2\da}$ is increased by one. 

The first graphs in each expansion term, that is subplots (a), (b) and (d) in Fig \ref{fig:TwoPhotonFeynman}, refer to single photon scattering events, where one of the photons go through the qubit with no interaction. Schematically, this fact is illustrated by a direct connection between the input photon $k_{i'}$ and the output photon $p_{j'}$. In each subsequent higher order term of the single photon scattering events, extra bubbles are introduced, essentially the same way as in Fig \ref{fig:OnePhotonFeynman}. The infinite series of single photon scattering events can be summed to give
\begin{align}
\label{eq:free_to_free_single}g^2 \frac{\delta(k_{i'}-p_{j'})}{H(z+\Omega/2;k_{i'})},
\end{align}
as illustrated in Fig \ref{fig:TwoPhotonFeynman_summed}(a). The hatched line in the figure refers to the summed propagator associated with the $H(z+\Omega/2;k_{i'})$ term.

Other terms in two-photon scattering also show a regular pattern. Looking at subplots (c), (e) and (f) of Fig \ref{fig:TwoPhotonFeynman} one can foresee the general pattern of internal bubbles to the right and left of the $k_{i'}$--$p_{j'}$ crossing. The summation of subplots (c), (e), (f) and all the other higher order terms with increasingly more bubbles to the right and left of the  $k_{i'}$--$p_{j'}$ crossing leads to Fig \ref{fig:TwoPhotonFeynman_summed}(b) which corresponds to
\begin{align}
&g^4\frac{1}{H(z+\Omega/2;p_{j'})} V_0(z+\Omega/2;p_{j'},k_{i'}) \frac{1}{H(z+\Omega/2;k_{i'})}, \nonumber \\
\label{eq:V0}&\text{where} \quad V_0(z;p,k) = \frac{1}{(z-\omega_p-\omega_k)}.
\end{align}
Similarly, the term in Fig \ref{fig:TwoPhotonFeynman}(g) and those with similar topology in higher order expansions of $T(z)$ but with internal bubbles on the left, center and right hand side of the structure in Fig \ref{fig:TwoPhotonFeynman}(g) lead to Fig \ref{fig:TwoPhotonFeynman_summed}(c) which is the Feynman plot referring to the term
\begin{align}
g^4 \frac{1}{H(z+\Omega/2;p_{j'})} V_1(z+\Omega/2;p_{j'},k_{i'}) \frac{1}{H(z+\Omega/2;k_{i'})},\nonumber \\ 
\label{eq:V1}
V_1(z;p,k) = g^2 \int_{-\pi}^\pi \dif v \frac{1}{(z-\omega_p-\omega_v)H(z;v)(z-\omega_v-\omega_k)}.
\end{align}
Indeed, there are an infinite number of such terms with subsequently larger number of crossing loops in cascade. For instance Fig \ref{fig:TwoPhotonFeynman_summed}(d) shows the case of two crossing loops with all internal bubbles added. Looking at Fig \ref{fig:TwoPhotonFeynman_summed}(c)--(d) one can easily predict the pattern of the infinite series of Feynman graphs with three or more crossing loops in cascade. For completeness \ref{fig:TwoPhotonFeynman_summed}(d) refers to the following term
\begin{widetext}
\begin{align}
\label{eq:V2}&g^4 \frac{1}{H(z+\Omega/2;p_{j'})} V_2(z+\Omega/2;p_{j'},k_{i'}) \frac{1}{H(z+\Omega/2;k_{i'})},\\
&V_2(z;p,k) = g^4\iint_{-\pi}^\pi \frac{\dif v_1 \dif v_2}{(z-\omega_p-\omega_{v_2})H(z;v_2)(z-\omega_{v_2}-\omega_{v_1})H(z;v_1)(z-\omega_{v_1}-\omega_{k})}.\nonumber
\end{align}
\end{widetext}

The summation of the terms in Fig \ref{fig:TwoPhotonFeynman_summed}(b--d) and the remaining infinite series of three or more cascade crossings---with all internal bubbles summed---lead to Fig  \ref{fig:TwoPhotonFeynman_summed}(e) where the cross hatched $U=\sum_{n=0}^\infty V_n$. The choice the letter $U$ is not a coincidence. Let us look at the singular integral equation for $U(z;p,k)$ given in \eqref{eq:U_sing_inteq}. One can represent $U(z;p,k)$ with an infinite Neumann series obtained by initially setting it equal to  $(z-\omega_p-\omega_k)^{-1}$ which is the inhomogeneous term on the right hand side of \eqref{eq:U_sing_inteq}. Through the use of this approximation in the integral on the right hand side of \eqref{eq:U_sing_inteq} we get a better approximation to $U(z;p,k)$. This process can be repeated infinitely to get
\begin{align}
\label{eq:U}U(z;p,k) = V_0(z;p,k) + V_1(z;p,k) + V_2(z;p,k) + \ldots \quad
\end{align}
where $V_0$, $V_1$, $V_2$ are as defined above, and the general form of $V_n$ can easily be obtained by iteration. We will refer to $V_0$ as the bare vertex term, and $V_1, V_2, \ldots$ as progressively higher order vertex correction terms. As noted in \cite{Pletyukhov2012} except for $V_0$ all higher order $V_n$ are identically equal to zero for linear dispersion and this can easily be derived by partial fraction expansion of the integrands in $V_n$ and the simpler form of $H(z;k)$ in the case of linear dispersion where $I(z)$ actually becomes independent of $z$. The same result can also be obtained via a path integral approach \cite{Ringel2013}. Also note that, the infinite series we obtain for the free to free scattering channel in the two photon case agrees with the results of the previous section, compare \eqref{eq:free_to_free_single}, \eqref{eq:V0}, \eqref{eq:V1}, \eqref{eq:V2} and \eqref{eq:U} with \eqref{eq:free_to_free_a}.

\begin{figure}
\includegraphics{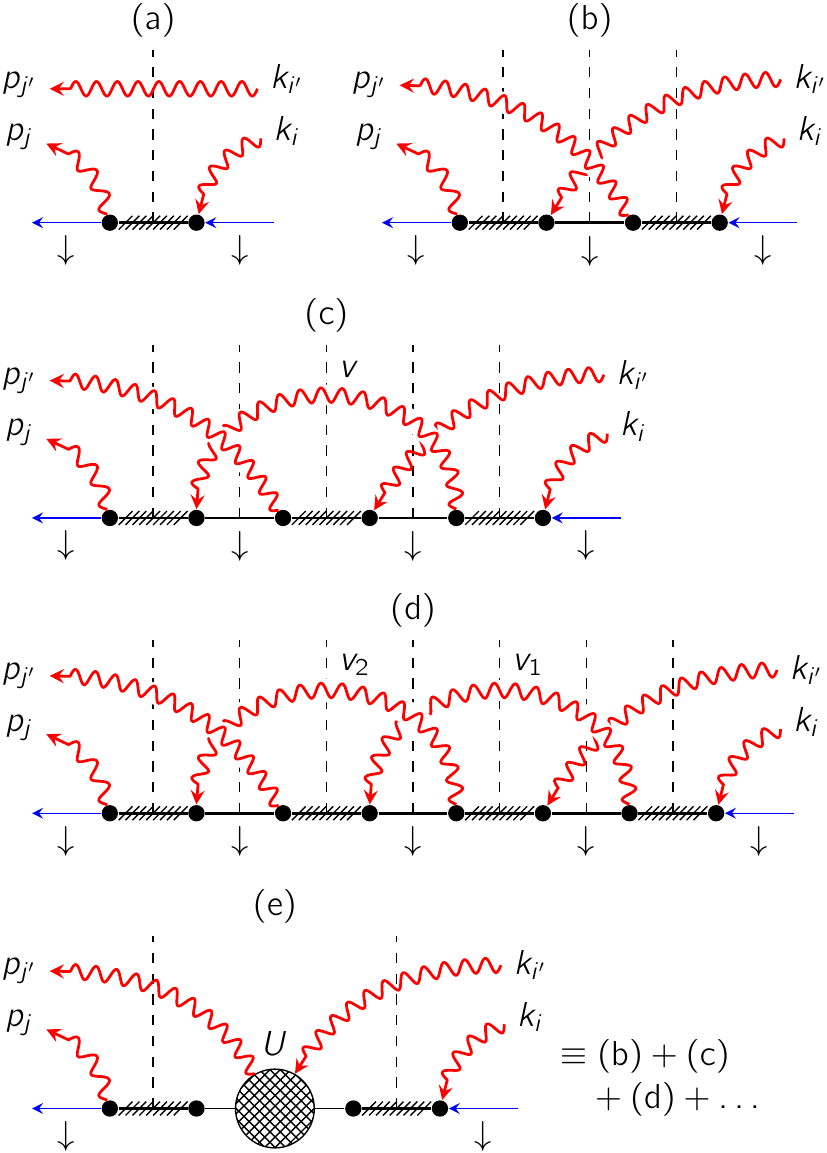}
\caption{(Color online) Feynman graphs for two-photon scattering after the internal bubbles are summed. Resummed propagators are shown with a hatched line. Integration variables in \protect\eqref{eq:V1} and \protect\eqref{eq:V2} are shown in subplots (c)-(d).} \label{fig:TwoPhotonFeynman_summed}
\end{figure}

We will now focus on \eqref{eq:free_to_free_a} and excavate all single photon events buried in it by taking steps similar to those for linear dispersion case \cite{Pletyukhov2012}. Note that, in the free to free scattering channel, at the on-shell energy of $z_\text{os}=\omega_{p_1}+\omega_{p_2}-\Omega/2=\omega_{k_1}+\omega_{k_2}-\Omega/2$, the lowest order process of Fig \ref{fig:TwoPhotonFeynman_summed}(b) described by \eqref{eq:V0} will be
\begin{align*}
&g^4\lim_{\eta \rightarrow 0^+}\frac{1}{H(\omega_{p_1}+\omega_{p_2}+\mi \eta;p_{j'})}\\ &\times\left(\frac{1}{\omega_{p_1}+\omega_{p_2}+\mi \eta-\omega_{p_{j'}}-\omega_{k_{i'}}}\right) \frac{1}{H(\omega_{k_1}+\omega_{k_2}+\mi \eta;k_{i'})},
\end{align*}
where we explicitly have shown the $\eta \rightarrow 0^+$ limit used in $S$-matrix definition as in \eqref{eq:S_multi}. From the definition of \eqref{eq:Hz}, we see that it is not possible to have $H(\omega_{p_1}+\omega_{p_2};p_j)=0$ for any $p_j \in [-\pi,\pi]$. Therefore, $H(z;k)$ terms do not lead to any poles for free to free scattering and $\eta = 0$ can safely be taken. However, the middle term in parentheses above does have a pole and through the use of the Sokhotski--Plemelj theorem \cite{Muskhelishvili1958,Kirilov1994} we get
\begin{align*}
&\lim_{\eta \rightarrow 0^+} \frac{1}{\omega_{p_1}+\omega_{p_2}+\mi \eta-\omega_{p_{j'}}-\omega_{k_{i'}}} \\
&= \lim_{\eta \rightarrow 0^+} \frac{1}{\omega_{p_j}-\omega_{k_{i'}}+\mi \eta} = \mathscr{P} \frac{1}{\omega_{p_j}-\omega_{k_{i'}}} -\mi \pi \delta(\omega_{p_j}-\omega_{k_{i'}}),
\end{align*}
where $\mathscr{P}$ stands for the principal value. We see that there is a Delta dirac function buried in $U(z;k)$. We will separate it away from $U$ and call the remaining term with the principal value as $U'$. As a result \eqref{eq:free_to_free_a} becomes
\begin{align*}
&\braopket{p_1 p_2 \da}{S}{k_1 k_2 \da}  = \braket{p_1 p_2\da}{k_1 k_2 \da} \\ 
&+\delta\bigl({\textstyle\sum_m \omega_{p_m}-\omega_{k_m}}\bigr)\sum_{m,n} \left(\frac{(- 2\pi \mi)g^4 U'(z_\text{pos};p_m,k_n)}{H(z_\text{pos};p_m)H(z_\text{pos};k_n)}\right)\nonumber \\
&+\sum_{m,n} \Bigl [ \frac{(-2\pi\mi g^2)\delta(\omega_{p_{m'}}-\omega_{k_{n'}}) \delta(p_m-k_n)}{H(z_\text{pos};p_m)} \nonumber \\
&\qquad+\frac{(-2\pi^2 g^4)\delta(\omega_{p_{m}}-\omega_{k_{n'}}) \delta(\omega_{p_{m'}}-\omega_{k_{n}})}{H(z_\text{pos};p_m) H(z_\text{pos};k_n)} \Bigr], \nonumber
\end{align*}
where the primed indices denote $m\neq m'$, $n\neq n'$. We used $\delta(\omega_{p_1}+\omega_{p_2}-\omega_{k_1}-\omega_{k_2}) \delta(p_m-k_n) = \delta(\omega_{p_{m'}}-\omega_{k_{n'}})\delta(p_m-k_n)$ and $\delta(\omega_{p_1}+\omega_{p_2}-\omega_{k_1}-\omega_{k_2}) \delta(\omega_{p_{m'}}-\omega_{k_{n}}) = \delta(\omega_{p_{m}}-\omega_{k_{n'}}) \delta(\omega_{p_{m'}}-\omega_{k_{n}})$ to write the expression above.

Through the use of \eqref{eq:dirac_delta} we convert all $\delta(\omega_{p_{m}}-\omega_{k_{n}})$ to $\delta({p_{m}}\pm {k_{n}})$, and by observing that 
\begin{align*}
\frac{-2\pi\mi g^2}{H(z_\text{pos};p_{m'}) 2J \abs{\sin p_m}}=r_{p_m},
\end{align*}
we factor out common terms. We employ the one-photon reflection and transmission coefficients in \eqref{eq:reflection} and \eqref{eq:transmission} to arrive at the final result
\begin{align}
&\braopket{p_1 p_2 \da}{S}{k_1 k_2 \da}  = \nonumber \\ 
&t_{p_1} t_{p_2} \bigl[\delta(p_1-k_1) \delta(p_2-k_2)+ \delta(p_1-k_2) \delta(p_2-k_1) \bigr] \nonumber \\
&+ r_{p_1} r_{p_2} \bigl[\delta(p_1+k_1) \delta(p_2+k_2)+ \delta(p_1+k_2) \delta(p_2+k_1) \bigr] \nonumber \\
\label{eq:free_to_free_S}&+ t_{p_1} r_{p_2} \bigl[\delta(p_1-k_1) \delta(p_2+k_2)+ \delta(p_1-k_2) \delta(p_2+k_1) \bigr] \\
&+ r_{p_1} t_{p_2} \bigl[\delta(p_1+k_1) \delta(p_2-k_2)+ \delta(p_1+k_2) \delta(p_2-k_1) \bigr] \nonumber \\
&- 2\pi\mi g^4 \delta\bigl({\textstyle\sum_m \omega_{p_m}-\omega_{k_m}}\bigr) \sum_{m,n} \left(\frac{U'(z_\text{pos};p_m,k_n)}{H(z_\text{pos};p_m)H(z_\text{pos};k_n)}\right).\nonumber
\end{align}
The equation above is reminiscent of the two-mode, two-photon scattering results in \cite{Shen2007}, generalized to a dispersive waveguide.

\begin{figure*}
\includegraphics{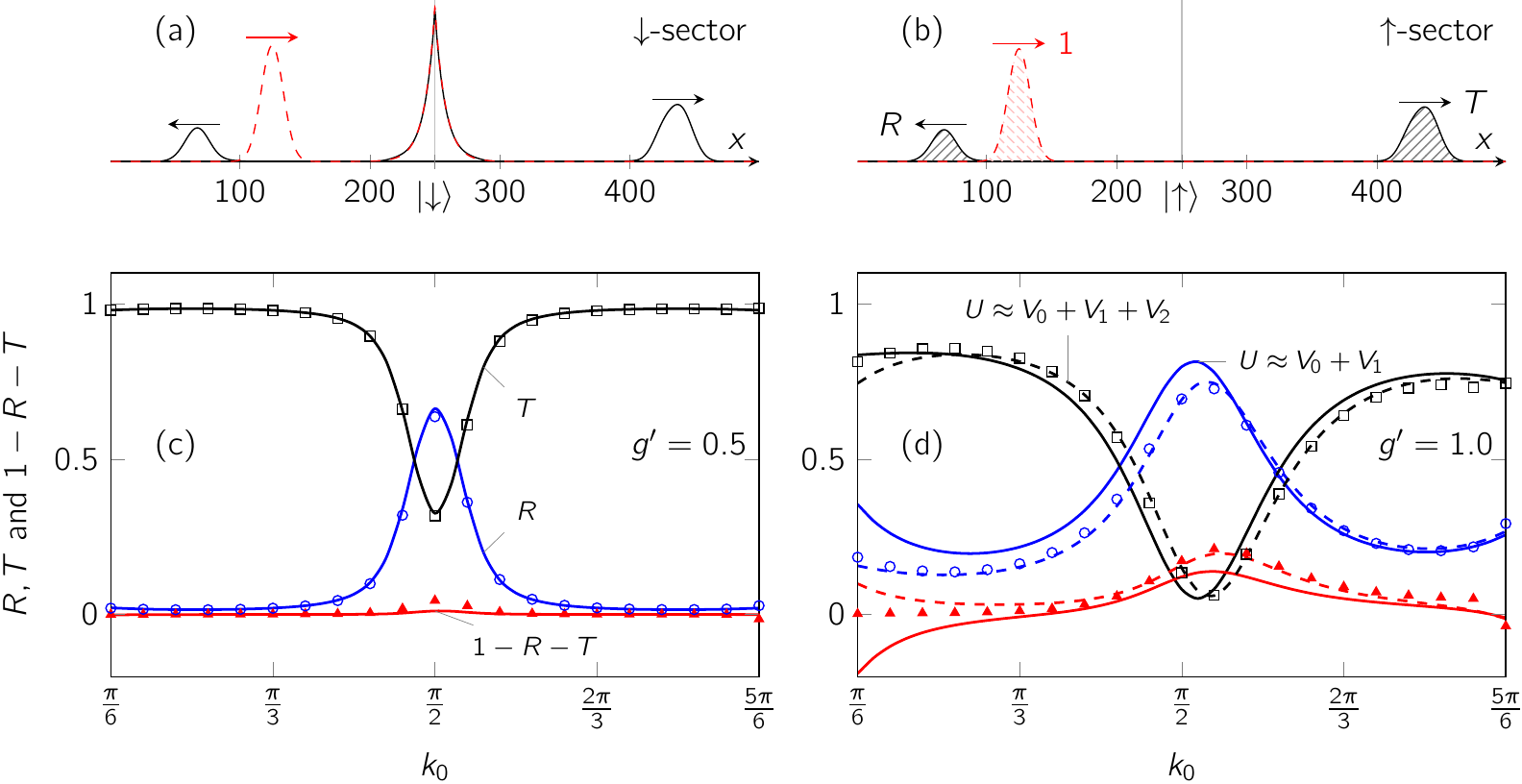}
\caption{(Color online) (a)-(b) Schematic of the numerical simulation setup. Initially, there is a single photon packet moving towards the atom-photon bound state $\ket{\Psi_-}$ (red). After the scattering there are pulses propagating away from the bound state (black). Integration, denoted by hatched areas, is done in the $\ua$-sector to get the reflection, $R$, and transmission, $T$, results. In order to have a unit probability incoming pulse as shown in (b) normalization by $p_b$ should be made. (c) Plot of $R$ (blue), $T$ (black) and $1-R-T$ (red) for $g'=0.5$ and Gaussian pulse parameter $s=12$, as a function of $k_0$. Symbols ($\circ$, $\Box$, $\blacktriangle$) show Krylov-subspace based numerical results, solid lines show the result of integration in \protect\eqref{eq:b_to_b_expand} where $U\approx V_0+V_1$ is taken. $\eta=10^{-6}$ in the integrations. (d) Similar to (c) but for $g'=1.0$. Solid lines are for the case $U\approx V_0+V_1$, dashed lines are for $U\approx V_0+V_1+V_2$. $\eta=10^{-4}$ in the integrations. Increasing the order of approximation for $U$ leads to a much better fit to numerical results.} \label{fig:BoundToBound}
\end{figure*}

\section{Numerical Results}\label{sec:results}

In this section, our aim is to verify that the equations we derived for multichannel scattering matrix elements are valid. To do so, we will be sending Gaussian photon packets of the form
\begin{align*}
f(x) = \frac{1}{(\pi s^2)^{1/4}}\exp\left(-\frac{(x-x_c)^2}{2s^2}+\mi k_0 x \right),
\end{align*}
towards the qubit either in its ground state, or in a bound state. Here, $s$ is the spatial width of the pulse, $x_c$ is the center location chosen to be away from the qubit by couple $s$ lengths and $k_0$ is the center wave vector. We will evolve the system described by \eqref{eq:Hd} in time, and wait until all the scattering is complete. The numerical approach we use for time evolution is based on a Krylov-subspace based method  \cite{Longo2009,Longo2010,Longo2011}. We will compare the results from our numerical approach with those of the equations we derived for multichannel scattering. All energy values are reported for $J=1$.

\subsection{Bound to Bound Calculations}
In our bound to bound scattering calculations, we are interested in the scattering of an incoming pulse, $f(k)$, off of a bound state. The input state is given by
\begin{align*}
\ket{\text{in}}=\int_{-\pi}^\pi \dif k f(k) \ket{k\Psi_\pm},
\end{align*}
with $\int_{-\pi}^\pi \dif k \abs{f(k)}^2 =1$.
We will look at the output in the same channel, i.e. channel 1, therefore, we will take the projections onto $\ket{p\Psi_\pm}$ to get the output state as
\begin{align*}
\ket{\text{out}}=\iint_{-\pi}^\pi \dif k \dif p f(k) \braopket{p\Psi_\pm}{S}{k\Psi_\pm}\ket{p\Psi_\pm}.
\end{align*}
We use \eqref{eq:b_to_b} and \eqref{eq:bound_p} to get
\begin{align}
\ket{\text{out}}&=\int_{-\pi}^\pi \dif k f(k) \left(t_{k\pm}^\odot\ket{k\Psi_\pm} + r_{k\pm}^\odot \ket{-k\Psi_\pm} \right) \nonumber \\
\label{eq:b_to_b_expand}&=\int_{-\pi}^\pi \dif k f(k) \sqrt{p_b}\left(t_{k\pm}^\odot\ket{k\ua} + r_{k\pm}^\odot \ket{-k\ua} \right) \\
& + \iint_{-\pi}^\pi \dif k \dif p f(k) \sqrt{p_b} g \Bigl(t_{k\pm}^\odot \frac{\ket{k p \da}}{\omega_\pm + 2 J \cos p} \nonumber \\ 
& \qquad \qquad + r_{k\pm}^\odot \frac{\ket{-k p \da}}{\omega_\pm + 2 J \cos p} \Bigr). \nonumber
\end{align}
In our numerical calculations, we prepare the initial state as a free Gaussian pulse propagating towards the bound state $\ket{\Psi_-}$ as sketched in Fig \ref{fig:BoundToBound}(a,b). The spatial form of the bound state is described in Appendix \ref{App:bound}. Our numerical approach returns all parts of the wave function. We call parts with the form $\ket{k_1k_2\da}$ as the $\da$-sector and parts with $\ket{k\ua}$ as the $\ua$-sector. After scattering, we integrate the pulses propagating to the left and to the right of the qubit in the $\ua$-sector to get the total reflection and transmission probabilities as shown in Fig \ref{fig:BoundToBound}(b). We note that we need to normalize the numerical data we get from the $\ua$-sector by $p_b$ due to the $\sqrt{p_b}$ factor in \eqref{eq:b_to_b_expand}. The reason we choose to use the $\ua$-sector is because the $\da$-sector additionally contains channel 1 to channel 0 transitions which cannot easily be separated from channel 1 to channel 1 transitions. On the other hand, $\ua$-sector only has contributions from channel 1.

In Fig \ref{fig:BoundToBound}(c) and (d) we illustrate bound to bound scattering results for $g'=0.5$ and $g'=1.0$, respectively. For calculating the solid curves in \ref{fig:BoundToBound}(c) via \eqref{eq:b_to_b}, we take the first two terms in the expansion of $U$ from \eqref{eq:U} and use $U\approx V_0 + V_1$ in \eqref{eq:b_to_b_expand}. As can be seen, there is a good overlap between the numerical results and the equations we derived.

When $g'$ is increased to $1.0$, the agreement between the numerical results and the model prediction with $U\approx V_0 + V_1$ (solid lines) deteriorates. However, when we add the next order term and set $U\approx V_0 + V_1 + V_2$ (dashed lines) we obtain a much better fit between the numerical results and our modeling formalism, see Fig \ref{fig:BoundToBound}(d). We end this subsection by noting that while calculating the integrals for $V_1$ and $V_2$ we keep a small but finite $\mi \eta$ value as reported in the figure caption.

\subsection{Free to Bound Calculations}

To calculate the probability of exciting the bound state when a photon packet with two free photons scatter off of a qubit in its ground state, we need to calculate the scattering matrix element for the following input state
\begin{align}
\label{eq:free_to_b_input}\ket{\text{in}}=\frac{1}{\sqrt{2}}\iint_{-\pi}^\pi \dif k_1 \dif k_2 f(k_1,k_2) \ket{k_1k_2\da},
\end{align}
with $\iint_{-\pi}^\pi \dif k_1 \dif k_2 \abs{f(k_1,k_2)}^2 =1$. Since $\braket{p_1p_2\da}{k_1k_2\da}$ has two pairs of Dirac delta functions, the factor $2^{-\frac{1}{2}}$ is required to have $\braket{\text{in}}{\text{in}}=1$.

The output we are interested in is a bound state thus we get
\begin{align}
&\ket{\text{out}_\pm}=\frac{1}{\sqrt{2}}\iiint \dif k_1 \dif k_2\dif p f(k_1,k_2) \braopket{p\Psi_\pm}{S}{k_1k_2\da}\ket{p\Psi_\pm} \nonumber \\
&=\frac{1}{\sqrt{2}}\iiint \dif k_1 \dif k_2\dif p f(k_1,k_2) \delta(\omega_{k_1}+\omega_{k_2}-\omega_{p}-\omega_{\pm}) \nonumber \\
\label{eq:free_to_b_int}&\quad \qquad \times T(\omega_{p},\omega_{\pm},\omega_{k_1},\omega_{k_2})\ket{p\Psi_\pm},
\end{align}
where $T(\omega_{p},\omega_{\pm},\omega_{k_1},\omega_{k_2})$ refers to \eqref{eq:free_to_b_S} without the Dirac delta function. Note that although $U(z;p,k)$ and $H(z;k)$ functions are used in \eqref{eq:free_to_b_S}, those functions only depend on the energies $\omega_{p/k}$ therefore our choice of arguments for the $T$ function is justified. The argument of the Dirac delta function in \eqref{eq:free_to_b_int} defines constant energy contours in the $k_1-k_2$ plane as illustrated in Fig \ref{fig:Change_Vars}(a). We make a change of variables $k(\omega_k)= \pm \arccos(-\frac{\omega_k}{2J})$. Inverse cosine function is restricted to the interval $[0,\pi]$ to make it single valued and the $\pm$ sign is used when $k \gtrless 0$ . As a result the two dimensional integral over $k_{i}$ variables is mapped to an integral in the energy variables $\omega_{k_i}$. In Fig \ref{fig:Change_Vars} points $A$, $B$, $C$ and $D$ are illustrated in the $k_i$ and $\omega_{k_i}$ planes. The circuit $ABCDA$ is a closed curve in the $k_i$ plane, it is composed of four overlapping lines in the $\omega_{k_i}$ plane. Because of the need to use a different sign in each of the four quadrants of the $k_i$ plane when changing variables from $k_i$ to $\omega_{k_i}$, we get the following expression
\begin{widetext}
\begin{align*}
\ket{\text{out}_\pm}&=\frac{1}{\sqrt{2}}\int \dif p \iint_{-2J}^{2J} \frac{\dif \omega_{k_1} \dif \omega_{k_2} \delta(\omega_{k_1}+\omega_{k_2}-\omega_{p}-\omega_{\pm})}{\sqrt{4J^2-\omega_{k_1}^2}\sqrt{4J^2-\omega_{k_2}^2}} T(\omega_{p},\omega_{\pm},\omega_{k_1},\omega_{k_2})\\
&\qquad \qquad \times  \Bigl\{ f\bigl(k(\omega_{k_1}),k(\omega_{k_2})\bigr) + f\bigl(-k(\omega_{k_1}),k(\omega_{k_2})\bigr)+ f\bigl(-k(\omega_{k_1}),-k(\omega_{k_2})\bigr) + f\bigl(k(\omega_{k_1}),-k(\omega_{k_2})\bigr)\Bigr\} \ket{p\Psi_\pm}.
\end{align*}
\end{widetext}
The Dirac delta function above forces the integral on a line with constant total energy as shown in Fig \ref{fig:Change_Vars}(b). We make another change of variables with $\omega_{k_{1,2}}=E/2 \pm \Delta$ or equivalently $E=\omega_{k_{1}}+\omega_{k_{2}}$, $\Delta=(\omega_{k_{1}}-\omega_{k_{2}})/2$ and obtain\footnote{These set of transformations are in line with the co-area formula \cite{Aubert2006,Evans2010,Hartmann2011}.}
\begin{align}
\ket{\text{out}_\pm}&=\frac{1}{\sqrt{2}}\int_{L_{p\pm}}^{H_{p\pm}} \dif p \int_{L_\Delta}^{H_\Delta} \dif \Delta \frac{T(\omega_{p},\omega_{\pm},\omega_{k_1},\omega_{k_2})}{\sqrt{4J^2-\omega_{k_1}^2}\sqrt{4J^2-\omega_{k_2}^2}} \nonumber \\ 
\label{eq:free_to_bound}&\times \Bigl\{ \sum_{\substack{s_1 = \pm 1 \\ s_2 = \pm 1}}f\bigl(s_1 k(\omega_{k_1}),s_2 k(\omega_{k_2})\bigr) \Bigr\} \ket{p\Psi_\pm}.
\end{align}
Here the lower and higher limits of integration for the $\Delta$ integral are given by $L_\Delta=\max(-2+\frac{E}{2},-2-\frac{E}{2})$ and $H_\Delta=\min(2-\frac{E}{2},2+\frac{E}{2})$ which correspond to the edges of the gray dashed square in Fig \ref{fig:Change_Vars}(b). For the $p$ integral, the limits of integration should be chosen such that $\ket{p\Psi_\pm}$ is an accessible state from two free photons with a total energy in the range $[-4J,4J]$ which implies that the energy of $\ket{p\Psi_\pm}$ should obey $-4J \le \omega_p+\omega_\pm \le 4J$ and since $\ket{p}$ is a free photon we have $-2J\le \omega_p\le 2J$ as well. These two conditions can be combined to give the energy interval $\omega_p \in [\max(-2J,-4J-\omega_\pm),\min(2J,4J-\omega_\pm)]$. $L_{p\pm}$ and $H_{p\pm}$ should be chosen to obey the energy interval for $\omega_p$, taking into consideration the dispersion relation $\omega_p=-2J\cos p$. As a result of the Dirac delta function, $E=\omega_p+\omega_\pm$ is used in the expressions for $\omega_{k_{1,2}}$ and the integral limits.

In Fig \ref{fig:free_to_prod}(a) we visualize the function $\abs{T(\omega_{p}=\omega_{k_1}+\omega_{k_2}-\omega_{+},\omega_{+},\omega_{k_1},\omega_{k_2})}^2+\abs{T(\omega_{p}=\omega_{k_1}+\omega_{k_2}-\omega_{-},\omega_{-},\omega_{k_1},\omega_{k_2})}^2$ which is indicative of the total trapping rate into two bound atom-photon states given by $\braket{\text{out}_-}{\text{out}_-} +\braket{\text{out}_+}{\text{out}_+}$. The visualization is done on a logarithmic coloring palette. There are resonances when $\omega_{k_1}+\omega_{k_2} = \omega_\pm$ as shown by the gray dashed lines and when $\omega_{k_1}+\omega_{k_2} = 0$. We numerically send two-photon Gaussian pulses with the same center wave-vector, $k_0$, for both of the photons. We scan $k_0$ value from $\pi/6$ to $5\pi/6$ as shown by the dashed circles which illustrate the two-photon wave packets. We time evolve the system, and measure the total excitation trapped in the $\ua$-sector, normalized by $p_b$. This value gives us the total trapping rate, which we also calculate from \eqref{eq:free_to_bound} as $\braket{\text{out}_-}{\text{out}_-} +\braket{\text{out}_+}{\text{out}_+}$. In Fig \ref{fig:free_to_prod}(b) we compare the results of  integration (blue curve) with those of numerical simulations (red curve). In the integrals we approximate $U\approx V_0+V_1$. The same approximation is also used in the definition of the $T$ function in Fig \ref{fig:free_to_prod}(a). We also solve \eqref{eq:U_sing_inteq} by the numerical scheme described in \cite{Atkinson2008} (gray curve). All three results have the same order of magnitude and show an increase in the trapping rate at resonances, highlighted by red dashed circles in Fig \ref{fig:free_to_prod}(a). The numerical values of integration are particularly in good agreement with those of Krylov-subspace based simulations. We did not increase the order of approximation for $U$ because of the long times required for multi-dimensional numerical integrals. However, we expect that the fit between the blue and red curves in Fig \ref{fig:free_to_prod}(b) would get even better as the order of approximation is increased. 

\subsection{Free to Free Calculations}

For the case of free to free scattering, input state is the same as \eqref{eq:free_to_b_input}. Output state becomes
\begin{align*}
\ket{\text{out}}=\frac{1}{2\sqrt{2}}\iiiint_{-\pi}^\pi & \dif k_1 \dif k_2 \dif p_1 \dif p_2 f(k_1,k_2)\\
& \times \braopket{p_1p_2\da}{S}{k_1k_2\da}\ket{p_1p_2\da},
\end{align*}
with the $\frac{1}{2}$ factor due to \eqref{eq:Identity2}. We proceed as in the free to bound subsection but use \eqref{eq:free_to_free_S} for the scattering matrix element. We apply the same change of variables to integrate out the Dirac delta function to get
\begin{widetext}
\begin{align}
\label{eq:free_to_free_result}\ket{\text{out}}=\frac{1}{\sqrt{2}}\iint_{-\pi}^\pi \dif p_1 \dif p_2 & \Biggl\{ f(p_1,p_2) t_{p_1} t_{p_2} + f(-p_1,-p_2) r_{p_1} r_{p_2} + f(p_1,-p_2) t_{p_1} r_{p_2} + f(-p_1,p_2) r_{p_1} t_{p_2} \\
& + \frac{1}{2}\int_{L_\Delta}^{H_\Delta} \dif \Delta \frac{B(\omega_{p_1},\omega_{p_2},\omega_{k_1},\omega_{k_2})}{\sqrt{4J^2-\omega_{k_1}^2}\sqrt{4J^2-\omega_{k_2}^2}} \sum_{\substack{s_1 = \pm 1\\ s_2 = \pm 1}}f\bigl(s_1 k(\omega_{k_1}),s_2 k(\omega_{k_2})\bigr) 
\Biggr\} \ket{p_1p_2\da}. \nonumber
\end{align}
\end{widetext}
The $B$ function is defined as the last line of \eqref{eq:free_to_free_S} without the Dirac delta term. $L_\Delta=\max(-2+\frac{E}{2},-2-\frac{E}{2})$ and $H_\Delta=\min(2-\frac{E}{2},2+\frac{E}{2})$ as before but the total photon energy is given by $E=\omega_{p_1}+\omega_{p_2}$.

\begin{figure}
\includegraphics{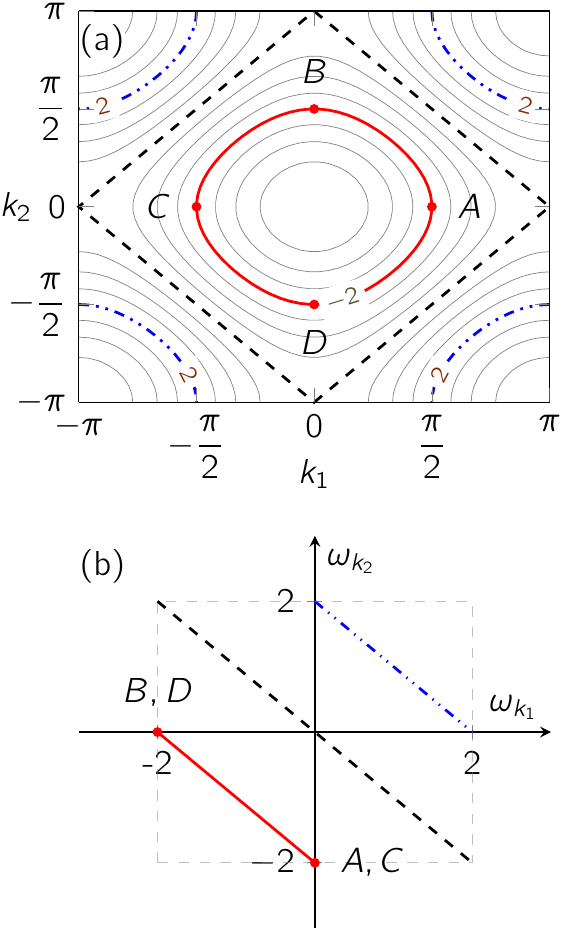}
\caption{(Color online) (a) Two dimensional isocontours of energy in the $k_1-k_2$ plane. (b) Same as (a) but in $\omega_{k_1}-\omega_{k_2}$ plane.} \label{fig:Change_Vars}
\end{figure}

In Fig \ref{fig:free_to_freeResults}(a) we plot the two-photon scattering spectrum in the $p_1-p_2$ plane via \eqref{eq:free_to_free_result} for a two-photon Gaussian pulse excitation with both photons centered around the wave-vector $k_0=2\pi/5$. The width of the Gaussian in real-space representation is $s=12$. Coupling parameter is $g'=0.5$, $\eta=10^{-6}$ is used in integrals, $\Omega=0$ is taken for the qubit. In the definition of the $B$ function in \eqref{eq:free_to_free_S}, $U\approx V_0$ is taken. The dashed line shows the fixed energy contour $E=2\omega_{k_0}$. The bright peaks around the contour refer to factorizable one-photon scattering events on the first line of \eqref{eq:free_to_free_result}, the `wings' around the bright peaks are due to the $B$ function. We label each quadrant as RR, LR, LL and RL denoting the propagating direction (right/left) of photons, analogous to \cite{Shen2007}. Right going photons are transmitted, left going ones are reflected. In Fig \ref{fig:free_to_freeResults}(b) we take the $\da$-sector part of the total wave function from Krylov-subspace simulations and translate it into the $p_1-p_2$ plane by an FFT. A comparison of the subplots (a) and (b) show that even the lowest order bare vertex approximation for $U$ is sufficient to get the overall scattering spectrum for two-photon scattering. 

\begin{figure}
\includegraphics{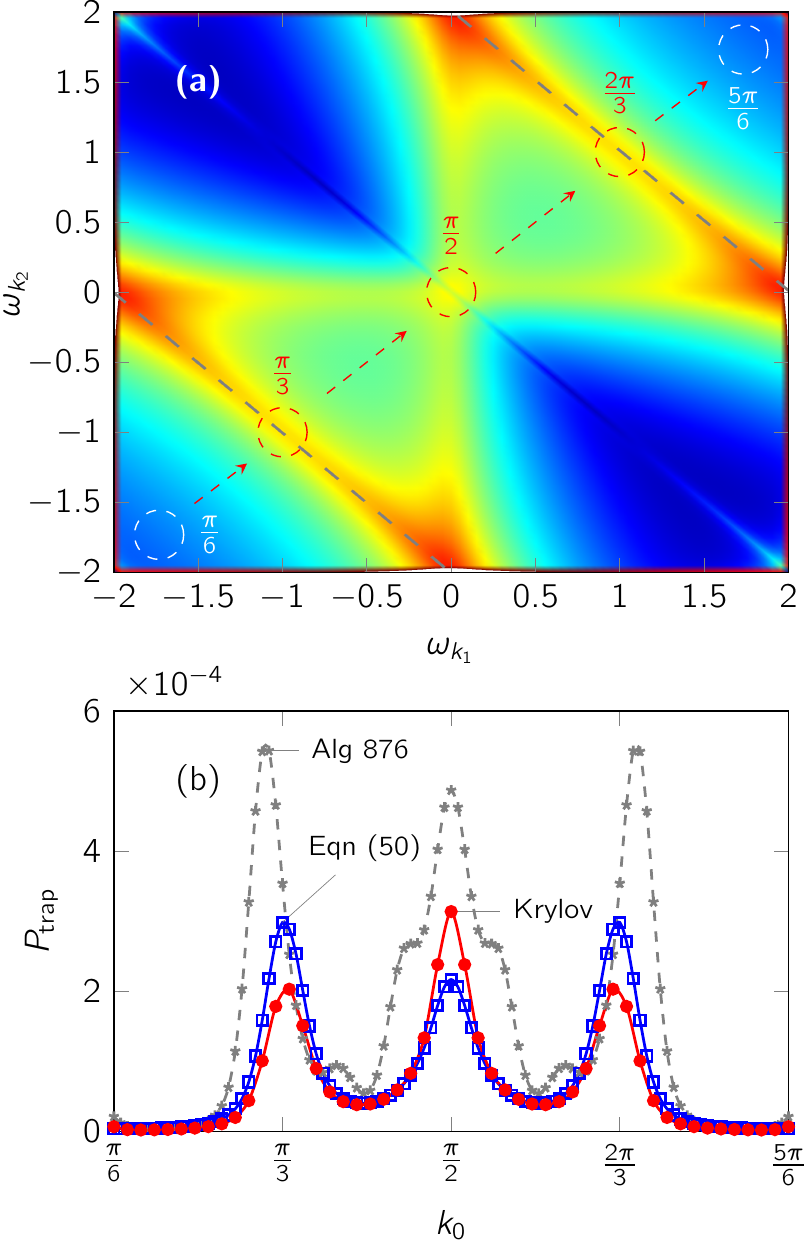}
\caption{(Color online) (a) Plot of $\abs{T(\omega_{p}=\omega_{k_1}+\omega_{k_2}-\omega_{+},\omega_{+},\omega_{k_1},\omega_{k_2})}^2+\abs{T(\omega_{p}=\omega_{k_1}+\omega_{k_2}-\omega_{-},\omega_{-},\omega_{k_1},\omega_{k_2})}^2$ in \protect\eqref{eq:free_to_bound} with $g'=0.5$. Logarithmic coloring is used. $U\approx V_0+V_1$ is taken. Integrals are done with $\eta=10^{-6}$. Dashed circles indicate the center wave-vector, $k_0$, of Gaussian packets. (b) Plot of total trapping rate $\braket{\text{out}_-}{\text{out}_-} +\braket{\text{out}_+}{\text{out}_+}$ as a function of $k_0$ obtained from Krylov-subspace based simulations (red, $\bullet$), the use of equation \protect\eqref{eq:free_to_bound} (blue, $\Box$) and the numerical algorithm described in \protect\cite{Atkinson2008} for solving integral equations (dashed gray, $\star$).} \label{fig:free_to_prod}
\end{figure}

\section{Discussion}\label{sec:discuss}
In the previous sections, we provided the infrastructure for multi-photon calculations in a dispersive photonic environment extending the techniques obtained for the Lee model. The particular form of the dispersion relationship within the Lee model is given by $\omega^2=k^2+m^2$ for a fixed $m$ value.\footnote{Incidentally, this dispersion relationship is very similar to that for modes in a rectangular waveguide, analyzed in \cite{Li2014} for the single excitation case.} Furthermore, in the Lee model, there are subtleties associated with the renormalization of energy levels, masses and the coupling constants to get a mapping to physically observable quantities \cite{Kallen1955}. Such complexities do not occur in the multi-mode Jaynes-Cummings Hamiltonian that we analyze in this manuscript. 

In the Lee model, it is possible to get a closed form formula for the solution of the integral equation for $U(z;p,k)$---one which is analogous to \eqref{eq:U_sing_inteq}---by using the branch-cuts of $\omega^2=k^2+m^2$ and complex analysis \cite{Amado1961,Bolsterli1968,Kazes1965,Kenschaft1964,Maxon1966,Pagnamenta1965,Pagnamenta1966,Sommerfield1965}. In \cite{Shi2009a} the time-independent Schr\"odinger equation within the two excitation sector is solved by constructing an ansatz based on the tight-binding dispersion relationship of the waveguide. Although few details are given in \cite{Shi2009a} regarding the method of solution, we verified that the bound-to-bound scattering matrix elements provided in \cite{Shi2009a} agree with  \eqref{eq:b_to_b}. Comparison of the two approaches is available in \cite{Supplemental}.

Our approach in calculating the $S$-matrix elements is based on Feynman diagrams, through which we obtain arbitrary order vertex correction terms---a method suggested in \cite{Pletyukhov2012}. Feynman diagram approach for approximating the matrix elements was considered starting from the original Lee model paper \cite{Lee1954} and was used to understand more complicated Hamiltonians with extra particles \cite{Bronzan1965}. For three \cite{Liu1968} or more \cite{Liu1970} excitations, Feynman diagram approach proved very useful in the absence of exact results within the Lee model. It is likely that a similar conclusion also holds for the multi-mode Jaynes-Cummings Hamiltonian. We think that it should be possible to extend the analysis we present to three or more photons interacting with a qubit, or to the case of multi-qubit systems with multiple photons bouncing between them---a scenario that is very relevant for designing quantum gates.

The link between Feynman diagrams and Neumann series expansion of the integral equation for $U$ was first reported in \cite{Hammer1988}, and we were heavily influenced by the diagrammatic description shown there to generate Fig \ref{fig:TwoPhotonFeynman_summed}.

While building the link between Feynman diagrams and the $S$-matrix elements, the passage through multi-channel scattering theory is crucial. If, for instance, instead of the channel dependent transition operators, $U_{fi}$ in \eqref{eq:S_multi}, the single channel equation in \eqref{eq:S1photon} is used, expressions obtained for the $S$-matrix terms diverge and do not agree with results obtained from Krylov-subspace based calculations.

In Sec.\ \ref{sec:results} we detail the steps needed to compare results from purely numerical approaches (e.g.\ obtained from the Krylov-subspace method) and analytical expressions (e.g.\ obtained through multi-channel scattering theory and Feynman diagrams). Dispersive effects lead to complications when considering multi-photon gaussian packets and one needs to be careful with the integrals.\footnote{In \cite{Supplemental} we provide the Mathematica code used to generate the data for the figures.} Although we present results for two-photon wave packets, the case of three or more photons can be considered in a similar vein, through a judicious application of the co-area formula \cite{Aubert2006,Evans2010,Hartmann2011}.

Recently, numerical results for scattering between photons and bound atom-photon states in a tight-binding lattice were reported in \cite{Li2015} through an equivalence to the Hubbard model. A similar mapping is also used in \cite{Roy2011a} to analyze two-photon scattering in a tight-binding lattice. However, the results reported in \cite{Roy2011a} involve current operators and it is not immediately obvious how to translate those results to ones comparable with the reflection, transmission or free to bound trapping probabilities reported in this manuscript. The appendix of \cite{Shi2015b} illustrates another similar mapping, one in which the qubit is replaced with a hardcore boson \cite{Shi2015}. Various matrix elements of $G(z)$ are obtained for an arbitrary dispersion relationship through the hardcore boson approach.

Scattering in the two-photon sector in a dispersive waveguide is very recently investigated in \cite{Schneider2016}. There, the results are obtained through a path integral approach with an equivalent Hamiltonian where the Pauli spin operators are mapped to auxiliary fermions. Whereas we chose to base our Feynman diagram description in terms of the matrix elements of the $T$ operator, in \cite{Schneider2016} diagrams are provided for equivalent matrix elements for $G(z)$. The relationship between linear and arbitrary dispersion profiles are described in detail in \cite{Schneider2016}. Additionally, band edge effects for $\omega \propto k^2$ are also investigated. However, there are no explicit formulas for scattering matrix elements, nor an independent numerical verification of the formalism in \cite{Schneider2016} and our results supplement their analysis.

Lastly, we would like to talk briefly on numerical issues related to scattering simulations involving bound states. In our Krylov-subspace based numerical investigations, we were not able to simulate cases with a low $g'$ value, due to the fact that the size of bound state becomes excessively large because of the slow decay of the exponential tails, c.f.\ Fig \ref{fig:BoundStateSpace}. On the other hand, when $g'$ is large, more and more correction terms to the $U$ function are required, with higher dimensions in numerical integrals. We therefore chose $g'=0.5$ and $g'=1.0$ values during various calculations, which was a compromise to get a bound state with a tightly bound photonic part as well as getting relatively quick convergence when making Feynman diagram based calculations. In our calculations we used the \texttt{NIntegrate} function of Mathematica system, however, there are other packages specifically designed for high dimensional integrals in Feynman diagrams, such as \textsc{cuba} \cite{Hahn2005}. Moreover, numerical algorithms specifically developed for handling the singularities in integrals exist as well  \cite{John1997}. We received many warning messages while using the code from \cite{Atkinson2008} when solving \eqref{eq:U_sing_inteq} numerically and had many convergence issues. Although in Fig \ref{fig:BoundToBound} we show that the results of \cite{Atkinson2008} have the same order of magnitude as other independent calculations, for free to bound case we could not get converging results from \cite{Atkinson2008}. This is not surprising, since \cite{Atkinson2008} was not designed for singular integral equations. Finally, the methods outlined in \cite{Richardson2004} combining interpolation and numerical integration in an iterative manner can provide alternative means to tackle \eqref{eq:U_sing_inteq}.

\begin{figure}
\includegraphics{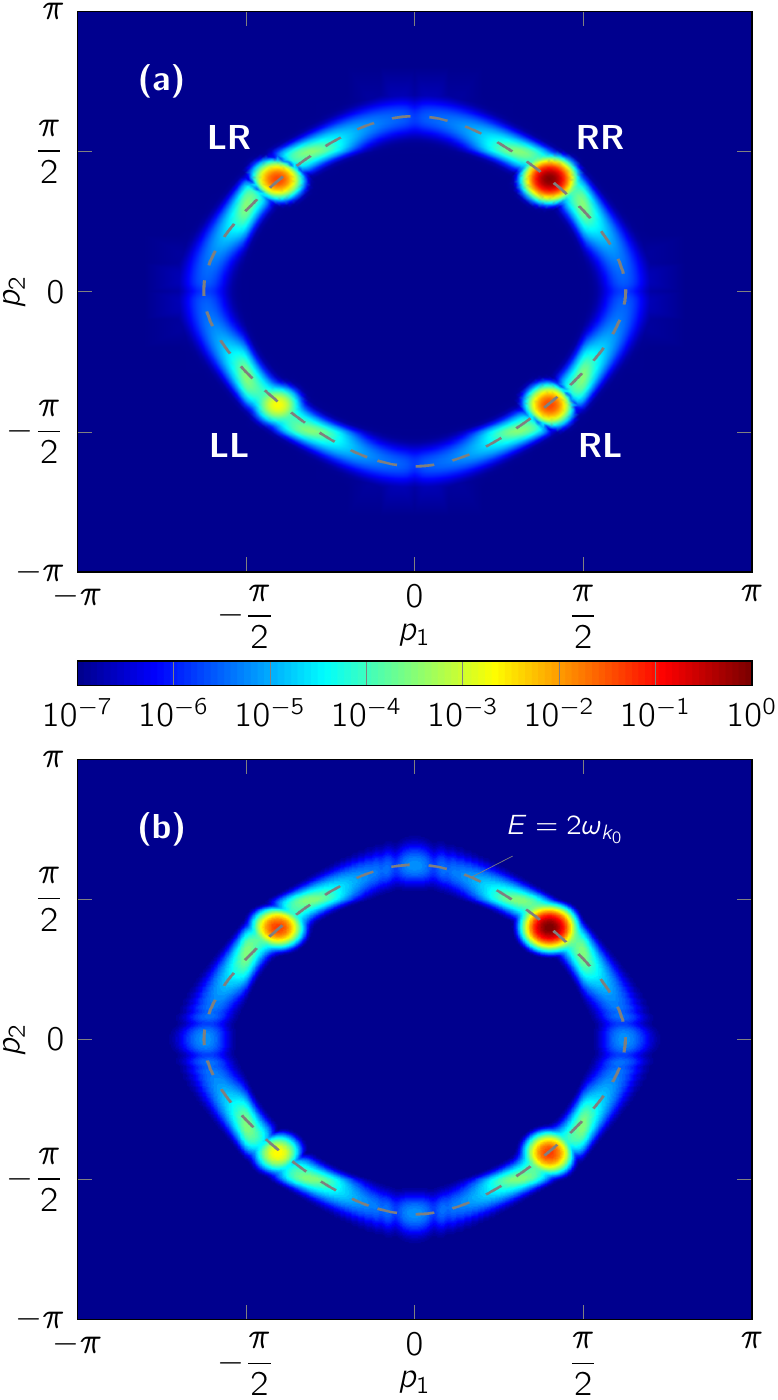}
\caption{(Color online) (a) Spectrum of free to free scattering obtained from \protect\eqref{eq:free_to_free_result} with the parameters listed in the text. (b) Spectrum of free to free scattering obtained from Krylov-subspace based simulations, by taking the FFT of two-photon scattering part of the wave function. Logarithmic colormap is used in both plots.} \label{fig:free_to_freeResults}
\end{figure}

\section{Conclusion}\label{sec:conclusion}

In this manuscript, we aimed to provide a synoptic vision of qubit-photon scattering where photons occupy the dispersive modes of a waveguide. We provided the general framework of one- and two-photon scattering in terms of the resolvent of the Hamiltonian, while linking the resolvent approach to a variety of cases such as the time evolution of states, definition of scattering matrix elements, multi-channel scattering in the presence of atom-photon bound modes and Feynman diagrams. We verified our formulas by independent numerical simulations and have shown their validity.

Our derivations follow the steps taken in the analysis of the Lee model, but for the particular case of photon-qubit interactions defined by a multi-mode Jaynes-Cummings Hamiltonian. Although elementary particles are represented as ``mere shadows of their true selves'' within the Lee model \cite{Amado1963}, the Jaynes-Cummings Hamiltonian has been verified in a number of experiments \cite{Lodahl2015}.

The energy levels and spatial profiles of photon-atom bound states play an important role in the scattering of free photons from such states. The energy level of the bound state was made to intersect the continuum of propagating states via a frequency dependent coupling of the qubit to the photons \cite{Longhi2007}. Localized eigenstates were shown to appear in a waveguide with linear dispersion and two qubits in them \cite{Gonzalez-Ballestero2013}. It would be of interest to study the dynamics of photon-atom bound states for two or more qubits \cite{Vega2014,Calajo2015,Shi2015b} in the presence of modal dispersion and frequency dependent photon-qubit coupling. Recent analyses show that a periodical drive of many atoms coupled to a nanophotonic waveguide can lead to the formation of topological states enabled via atom-atom interactions obtained through the help of the bound states \cite{Grass2015}. Scattering of one- and two-photon pulses from such systems could lead to novel effects that can have experimental significance in waveguides carved into photonic band-gap structures where precise control over dispersion is possible.

Correlations among scattered photons were shown to depend on the dispersive characteristics of the waveguide \cite{Moeferdt2013}. An analysis of pulse shape effects \cite{Nysteen2015b} on the correlations among two or more photons \cite{Shen2015a,Xu2015} scattering off of qubits (or atoms with a more complicated energy level structure \cite{Obi2015}) in a dispersive waveguide, can lead to useful results for quantum information and quantum communication proposals \cite{Chang2014} where universal gate designs based on qubit integrated nanophotonic waveguides have been presented \cite{Paulisch2015}.

\begin{acknowledgments}
The author would like to thank H{\"u}meyra \c{C}a\u{g}layan for encouraging him to look into the effects of modal dispersion on photon-qubit interactions in waveguides. The author also thanks Emre Mengi for his valuable comments on the numerical solution of integral equations.

This manuscript is dedicated to all those who lost their lives and those who got injured during the vicious bomb attack at the peace rally in Ankara, on Oct 10, 2015.
\end{acknowledgments}

\appendix

\section{Properties of \texorpdfstring{\boldmath$I(z)$}{I(z)}}\label{App:Iz}
The definition of $I(z)$ is given by \eqref{eq:Iz}. For $\abs{z}>2J$ there is no singularity in the integrand and the integral can be calculated by using the residue theorem. For $\abs{z}<2J$ with $\Real(z)=0$, one needs to add an infinitesimally small imaginary part to $z$ to get\footnote{See Sec 5.3.1 in \cite{Economou2006}.}
\begin{align*}
I(z) = \lim_{\eta\rightarrow 0^+} \int_{-\pi}^\pi \dif k \frac{1}{z - \omega_k +\mi \eta},
\end{align*}
so that the integral is well defined. We again use the residue theorem, choose the pole in the upper half part of the complex plane enclosed within the contour composed of joining the points $(-\pi,\mi \infty),(-\pi,0),(\pi,0),(\pi,\mi \infty)$ where we specify the real and imaginary parts of points as 2D coordinates. The end result is the following expression for $I(z)$
\begin{align}
\label{eq:Izexpand}I(z)= \begin{cases}
\displaystyle\frac{-2 \pi}{\sqrt{z^2-4 J^2}}& z<-2 J \\
\displaystyle\frac{-2 \pi \mi}{\sqrt{4 J^2-z^2}}& -2 J < z< 2 J \\
\displaystyle\frac{+2 \pi}{\sqrt{z^2-4 J^2}}& 2J <z.
\end{cases}
\end{align}

\section{Properties of the Bound States} \label{App:bound}
In this appendix we will briefly summarize the properties of the atom-photon bound state for the case when $\Omega=0$. Chapter 6 of \cite{Economou2006} and \cite{Lombardo2014} provide more detailed calculations on the bound states. As noted in the main text, the bound states are associated with the poles of $G_1(z)$. We use \eqref{eq:G1} and \eqref{eq:Izexpand} to get the bound state energies, $\omega_\pm$, for $\Omega=0$ as
\begin{align*}
\omega_+ =-\omega_- = \sqrt{2J^2+\sqrt{4J^4+g'^4}}.
\end{align*}
From the theory of Green's functions we can write
\begin{align*}
\braket{\ua}{\Psi_\pm}\braket{\Psi_\pm}{\ua}=\res\left(\braopket{\ua}{G(z)}{\ua},\omega_\pm\right)=p_b.
\end{align*}
The residues of $G_1(z)$ at $\omega_\pm$ are the same and are given by
\begin{align*}
p_b=\frac{g'^4}{2\omega_\pm^2(\omega_\pm^2-2J^2)},
\end{align*}
with which \eqref{eq:H_residue} can also be shown. Similarly, in order to calculate the photon part of the bound state in $\ket{k}$ representation we need
\begin{align*}
&\braket{p\da}{\Psi_\pm}\braket{\Psi_\pm}{k\da}=\res\left(\braopket{p\da}{G(z)}{k\da},\omega_\pm\right) \\
&=\frac{g^2 \res\bigl(G_1(z),\omega_\pm\bigr)}{(\omega_\pm-\omega_k)(\omega_\pm-\omega_p)} =\frac{g^2 p_b }{(\omega_\pm-\omega_k)(\omega_\pm-\omega_p)},
\end{align*}
where in the last line we used \eqref{eq:G4}. Putting all these results together gives us the bound state as
\begin{align}
\label{eq:bound_p}\ket{\Psi_\pm} = \sqrt{p_b} \ket{\ua} + \sqrt{p_b} g \int_{-\pi}^\pi \dif k \frac{\ket{k\da}}{\omega_\pm+2 J \cos k}.
\end{align}
It can explicitly be shown through integration that $\braket{\Psi_\pm}{\Psi_\pm}=1$, written as \eqref{eq:Psi_norm}.

In order to convert \eqref{eq:bound_p} to real-space representation, we use the relationship
\begin{align*}
\ket{x} = \frac{1}{\sqrt{2\pi}}\int_{-\pi}^\pi \dif k \, \me^{-\mi k x}\ket{k},
\end{align*}
obtained in the continuum limit, $N\rightarrow\infty$ \cite{Lombardo2014}. As a result we have
\begin{align*}
\braket{x\da}{\Psi_\pm}=g' \sqrt{p_b} \int_{-\pi}^\pi \dif k \frac{\me^{\mi k x}}{\omega_\pm+2J \cos k}.
\end{align*}
We evaluate the integral above using the residue theorem to get
\begin{align*}
\braket{x\da}{\Psi_+} = \frac{g' \sqrt{p_b}}{\sqrt{\omega_+^2-4J^2}} \left(-\frac{\omega_+}{2J}+\sqrt{\frac{\omega_+^2}{4J^2}-1} \right)^{\abs{x}},
\intertext{and}
\braket{x\da}{\Psi_-} = \frac{-g' \sqrt{p_b}}{\sqrt{\omega_+^2-4J^2}} \left(\frac{\omega_+}{2J}-\sqrt{\frac{\omega_+^2}{4J^2}-1} \right)^{\abs{x}}.
\end{align*}
These real-space functions denote localized photons around the qubit at $x=0$ as plotted in Fig \ref{fig:BoundStateSpace}. Note that although we took the continuum limit for the $k$ variable, the $x$ variable is still discrete, and summations should be used to show $\braket{\Psi_\pm}{\Psi_\pm}=1$ in the $\ket{x}$ representation. Lastly, the definitions of the square roots on p. 88 of \cite{Economou2006} are unconventional, nevertheless the results there agree with the results in this appendix, written in terms of the conventional positive square root definition.

\section{Details for the Calculation of \texorpdfstring{\boldmath $e(t)$}{e(t)}} \label{App:spontaneous}
In this appendix we will provide an expression for $e(t)$, the inverse Laplace transform of \eqref{eq:time_evolve}. We split \eqref{eq:time_evolve} into two parts. The first part can be inverted via partial fraction expansion to get
\begin{align}
\label{eq:laplace1}&\mathscr{L}^{-1}\left\{ \frac{s(s^2+4J^2)}{s^4+4J^2s^2-g'^4}\right\}(t) \\ 
&= \frac{\sqrt{4J^4+g'^4}+2J^2}{2 \sqrt{4J^4+g'^4}} \cosh s_1 t +\frac{\sqrt{4J^4+g'^4}-2J^2}{2 \sqrt{4J^4+g'^4}} \cos s_2 t, \nonumber
\end{align}
where 
\begin{align*}
s_1  = \sqrt{\sqrt{4J^4+g'^4}-2J^2}, \quad
s_2  = \sqrt{\sqrt{4J^4+g'^4}+2J^2}.
\end{align*}
In order to invert the second part, we make the substitution $v=\sqrt{s^2+4J^2}$ and get
\begin{align*}
\mathscr{L}^{-1}\left\{ \frac{-g'^2 v}{v^4-4J^2v^2-g'^4} \right\}(t).
\end{align*}
We apply partial fraction expansion in terms of $v$ to the expression above. We then re-write the result in terms of $s$ to arrive at
\begin{align*}
\frac{-g'^2}{2\sqrt{4J^4+g'^4}}\mathscr{L}^{-1}\Bigl\{ \frac{1}{\sqrt{s^2+4J^2}} \Biggl[\frac{s_1^2}{s^2+s_2^2}+\frac{s_2^2}{s^2-s_1^2} \Biggr]\Bigr\}(t).
\end{align*}
We use the property that the multiplication in Laplace domain corresponds to convolution in time domain to write the inverse Laplace transform of the second part as
\begin{align}
\label{eq:laplace2}&=\frac{-g'^2}{2\sqrt{4J^4+g'^4}}\\
&\times\int_0^t \dif \tau \, J_0(2 J \tau)\left[ \frac{s_1^2}{s_2} \sin s_2(t-\tau) + \frac{s_2^2}{s_1} \sinh s_1(t-\tau) \right], \nonumber
\end{align}
where $J_0(t) = \mathscr{L}^{-1}\bigl\{(s^2+1)^{-1/2}\bigr\}$ is the Bessel function of the first kind of order 0. Addition of \eqref{eq:laplace1} and \eqref{eq:laplace2} gives us $e(t)$. We used high precision arithmetic in the numerical integration that led to the generation of Fig \ref{fig:BoundStateTime}.

%\bibliography{./dispersive_waveguideQED}

\begin{thebibliography}{95}%
\makeatletter
\providecommand \@ifxundefined [1]{%
 \@ifx{#1\undefined}
}%
\providecommand \@ifnum [1]{%
 \ifnum #1\expandafter \@firstoftwo
 \else \expandafter \@secondoftwo
 \fi
}%
\providecommand \@ifx [1]{%
 \ifx #1\expandafter \@firstoftwo
 \else \expandafter \@secondoftwo
 \fi
}%
\providecommand \natexlab [1]{#1}%
\providecommand \enquote  [1]{``#1''}%
\providecommand \bibnamefont  [1]{#1}%
\providecommand \bibfnamefont [1]{#1}%
\providecommand \citenamefont [1]{#1}%
\providecommand \href@noop [0]{\@secondoftwo}%
\providecommand \href [0]{\begingroup \@sanitize@url \@href}%
\providecommand \@href[1]{\@@startlink{#1}\@@href}%
\providecommand \@@href[1]{\endgroup#1\@@endlink}%
\providecommand \@sanitize@url [0]{\catcode `\\12\catcode `\$12\catcode
  `\&12\catcode `\#12\catcode `\^12\catcode `\_12\catcode `\%12\relax}%
\providecommand \@@startlink[1]{}%
\providecommand \@@endlink[0]{}%
\providecommand \url  [0]{\begingroup\@sanitize@url \@url }%
\providecommand \@url [1]{\endgroup\@href {#1}{\urlprefix }}%
\providecommand \urlprefix  [0]{URL }%
\providecommand \Eprint [0]{\href }%
\providecommand \doibase [0]{http://dx.doi.org/}%
\providecommand \selectlanguage [0]{\@gobble}%
\providecommand \bibinfo  [0]{\@secondoftwo}%
\providecommand \bibfield  [0]{\@secondoftwo}%
\providecommand \translation [1]{[#1]}%
\providecommand \BibitemOpen [0]{}%
\providecommand \bibitemStop [0]{}%
\providecommand \bibitemNoStop [0]{.\EOS\space}%
\providecommand \EOS [0]{\spacefactor3000\relax}%
\providecommand \BibitemShut  [1]{\csname bibitem#1\endcsname}%
\let\auto@bib@innerbib\@empty
%</preamble>
\bibitem [{\citenamefont {O'Shea}\ \emph {et~al.}(2013)\citenamefont {O'Shea},
  \citenamefont {Junge}, \citenamefont {Volz},\ and\ \citenamefont
  {Rauschenbeutel}}]{OShea2013}%
  \BibitemOpen
  \bibfield  {author} {\bibinfo {author} {\bibfnamefont {D.}~\bibnamefont
  {O'Shea}}, \bibinfo {author} {\bibfnamefont {C.}~\bibnamefont {Junge}},
  \bibinfo {author} {\bibfnamefont {J.}~\bibnamefont {Volz}}, \ and\ \bibinfo
  {author} {\bibfnamefont {A.}~\bibnamefont {Rauschenbeutel}},\ }\bibfield
  {title} {\emph {\bibinfo {title} {Fiber-optical switch controlled by a single
  atom},\ }}\href {\doibase 10.1103/PhysRevLett.111.193601} {\bibfield
  {journal} {\bibinfo  {journal} {Phys. Rev. Lett.}\ }\textbf {\bibinfo
  {volume} {111}},\ \bibinfo {pages} {193601} (\bibinfo {year}
  {2013})}\BibitemShut {NoStop}%
\bibitem [{\citenamefont {Makhonin}\ \emph {et~al.}(2014)\citenamefont
  {Makhonin}, \citenamefont {Dixon}, \citenamefont {Coles}, \citenamefont
  {Royall}, \citenamefont {Luxmoore}, \citenamefont {Clarke}, \citenamefont
  {Hugues}, \citenamefont {Skolnick},\ and\ \citenamefont
  {Fox}}]{Makhonin2014}%
  \BibitemOpen
  \bibfield  {author} {\bibinfo {author} {\bibfnamefont {M.~N.}\ \bibnamefont
  {Makhonin}}, \bibinfo {author} {\bibfnamefont {J.~E.}\ \bibnamefont {Dixon}},
  \bibinfo {author} {\bibfnamefont {R.~J.}\ \bibnamefont {Coles}}, \bibinfo
  {author} {\bibfnamefont {B.}~\bibnamefont {Royall}}, \bibinfo {author}
  {\bibfnamefont {I.~J.}\ \bibnamefont {Luxmoore}}, \bibinfo {author}
  {\bibfnamefont {E.}~\bibnamefont {Clarke}}, \bibinfo {author} {\bibfnamefont
  {M.}~\bibnamefont {Hugues}}, \bibinfo {author} {\bibfnamefont {M.~S.}\
  \bibnamefont {Skolnick}}, \ and\ \bibinfo {author} {\bibfnamefont {A.~M.}\
  \bibnamefont {Fox}},\ }\bibfield  {title} {\emph {\bibinfo {title} {Waveguide
  coupled resonance fluorescence from on-chip quantum emitter},\ }}\href
  {\doibase 10.1021/nl5032937} {\bibfield  {journal} {\bibinfo  {journal} {Nano
  Letters}\ }\textbf {\bibinfo {volume} {14}},\ \bibinfo {pages} {6997}
  (\bibinfo {year} {2014})}\BibitemShut {NoStop}%
\bibitem [{\citenamefont {Goban}\ \emph {et~al.}(2015)\citenamefont {Goban},
  \citenamefont {Hung}, \citenamefont {Hood}, \citenamefont {Yu}, \citenamefont
  {Muniz}, \citenamefont {Painter},\ and\ \citenamefont {Kimble}}]{Goban2015}%
  \BibitemOpen
  \bibfield  {author} {\bibinfo {author} {\bibfnamefont {A.}~\bibnamefont
  {Goban}}, \bibinfo {author} {\bibfnamefont {C.-L.}\ \bibnamefont {Hung}},
  \bibinfo {author} {\bibfnamefont {J.~D.}\ \bibnamefont {Hood}}, \bibinfo
  {author} {\bibfnamefont {S.-P.}\ \bibnamefont {Yu}}, \bibinfo {author}
  {\bibfnamefont {J.~A.}\ \bibnamefont {Muniz}}, \bibinfo {author}
  {\bibfnamefont {O.}~\bibnamefont {Painter}}, \ and\ \bibinfo {author}
  {\bibfnamefont {H.~J.}\ \bibnamefont {Kimble}},\ }\bibfield  {title} {\emph
  {\bibinfo {title} {Superradiance for atoms trapped along a photonic crystal
  waveguide},\ }}\href {\doibase 10.1103/PhysRevLett.115.063601} {\bibfield
  {journal} {\bibinfo  {journal} {Phys. Rev. Lett.}\ }\textbf {\bibinfo
  {volume} {115}},\ \bibinfo {pages} {063601} (\bibinfo {year}
  {2015})}\BibitemShut {NoStop}%
\bibitem [{\citenamefont {Kress}\ \emph {et~al.}(2015)\citenamefont {Kress},
  \citenamefont {Antolinez}, \citenamefont {Richner}, \citenamefont {Jayanti},
  \citenamefont {Kim}, \citenamefont {Prins}, \citenamefont {Riedinger},
  \citenamefont {Fischer}, \citenamefont {Meyer}, \citenamefont {McPeak},
  \citenamefont {Poulikakos},\ and\ \citenamefont {Norris}}]{Kress2015}%
  \BibitemOpen
  \bibfield  {author} {\bibinfo {author} {\bibfnamefont {S.~J.~P.}\
  \bibnamefont {Kress}}, \bibinfo {author} {\bibfnamefont {F.~V.}\ \bibnamefont
  {Antolinez}}, \bibinfo {author} {\bibfnamefont {P.}~\bibnamefont {Richner}},
  \bibinfo {author} {\bibfnamefont {S.~V.}\ \bibnamefont {Jayanti}}, \bibinfo
  {author} {\bibfnamefont {D.~K.}\ \bibnamefont {Kim}}, \bibinfo {author}
  {\bibfnamefont {F.}~\bibnamefont {Prins}}, \bibinfo {author} {\bibfnamefont
  {A.}~\bibnamefont {Riedinger}}, \bibinfo {author} {\bibfnamefont {M.~P.~C.}\
  \bibnamefont {Fischer}}, \bibinfo {author} {\bibfnamefont {S.}~\bibnamefont
  {Meyer}}, \bibinfo {author} {\bibfnamefont {K.~M.}\ \bibnamefont {McPeak}},
  \bibinfo {author} {\bibfnamefont {D.}~\bibnamefont {Poulikakos}}, \ and\
  \bibinfo {author} {\bibfnamefont {D.~J.}\ \bibnamefont {Norris}},\ }\bibfield
   {title} {\emph {\bibinfo {title} {Wedge waveguides and resonators for
  quantum plasmonics},\ }}\href {\doibase 10.1021/acs.nanolett.5b03051}
  {\bibfield  {journal} {\bibinfo  {journal} {Nano Letters}\ }\textbf {\bibinfo
  {volume} {15}},\ \bibinfo {pages} {6267} (\bibinfo {year}
  {2015})}\BibitemShut {NoStop}%
\bibitem [{\citenamefont {S{\"o}llner}\ \emph {et~al.}(2015)\citenamefont
  {S{\"o}llner}, \citenamefont {Mahmoodian}, \citenamefont {Hansen},
  \citenamefont {Midolo}, \citenamefont {Javadi}, \citenamefont
  {Kir{\v{s}}ansk{\.e}}, \citenamefont {Pregnolato}, \citenamefont {El-Ella},
  \citenamefont {Lee}, \citenamefont {Song}, \citenamefont {Stobbe},\ and\
  \citenamefont {Lodahl}}]{Sollner2015}%
  \BibitemOpen
  \bibfield  {author} {\bibinfo {author} {\bibfnamefont {I.}~\bibnamefont
  {S{\"o}llner}}, \bibinfo {author} {\bibfnamefont {S.}~\bibnamefont
  {Mahmoodian}}, \bibinfo {author} {\bibfnamefont {S.~L.}\ \bibnamefont
  {Hansen}}, \bibinfo {author} {\bibfnamefont {L.}~\bibnamefont {Midolo}},
  \bibinfo {author} {\bibfnamefont {A.}~\bibnamefont {Javadi}}, \bibinfo
  {author} {\bibfnamefont {G.}~\bibnamefont {Kir{\v{s}}ansk{\.e}}}, \bibinfo
  {author} {\bibfnamefont {T.}~\bibnamefont {Pregnolato}}, \bibinfo {author}
  {\bibfnamefont {H.}~\bibnamefont {El-Ella}}, \bibinfo {author} {\bibfnamefont
  {E.~H.}\ \bibnamefont {Lee}}, \bibinfo {author} {\bibfnamefont {J.~D.}\
  \bibnamefont {Song}}, \bibinfo {author} {\bibfnamefont {S.}~\bibnamefont
  {Stobbe}}, \ and\ \bibinfo {author} {\bibfnamefont {P.}~\bibnamefont
  {Lodahl}},\ }\bibfield  {title} {\emph {\bibinfo {title} {Deterministic
  photon-emitter coupling in chiral photonic circuits},\ }}\href
  {http://dx.doi.org/10.1038/nnano.2015.159} {\bibfield  {journal} {\bibinfo
  {journal} {Nat Nano}\ }\textbf {\bibinfo {volume} {10}},\ \bibinfo {pages}
  {775} (\bibinfo {year} {2015})}\BibitemShut {NoStop}%
\bibitem [{\citenamefont {Lodahl}\ \emph {et~al.}(2015)\citenamefont {Lodahl},
  \citenamefont {Mahmoodian},\ and\ \citenamefont {Stobbe}}]{Lodahl2015}%
  \BibitemOpen
  \bibfield  {author} {\bibinfo {author} {\bibfnamefont {P.}~\bibnamefont
  {Lodahl}}, \bibinfo {author} {\bibfnamefont {S.}~\bibnamefont {Mahmoodian}},
  \ and\ \bibinfo {author} {\bibfnamefont {S.}~\bibnamefont {Stobbe}},\
  }\bibfield  {title} {\emph {\bibinfo {title} {Interfacing single photons and
  single quantum dots with photonic nanostructures},\ }}\href {\doibase
  10.1103/RevModPhys.87.347} {\bibfield  {journal} {\bibinfo  {journal} {Rev.
  Mod. Phys.}\ }\textbf {\bibinfo {volume} {87}},\ \bibinfo {pages} {347}
  (\bibinfo {year} {2015})}\BibitemShut {NoStop}%
\bibitem [{\citenamefont {Tame}\ \emph {et~al.}(2013)\citenamefont {Tame},
  \citenamefont {McEnery}, \citenamefont {Ozdemir}, \citenamefont {Lee},
  \citenamefont {Maier},\ and\ \citenamefont {Kim}}]{Tame2013}%
  \BibitemOpen
  \bibfield  {author} {\bibinfo {author} {\bibfnamefont {M.~S.}\ \bibnamefont
  {Tame}}, \bibinfo {author} {\bibfnamefont {K.~R.}\ \bibnamefont {McEnery}},
  \bibinfo {author} {\bibfnamefont {S.~K.}\ \bibnamefont {Ozdemir}}, \bibinfo
  {author} {\bibfnamefont {J.}~\bibnamefont {Lee}}, \bibinfo {author}
  {\bibfnamefont {S.~A.}\ \bibnamefont {Maier}}, \ and\ \bibinfo {author}
  {\bibfnamefont {M.~S.}\ \bibnamefont {Kim}},\ }\bibfield  {title} {\emph
  {\bibinfo {title} {Quantum plasmonics},\ }}\href
  {http://dx.doi.org/10.1038/nphys2615} {\bibfield  {journal} {\bibinfo
  {journal} {Nat Phys}\ }\textbf {\bibinfo {volume} {9}},\ \bibinfo {pages}
  {329} (\bibinfo {year} {2013})}\BibitemShut {NoStop}%
\bibitem [{\citenamefont {Walmsley}(2015)}]{Walmsley2015}%
  \BibitemOpen
  \bibfield  {author} {\bibinfo {author} {\bibfnamefont {I.~A.}\ \bibnamefont
  {Walmsley}},\ }\bibfield  {title} {\emph {\bibinfo {title} {Quantum optics:
  Science and technology in a new light},\ }}\href {\doibase
  10.1126/science.aab0097} {\bibfield  {journal} {\bibinfo  {journal}
  {Science}\ }\textbf {\bibinfo {volume} {348}},\ \bibinfo {pages} {525}
  (\bibinfo {year} {2015})}\BibitemShut {NoStop}%
\bibitem [{\citenamefont {Douglas}\ \emph {et~al.}(2015)\citenamefont
  {Douglas}, \citenamefont {Habibian}, \citenamefont {Hung}, \citenamefont
  {Gorshkov}, \citenamefont {Kimble},\ and\ \citenamefont
  {Chang}}]{Douglas2015}%
  \BibitemOpen
  \bibfield  {author} {\bibinfo {author} {\bibfnamefont {S.~J.}\ \bibnamefont
  {Douglas}}, \bibinfo {author} {\bibfnamefont {H.}~\bibnamefont {Habibian}},
  \bibinfo {author} {\bibfnamefont {C.~L.}\ \bibnamefont {Hung}}, \bibinfo
  {author} {\bibfnamefont {A.~V.}\ \bibnamefont {Gorshkov}}, \bibinfo {author}
  {\bibfnamefont {H.~J.}\ \bibnamefont {Kimble}}, \ and\ \bibinfo {author}
  {\bibfnamefont {D.~E.}\ \bibnamefont {Chang}},\ }\bibfield  {title} {\emph
  {\bibinfo {title} {Quantum many-body models with cold atoms coupled to
  photonic crystals},\ }}\href {http://dx.doi.org/10.1038/nphoton.2015.57}
  {\bibfield  {journal} {\bibinfo  {journal} {Nat Photon}\ }\textbf {\bibinfo
  {volume} {9}},\ \bibinfo {pages} {326} (\bibinfo {year} {2015})}\BibitemShut
  {NoStop}%
\bibitem [{\citenamefont {Chang}\ \emph {et~al.}(2014)\citenamefont {Chang},
  \citenamefont {Vuletic},\ and\ \citenamefont {Lukin}}]{Chang2014}%
  \BibitemOpen
  \bibfield  {author} {\bibinfo {author} {\bibfnamefont {D.~E.}\ \bibnamefont
  {Chang}}, \bibinfo {author} {\bibfnamefont {V.}~\bibnamefont {Vuletic}}, \
  and\ \bibinfo {author} {\bibfnamefont {M.~D.}\ \bibnamefont {Lukin}},\
  }\bibfield  {title} {\emph {\bibinfo {title} {Quantum nonlinear optics ---
  photon by photon},\ }}\href {http://dx.doi.org/10.1038/nphoton.2014.192}
  {\bibfield  {journal} {\bibinfo  {journal} {Nat Photon}\ }\textbf {\bibinfo
  {volume} {8}},\ \bibinfo {pages} {685} (\bibinfo {year} {2014})}\BibitemShut
  {NoStop}%
\bibitem [{\citenamefont {John}\ and\ \citenamefont {Wang}(1991)}]{John1991}%
  \BibitemOpen
  \bibfield  {author} {\bibinfo {author} {\bibfnamefont {S.}~\bibnamefont
  {John}}\ and\ \bibinfo {author} {\bibfnamefont {J.}~\bibnamefont {Wang}},\
  }\bibfield  {title} {\emph {\bibinfo {title} {Quantum optics of localized
  light in a photonic band gap},\ }}\href {\doibase 10.1103/PhysRevB.43.12772}
  {\bibfield  {journal} {\bibinfo  {journal} {Phys. Rev. B}\ }\textbf {\bibinfo
  {volume} {43}},\ \bibinfo {pages} {12772} (\bibinfo {year}
  {1991})}\BibitemShut {NoStop}%
\bibitem [{\citenamefont {John}\ and\ \citenamefont {Quang}(1994)}]{John1994}%
  \BibitemOpen
  \bibfield  {author} {\bibinfo {author} {\bibfnamefont {S.}~\bibnamefont
  {John}}\ and\ \bibinfo {author} {\bibfnamefont {T.}~\bibnamefont {Quang}},\
  }\bibfield  {title} {\emph {\bibinfo {title} {Spontaneous emission near the
  edge of a photonic band gap},\ }}\href {\doibase 10.1103/PhysRevA.50.1764}
  {\bibfield  {journal} {\bibinfo  {journal} {Phys. Rev. A}\ }\textbf {\bibinfo
  {volume} {50}},\ \bibinfo {pages} {1764} (\bibinfo {year}
  {1994})}\BibitemShut {NoStop}%
\bibitem [{\citenamefont {Kofman}\ \emph {et~al.}(1994)\citenamefont {Kofman},
  \citenamefont {Kurizki},\ and\ \citenamefont {Sherman}}]{Kofman1994}%
  \BibitemOpen
  \bibfield  {author} {\bibinfo {author} {\bibfnamefont {A.~G.}\ \bibnamefont
  {Kofman}}, \bibinfo {author} {\bibfnamefont {G.}~\bibnamefont {Kurizki}}, \
  and\ \bibinfo {author} {\bibfnamefont {B.}~\bibnamefont {Sherman}},\
  }\bibfield  {title} {\emph {\bibinfo {title} {Spontaneous and induced atomic
  decay in photonic band structures},\ }}\href {\doibase
  10.1080/09500349414550381} {\bibfield  {journal} {\bibinfo  {journal}
  {Journal of Modern Optics}\ }\textbf {\bibinfo {volume} {41}},\ \bibinfo
  {pages} {353} (\bibinfo {year} {1994})}\BibitemShut {NoStop}%
\bibitem [{\citenamefont {Gaveau}\ and\ \citenamefont
  {Schulman}(1995)}]{Gaveau1995}%
  \BibitemOpen
  \bibfield  {author} {\bibinfo {author} {\bibfnamefont {B.}~\bibnamefont
  {Gaveau}}\ and\ \bibinfo {author} {\bibfnamefont {L.~S.}\ \bibnamefont
  {Schulman}},\ }\bibfield  {title} {\emph {\bibinfo {title} {Limited quantum
  decay},\ }}\href {http://stacks.iop.org/0305-4470/28/i=24/a=029} {\bibfield
  {journal} {\bibinfo  {journal} {Journal of Physics A: Mathematical and
  General}\ }\textbf {\bibinfo {volume} {28}},\ \bibinfo {pages} {7359}
  (\bibinfo {year} {1995})}\BibitemShut {NoStop}%
\bibitem [{\citenamefont {Lambropoulos}\ \emph {et~al.}(2000)\citenamefont
  {Lambropoulos}, \citenamefont {Nikolopoulos}, \citenamefont {Nielsen},\ and\
  \citenamefont {Bay}}]{Lambropoulos2000}%
  \BibitemOpen
  \bibfield  {author} {\bibinfo {author} {\bibfnamefont {P.}~\bibnamefont
  {Lambropoulos}}, \bibinfo {author} {\bibfnamefont {G.~M.}\ \bibnamefont
  {Nikolopoulos}}, \bibinfo {author} {\bibfnamefont {T.~R.}\ \bibnamefont
  {Nielsen}}, \ and\ \bibinfo {author} {\bibfnamefont {S.}~\bibnamefont
  {Bay}},\ }\bibfield  {title} {\emph {\bibinfo {title} {Fundamental quantum
  optics in structured reservoirs},\ }}\href
  {http://stacks.iop.org/0034-4885/63/i=4/a=201} {\bibfield  {journal}
  {\bibinfo  {journal} {Reports on Progress in Physics}\ }\textbf {\bibinfo
  {volume} {63}},\ \bibinfo {pages} {455} (\bibinfo {year} {2000})}\BibitemShut
  {NoStop}%
\bibitem [{\citenamefont {Vats}\ \emph {et~al.}(2002)\citenamefont {Vats},
  \citenamefont {John},\ and\ \citenamefont {Busch}}]{Vats2002}%
  \BibitemOpen
  \bibfield  {author} {\bibinfo {author} {\bibfnamefont {N.}~\bibnamefont
  {Vats}}, \bibinfo {author} {\bibfnamefont {S.}~\bibnamefont {John}}, \ and\
  \bibinfo {author} {\bibfnamefont {K.}~\bibnamefont {Busch}},\ }\bibfield
  {title} {\emph {\bibinfo {title} {Theory of fluorescence in photonic
  crystals},\ }}\href {\doibase 10.1103/PhysRevA.65.043808} {\bibfield
  {journal} {\bibinfo  {journal} {Phys. Rev. A}\ }\textbf {\bibinfo {volume}
  {65}},\ \bibinfo {pages} {043808} (\bibinfo {year} {2002})}\BibitemShut
  {NoStop}%
\bibitem [{\citenamefont {Kofman}\ and\ \citenamefont
  {Kurizki}(1996)}]{Kofman1996}%
  \BibitemOpen
  \bibfield  {author} {\bibinfo {author} {\bibfnamefont {A.~G.}\ \bibnamefont
  {Kofman}}\ and\ \bibinfo {author} {\bibfnamefont {G.}~\bibnamefont
  {Kurizki}},\ }\bibfield  {title} {\emph {\bibinfo {title} {Quantum zeno
  effect on atomic excitation decay in resonators},\ }}\href {\doibase
  10.1103/PhysRevA.54.R3750} {\bibfield  {journal} {\bibinfo  {journal} {Phys.
  Rev. A}\ }\textbf {\bibinfo {volume} {54}},\ \bibinfo {pages} {R3750}
  (\bibinfo {year} {1996})}\BibitemShut {NoStop}%
\bibitem [{\citenamefont {Shen}\ and\ \citenamefont {Fan}(2007)}]{Shen2007}%
  \BibitemOpen
  \bibfield  {author} {\bibinfo {author} {\bibfnamefont {J.~T.}\ \bibnamefont
  {Shen}}\ and\ \bibinfo {author} {\bibfnamefont {S.}~\bibnamefont {Fan}},\
  }\bibfield  {title} {\emph {\bibinfo {title} {{Strongly correlated
  multiparticle transport in one dimension through a quantum impurity}},\
  }}\href {\doibase 10.1103/PhysRevA.76.062709} {\bibfield  {journal} {\bibinfo
   {journal} {Physical Review A}\ }\textbf {\bibinfo {volume} {76}},\ \bibinfo
  {pages} {062709} (\bibinfo {year} {2007})}\BibitemShut {NoStop}%
\bibitem [{\citenamefont {Shi}\ and\ \citenamefont
  {Sun}(2009{\natexlab{a}})}]{Shi2009}%
  \BibitemOpen
  \bibfield  {author} {\bibinfo {author} {\bibfnamefont {T.}~\bibnamefont
  {Shi}}\ and\ \bibinfo {author} {\bibfnamefont {C.~P.}\ \bibnamefont {Sun}},\
  }\bibfield  {title} {\emph {\bibinfo {title} {{Lehmann-Symanzik-Zimmermann}
  reduction approach to multiphoton scattering in coupled-resonator arrays},\
  }}\href {\doibase 10.1103/PhysRevB.79.205111} {\bibfield  {journal} {\bibinfo
   {journal} {Phys. Rev. B}\ }\textbf {\bibinfo {volume} {79}},\ \bibinfo
  {pages} {205111} (\bibinfo {year} {2009}{\natexlab{a}})}\BibitemShut
  {NoStop}%
\bibitem [{\citenamefont {Fan}\ \emph {et~al.}(2010)\citenamefont {Fan},
  \citenamefont {Kocaba\c{s}},\ and\ \citenamefont {Shen}}]{Fan2010}%
  \BibitemOpen
  \bibfield  {author} {\bibinfo {author} {\bibfnamefont {S.}~\bibnamefont
  {Fan}}, \bibinfo {author} {\bibfnamefont {{\c{S}}.~E.}\ \bibnamefont
  {Kocaba\c{s}}}, \ and\ \bibinfo {author} {\bibfnamefont {J.-T.}\ \bibnamefont
  {Shen}},\ }\bibfield  {title} {\emph {\bibinfo {title} {Input-output
  formalism for few-photon transport in one-dimensional nanophotonic waveguides
  coupled to a qubit},\ }}\href {\doibase 10.1103/PhysRevA.82.063821}
  {\bibfield  {journal} {\bibinfo  {journal} {Phys. Rev. A}\ }\textbf {\bibinfo
  {volume} {82}},\ \bibinfo {pages} {063821} (\bibinfo {year}
  {2010})}\BibitemShut {NoStop}%
\bibitem [{\citenamefont {Pletyukhov}\ and\ \citenamefont
  {Gritsev}(2012)}]{Pletyukhov2012}%
  \BibitemOpen
  \bibfield  {author} {\bibinfo {author} {\bibfnamefont {M.}~\bibnamefont
  {Pletyukhov}}\ and\ \bibinfo {author} {\bibfnamefont {V.}~\bibnamefont
  {Gritsev}},\ }\bibfield  {title} {\emph {\bibinfo {title} {Scattering of
  massless particles in one-dimensional chiral channel},\ }}\href
  {http://stacks.iop.org/1367-2630/14/i=9/a=095028} {\bibfield  {journal}
  {\bibinfo  {journal} {New Journal of Physics}\ }\textbf {\bibinfo {volume}
  {14}},\ \bibinfo {pages} {095028} (\bibinfo {year} {2012})}\BibitemShut
  {NoStop}%
\bibitem [{\citenamefont {Chen}\ \emph {et~al.}(2011)\citenamefont {Chen},
  \citenamefont {Wubs}, \citenamefont {M{\o}rk},\ and\ \citenamefont
  {Koenderink}}]{Chen2011a}%
  \BibitemOpen
  \bibfield  {author} {\bibinfo {author} {\bibfnamefont {Y.}~\bibnamefont
  {Chen}}, \bibinfo {author} {\bibfnamefont {M.}~\bibnamefont {Wubs}}, \bibinfo
  {author} {\bibfnamefont {J.}~\bibnamefont {M{\o}rk}}, \ and\ \bibinfo
  {author} {\bibfnamefont {A.~F.}\ \bibnamefont {Koenderink}},\ }\bibfield
  {title} {\emph {\bibinfo {title} {Coherent single-photon absorption by single
  emitters coupled to one-dimensional nanophotonic waveguides},\ }}\href
  {http://stacks.iop.org/1367-2630/13/i=10/a=103010} {\bibfield  {journal}
  {\bibinfo  {journal} {New Journal of Physics}\ }\textbf {\bibinfo {volume}
  {13}},\ \bibinfo {pages} {103010} (\bibinfo {year} {2011})}\BibitemShut
  {NoStop}%
\bibitem [{\citenamefont {Chumak}\ and\ \citenamefont
  {Stolyarov}(2014)}]{Chumak2014}%
  \BibitemOpen
  \bibfield  {author} {\bibinfo {author} {\bibfnamefont {O.~O.}\ \bibnamefont
  {Chumak}}\ and\ \bibinfo {author} {\bibfnamefont {E.~V.}\ \bibnamefont
  {Stolyarov}},\ }\bibfield  {title} {\emph {\bibinfo {title} {Photon
  distribution function for propagation of two-photon pulses in waveguide-qubit
  systems},\ }}\href {\doibase 10.1103/PhysRevA.90.063832} {\bibfield
  {journal} {\bibinfo  {journal} {Phys. Rev. A}\ }\textbf {\bibinfo {volume}
  {90}},\ \bibinfo {pages} {063832} (\bibinfo {year} {2014})}\BibitemShut
  {NoStop}%
\bibitem [{\citenamefont {Gonzalez-Ballestero}\ \emph
  {et~al.}(2013)\citenamefont {Gonzalez-Ballestero}, \citenamefont
  {Garc\'{i}a-Vidal},\ and\ \citenamefont {Moreno}}]{Gonzalez-Ballestero2013}%
  \BibitemOpen
  \bibfield  {author} {\bibinfo {author} {\bibfnamefont {C.}~\bibnamefont
  {Gonzalez-Ballestero}}, \bibinfo {author} {\bibfnamefont {F.~J.}\
  \bibnamefont {Garc\'{i}a-Vidal}}, \ and\ \bibinfo {author} {\bibfnamefont
  {E.}~\bibnamefont {Moreno}},\ }\bibfield  {title} {\emph {\bibinfo {title}
  {Non-markovian effects in waveguide-mediated entanglement},\ }}\href
  {http://stacks.iop.org/1367-2630/15/i=7/a=073015} {\bibfield  {journal}
  {\bibinfo  {journal} {New Journal of Physics}\ }\textbf {\bibinfo {volume}
  {15}},\ \bibinfo {pages} {073015} (\bibinfo {year} {2013})}\BibitemShut
  {NoStop}%
\bibitem [{\citenamefont {Laakso}\ and\ \citenamefont
  {Pletyukhov}(2014)}]{Laakso2014}%
  \BibitemOpen
  \bibfield  {author} {\bibinfo {author} {\bibfnamefont {M.}~\bibnamefont
  {Laakso}}\ and\ \bibinfo {author} {\bibfnamefont {M.}~\bibnamefont
  {Pletyukhov}},\ }\bibfield  {title} {\emph {\bibinfo {title} {Scattering of
  two photons from two distant qubits: Exact solution},\ }}\href {\doibase
  10.1103/PhysRevLett.113.183601} {\bibfield  {journal} {\bibinfo  {journal}
  {Phys. Rev. Lett.}\ }\textbf {\bibinfo {volume} {113}},\ \bibinfo {pages}
  {183601} (\bibinfo {year} {2014})}\BibitemShut {NoStop}%
\bibitem [{\citenamefont {Fang}\ and\ \citenamefont
  {Baranger}(2015)}]{Fang2015a}%
  \BibitemOpen
  \bibfield  {author} {\bibinfo {author} {\bibfnamefont {Y.-L.~L.}\
  \bibnamefont {Fang}}\ and\ \bibinfo {author} {\bibfnamefont {H.~U.}\
  \bibnamefont {Baranger}},\ }\bibfield  {title} {\emph {\bibinfo {title}
  {Waveguide {QED}: Power spectra and correlations of two photons scattered off
  multiple distant qubits and a mirror},\ }}\href {\doibase
  10.1103/PhysRevA.91.053845} {\bibfield  {journal} {\bibinfo  {journal} {Phys.
  Rev. A}\ }\textbf {\bibinfo {volume} {91}},\ \bibinfo {pages} {053845}
  (\bibinfo {year} {2015})}\BibitemShut {NoStop}%
\bibitem [{\citenamefont {Shi}\ \emph {et~al.}(2015{\natexlab{a}})\citenamefont
  {Shi}, \citenamefont {Chang},\ and\ \citenamefont {Cirac}}]{Shi2015}%
  \BibitemOpen
  \bibfield  {author} {\bibinfo {author} {\bibfnamefont {T.}~\bibnamefont
  {Shi}}, \bibinfo {author} {\bibfnamefont {D.~E.}\ \bibnamefont {Chang}}, \
  and\ \bibinfo {author} {\bibfnamefont {J.~I.}\ \bibnamefont {Cirac}},\
  }\bibfield  {title} {\emph {\bibinfo {title} {Multiphoton-scattering theory
  and generalized master equations},\ }}\href {\doibase
  10.1103/PhysRevA.92.053834} {\bibfield  {journal} {\bibinfo  {journal} {Phys.
  Rev. A}\ }\textbf {\bibinfo {volume} {92}},\ \bibinfo {pages} {053834}
  (\bibinfo {year} {2015}{\natexlab{a}})}\BibitemShut {NoStop}%
\bibitem [{\citenamefont {Longo}\ \emph {et~al.}(2009)\citenamefont {Longo},
  \citenamefont {Schmitteckert},\ and\ \citenamefont {Busch}}]{Longo2009}%
  \BibitemOpen
  \bibfield  {author} {\bibinfo {author} {\bibfnamefont {P.}~\bibnamefont
  {Longo}}, \bibinfo {author} {\bibfnamefont {P.}~\bibnamefont
  {Schmitteckert}}, \ and\ \bibinfo {author} {\bibfnamefont {K.}~\bibnamefont
  {Busch}},\ }\bibfield  {title} {\emph {\bibinfo {title} {Dynamics of photon
  transport through quantum impurities in dispersion-engineered one-dimensional
  systems},\ }}\href {http://stacks.iop.org/1464-4258/11/i=11/a=114009}
  {\bibfield  {journal} {\bibinfo  {journal} {Journal of Optics A: Pure and
  Applied Optics}\ }\textbf {\bibinfo {volume} {11}},\ \bibinfo {pages}
  {114009} (\bibinfo {year} {2009})}\BibitemShut {NoStop}%
\bibitem [{\citenamefont {Longo}\ \emph {et~al.}(2010)\citenamefont {Longo},
  \citenamefont {Schmitteckert},\ and\ \citenamefont {Busch}}]{Longo2010}%
  \BibitemOpen
  \bibfield  {author} {\bibinfo {author} {\bibfnamefont {P.}~\bibnamefont
  {Longo}}, \bibinfo {author} {\bibfnamefont {P.}~\bibnamefont
  {Schmitteckert}}, \ and\ \bibinfo {author} {\bibfnamefont {K.}~\bibnamefont
  {Busch}},\ }\bibfield  {title} {\emph {\bibinfo {title} {Few-photon transport
  in low-dimensional systems: Interaction-induced radiation trapping},\ }}\href
  {\doibase 10.1103/PhysRevLett.104.023602} {\bibfield  {journal} {\bibinfo
  {journal} {Phys. Rev. Lett.}\ }\textbf {\bibinfo {volume} {104}},\ \bibinfo
  {pages} {023602} (\bibinfo {year} {2010})}\BibitemShut {NoStop}%
\bibitem [{\citenamefont {Longo}\ \emph {et~al.}(2011)\citenamefont {Longo},
  \citenamefont {Schmitteckert},\ and\ \citenamefont {Busch}}]{Longo2011}%
  \BibitemOpen
  \bibfield  {author} {\bibinfo {author} {\bibfnamefont {P.}~\bibnamefont
  {Longo}}, \bibinfo {author} {\bibfnamefont {P.}~\bibnamefont
  {Schmitteckert}}, \ and\ \bibinfo {author} {\bibfnamefont {K.}~\bibnamefont
  {Busch}},\ }\bibfield  {title} {\emph {\bibinfo {title} {Few-photon transport
  in low-dimensional systems},\ }}\href {\doibase 10.1103/PhysRevA.83.063828}
  {\bibfield  {journal} {\bibinfo  {journal} {Phys. Rev. A}\ }\textbf {\bibinfo
  {volume} {83}},\ \bibinfo {pages} {063828} (\bibinfo {year}
  {2011})}\BibitemShut {NoStop}%
\bibitem [{\citenamefont {Prior}\ \emph {et~al.}(2013)\citenamefont {Prior},
  \citenamefont {de~Vega}, \citenamefont {Chin}, \citenamefont {Huelga},\ and\
  \citenamefont {Plenio}}]{Prior2013}%
  \BibitemOpen
  \bibfield  {author} {\bibinfo {author} {\bibfnamefont {J.}~\bibnamefont
  {Prior}}, \bibinfo {author} {\bibfnamefont {I.}~\bibnamefont {de~Vega}},
  \bibinfo {author} {\bibfnamefont {A.~W.}\ \bibnamefont {Chin}}, \bibinfo
  {author} {\bibfnamefont {S.~F.}\ \bibnamefont {Huelga}}, \ and\ \bibinfo
  {author} {\bibfnamefont {M.~B.}\ \bibnamefont {Plenio}},\ }\bibfield  {title}
  {\emph {\bibinfo {title} {Quantum dynamics in photonic crystals},\ }}\href
  {\doibase 10.1103/PhysRevA.87.013428} {\bibfield  {journal} {\bibinfo
  {journal} {Phys. Rev. A}\ }\textbf {\bibinfo {volume} {87}},\ \bibinfo
  {pages} {013428} (\bibinfo {year} {2013})}\BibitemShut {NoStop}%
\bibitem [{\citenamefont {Sanchez-Burillo}\ \emph {et~al.}(2014)\citenamefont
  {Sanchez-Burillo}, \citenamefont {Zueco}, \citenamefont {Garcia-Ripoll},\
  and\ \citenamefont {Martin-Moreno}}]{Sanchez-Burillo2014}%
  \BibitemOpen
  \bibfield  {author} {\bibinfo {author} {\bibfnamefont {E.}~\bibnamefont
  {Sanchez-Burillo}}, \bibinfo {author} {\bibfnamefont {D.}~\bibnamefont
  {Zueco}}, \bibinfo {author} {\bibfnamefont {J.~J.}\ \bibnamefont
  {Garcia-Ripoll}}, \ and\ \bibinfo {author} {\bibfnamefont {L.}~\bibnamefont
  {Martin-Moreno}},\ }\bibfield  {title} {\emph {\bibinfo {title} {Scattering
  in the ultrastrong regime: Nonlinear optics with one photon},\ }}\href
  {\doibase 10.1103/PhysRevLett.113.263604} {\bibfield  {journal} {\bibinfo
  {journal} {Phys. Rev. Lett.}\ }\textbf {\bibinfo {volume} {113}},\ \bibinfo
  {pages} {263604} (\bibinfo {year} {2014})}\BibitemShut {NoStop}%
\bibitem [{\citenamefont {Caneva}\ \emph {et~al.}(2015)\citenamefont {Caneva},
  \citenamefont {Manzoni}, \citenamefont {Shi}, \citenamefont {Douglas},
  \citenamefont {Cirac},\ and\ \citenamefont {Chang}}]{Caneva2015}%
  \BibitemOpen
  \bibfield  {author} {\bibinfo {author} {\bibfnamefont {T.}~\bibnamefont
  {Caneva}}, \bibinfo {author} {\bibfnamefont {M.~T.}\ \bibnamefont {Manzoni}},
  \bibinfo {author} {\bibfnamefont {T.}~\bibnamefont {Shi}}, \bibinfo {author}
  {\bibfnamefont {J.~S.}\ \bibnamefont {Douglas}}, \bibinfo {author}
  {\bibfnamefont {J.~I.}\ \bibnamefont {Cirac}}, \ and\ \bibinfo {author}
  {\bibfnamefont {D.~E.}\ \bibnamefont {Chang}},\ }\bibfield  {title} {\emph
  {\bibinfo {title} {Quantum dynamics of propagating photons with strong
  interactions: a generalized input-output formalism},\ }}\href
  {http://stacks.iop.org/1367-2630/17/i=11/a=113001} {\bibfield  {journal}
  {\bibinfo  {journal} {New Journal of Physics}\ }\textbf {\bibinfo {volume}
  {17}},\ \bibinfo {pages} {113001} (\bibinfo {year} {2015})}\BibitemShut
  {NoStop}%
\bibitem [{\citenamefont {Nysteen}\ \emph
  {et~al.}(2015{\natexlab{a}})\citenamefont {Nysteen}, \citenamefont
  {Kristensen}, \citenamefont {McCutcheon}, \citenamefont {Kaer},\ and\
  \citenamefont {M{\o}rk}}]{Nysteen2015}%
  \BibitemOpen
  \bibfield  {author} {\bibinfo {author} {\bibfnamefont {A.}~\bibnamefont
  {Nysteen}}, \bibinfo {author} {\bibfnamefont {P.~T.}\ \bibnamefont
  {Kristensen}}, \bibinfo {author} {\bibfnamefont {D.~P.~S.}\ \bibnamefont
  {McCutcheon}}, \bibinfo {author} {\bibfnamefont {P.}~\bibnamefont {Kaer}}, \
  and\ \bibinfo {author} {\bibfnamefont {J.}~\bibnamefont {M{\o}rk}},\
  }\bibfield  {title} {\emph {\bibinfo {title} {Scattering of two photons on a
  quantum emitter in a one-dimensional waveguide: exact dynamics and induced
  correlations},\ }}\href {http://stacks.iop.org/1367-2630/17/i=2/a=023030}
  {\bibfield  {journal} {\bibinfo  {journal} {New Journal of Physics}\ }\textbf
  {\bibinfo {volume} {17}},\ \bibinfo {pages} {023030} (\bibinfo {year}
  {2015}{\natexlab{a}})}\BibitemShut {NoStop}%
\bibitem [{\citenamefont {Longhi}(2007)}]{Longhi2007}%
  \BibitemOpen
  \bibfield  {author} {\bibinfo {author} {\bibfnamefont {S.}~\bibnamefont
  {Longhi}},\ }\bibfield  {title} {\emph {\bibinfo {title} {Bound states in the
  continuum in a single-level {Fano-Anderson} model},\ }}\href {\doibase
  10.1140/epjb/e2007-00143-2} {\bibfield  {journal} {\bibinfo  {journal} {The
  European Physical Journal B}\ }\textbf {\bibinfo {volume} {57}},\ \bibinfo
  {pages} {45} (\bibinfo {year} {2007})}\BibitemShut {NoStop}%
\bibitem [{\citenamefont {Zhou}\ \emph {et~al.}(2008)\citenamefont {Zhou},
  \citenamefont {Gong}, \citenamefont {Liu}, \citenamefont {Sun},\ and\
  \citenamefont {Nori}}]{Zhou2008}%
  \BibitemOpen
  \bibfield  {author} {\bibinfo {author} {\bibfnamefont {L.}~\bibnamefont
  {Zhou}}, \bibinfo {author} {\bibfnamefont {Z.~R.}\ \bibnamefont {Gong}},
  \bibinfo {author} {\bibfnamefont {Y.-x.}\ \bibnamefont {Liu}}, \bibinfo
  {author} {\bibfnamefont {C.~P.}\ \bibnamefont {Sun}}, \ and\ \bibinfo
  {author} {\bibfnamefont {F.}~\bibnamefont {Nori}},\ }\bibfield  {title}
  {\emph {\bibinfo {title} {Controllable scattering of a single photon inside a
  one-dimensional resonator waveguide},\ }}\href {\doibase
  10.1103/PhysRevLett.101.100501} {\bibfield  {journal} {\bibinfo  {journal}
  {Phys. Rev. Lett.}\ }\textbf {\bibinfo {volume} {101}},\ \bibinfo {pages}
  {100501} (\bibinfo {year} {2008})}\BibitemShut {NoStop}%
\bibitem [{\citenamefont {Roy}(2011)}]{Roy2011a}%
  \BibitemOpen
  \bibfield  {author} {\bibinfo {author} {\bibfnamefont {D.}~\bibnamefont
  {Roy}},\ }\bibfield  {title} {\emph {\bibinfo {title} {Correlated few-photon
  transport in one-dimensional waveguides: Linear and nonlinear dispersions},\
  }}\href {\doibase 10.1103/PhysRevA.83.043823} {\bibfield  {journal} {\bibinfo
   {journal} {Phys. Rev. A}\ }\textbf {\bibinfo {volume} {83}},\ \bibinfo
  {pages} {043823} (\bibinfo {year} {2011})}\BibitemShut {NoStop}%
\bibitem [{\citenamefont {Lombardo}\ \emph {et~al.}(2014)\citenamefont
  {Lombardo}, \citenamefont {Ciccarello},\ and\ \citenamefont
  {Palma}}]{Lombardo2014}%
  \BibitemOpen
  \bibfield  {author} {\bibinfo {author} {\bibfnamefont {F.}~\bibnamefont
  {Lombardo}}, \bibinfo {author} {\bibfnamefont {F.}~\bibnamefont
  {Ciccarello}}, \ and\ \bibinfo {author} {\bibfnamefont {G.~M.}\ \bibnamefont
  {Palma}},\ }\bibfield  {title} {\emph {\bibinfo {title} {Photon localization
  versus population trapping in a coupled-cavity array},\ }}\href {\doibase
  10.1103/PhysRevA.89.053826} {\bibfield  {journal} {\bibinfo  {journal} {Phys.
  Rev. A}\ }\textbf {\bibinfo {volume} {89}},\ \bibinfo {pages} {053826}
  (\bibinfo {year} {2014})}\BibitemShut {NoStop}%
\bibitem [{\citenamefont {Wang}\ \emph {et~al.}(2014)\citenamefont {Wang},
  \citenamefont {Zhou}, \citenamefont {Li},\ and\ \citenamefont
  {Sun}}]{Wang2014}%
  \BibitemOpen
  \bibfield  {author} {\bibinfo {author} {\bibfnamefont {Z.~H.}\ \bibnamefont
  {Wang}}, \bibinfo {author} {\bibfnamefont {L.}~\bibnamefont {Zhou}}, \bibinfo
  {author} {\bibfnamefont {Y.}~\bibnamefont {Li}}, \ and\ \bibinfo {author}
  {\bibfnamefont {C.~P.}\ \bibnamefont {Sun}},\ }\bibfield  {title} {\emph
  {\bibinfo {title} {Controllable single-photon frequency converter via a
  one-dimensional waveguide},\ }}\href {\doibase 10.1103/PhysRevA.89.053813}
  {\bibfield  {journal} {\bibinfo  {journal} {Phys. Rev. A}\ }\textbf {\bibinfo
  {volume} {89}},\ \bibinfo {pages} {053813} (\bibinfo {year}
  {2014})}\BibitemShut {NoStop}%
\bibitem [{\citenamefont {Biondi}\ \emph {et~al.}(2014)\citenamefont {Biondi},
  \citenamefont {Schmidt}, \citenamefont {Blatter},\ and\ \citenamefont
  {T\"ureci}}]{Biondi2014}%
  \BibitemOpen
  \bibfield  {author} {\bibinfo {author} {\bibfnamefont {M.}~\bibnamefont
  {Biondi}}, \bibinfo {author} {\bibfnamefont {S.}~\bibnamefont {Schmidt}},
  \bibinfo {author} {\bibfnamefont {G.}~\bibnamefont {Blatter}}, \ and\
  \bibinfo {author} {\bibfnamefont {H.~E.}\ \bibnamefont {T\"ureci}},\
  }\bibfield  {title} {\emph {\bibinfo {title} {Self-protected polariton states
  in photonic quantum metamaterials},\ }}\href {\doibase
  10.1103/PhysRevA.89.025801} {\bibfield  {journal} {\bibinfo  {journal} {Phys.
  Rev. A}\ }\textbf {\bibinfo {volume} {89}},\ \bibinfo {pages} {025801}
  (\bibinfo {year} {2014})}\BibitemShut {NoStop}%
\bibitem [{\citenamefont {Huang}\ \emph {et~al.}(2013)\citenamefont {Huang},
  \citenamefont {Shi}, \citenamefont {Sun},\ and\ \citenamefont
  {Nori}}]{Huang2013b}%
  \BibitemOpen
  \bibfield  {author} {\bibinfo {author} {\bibfnamefont {J.-F.}\ \bibnamefont
  {Huang}}, \bibinfo {author} {\bibfnamefont {T.}~\bibnamefont {Shi}}, \bibinfo
  {author} {\bibfnamefont {C.~P.}\ \bibnamefont {Sun}}, \ and\ \bibinfo
  {author} {\bibfnamefont {F.}~\bibnamefont {Nori}},\ }\bibfield  {title}
  {\emph {\bibinfo {title} {Controlling single-photon transport in waveguides
  with finite cross section},\ }}\href {\doibase 10.1103/PhysRevA.88.013836}
  {\bibfield  {journal} {\bibinfo  {journal} {Phys. Rev. A}\ }\textbf {\bibinfo
  {volume} {88}},\ \bibinfo {pages} {013836} (\bibinfo {year}
  {2013})}\BibitemShut {NoStop}%
\bibitem [{\citenamefont {Li}\ \emph {et~al.}(2014)\citenamefont {Li},
  \citenamefont {Zhou},\ and\ \citenamefont {Sun}}]{Li2014}%
  \BibitemOpen
  \bibfield  {author} {\bibinfo {author} {\bibfnamefont {Q.}~\bibnamefont
  {Li}}, \bibinfo {author} {\bibfnamefont {L.}~\bibnamefont {Zhou}}, \ and\
  \bibinfo {author} {\bibfnamefont {C.~P.}\ \bibnamefont {Sun}},\ }\bibfield
  {title} {\emph {\bibinfo {title} {Waveguide quantum electrodynamics:
  Controllable channel from quantum interference},\ }}\href {\doibase
  10.1103/PhysRevA.89.063810} {\bibfield  {journal} {\bibinfo  {journal} {Phys.
  Rev. A}\ }\textbf {\bibinfo {volume} {89}},\ \bibinfo {pages} {063810}
  (\bibinfo {year} {2014})}\BibitemShut {NoStop}%
\bibitem [{\citenamefont {de~Vega}(2014)}]{Vega2014}%
  \BibitemOpen
  \bibfield  {author} {\bibinfo {author} {\bibfnamefont {I.}~\bibnamefont
  {de~Vega}},\ }\bibfield  {title} {\emph {\bibinfo {title} {Lattice mapping
  for many-body open quantum systems and its application to atoms in photonic
  crystals},\ }}\href {\doibase 10.1103/PhysRevA.90.043806} {\bibfield
  {journal} {\bibinfo  {journal} {Phys. Rev. A}\ }\textbf {\bibinfo {volume}
  {90}},\ \bibinfo {pages} {043806} (\bibinfo {year} {2014})}\BibitemShut
  {NoStop}%
\bibitem [{\citenamefont {Calajo}\ \emph {et~al.}(2015)\citenamefont {Calajo},
  \citenamefont {Ciccarello}, \citenamefont {Chang},\ and\ \citenamefont
  {Rabl}}]{Calajo2015}%
  \BibitemOpen
  \bibfield  {author} {\bibinfo {author} {\bibfnamefont {G.}~\bibnamefont
  {Calajo}}, \bibinfo {author} {\bibfnamefont {F.}~\bibnamefont {Ciccarello}},
  \bibinfo {author} {\bibfnamefont {D.}~\bibnamefont {Chang}}, \ and\ \bibinfo
  {author} {\bibfnamefont {P.}~\bibnamefont {Rabl}},\ }\bibfield  {title}
  {\emph {\bibinfo {title} {Atom-field dressed states in slow-light waveguide
  {QED}},\ }}\href@noop {} {\  (\bibinfo {year} {2015})},\ \Eprint
  {http://arxiv.org/abs/1512.04946} {1512.04946} \BibitemShut {NoStop}%
\bibitem [{\citenamefont {Shi}\ \emph {et~al.}(2015{\natexlab{b}})\citenamefont
  {Shi}, \citenamefont {Wu}, \citenamefont {Gonzalez-Tudela},\ and\
  \citenamefont {Cirac}}]{Shi2015b}%
  \BibitemOpen
  \bibfield  {author} {\bibinfo {author} {\bibfnamefont {T.}~\bibnamefont
  {Shi}}, \bibinfo {author} {\bibfnamefont {Y.-H.}\ \bibnamefont {Wu}},
  \bibinfo {author} {\bibfnamefont {A.}~\bibnamefont {Gonzalez-Tudela}}, \ and\
  \bibinfo {author} {\bibfnamefont {J.~I.}\ \bibnamefont {Cirac}},\ }\bibfield
  {title} {\emph {\bibinfo {title} {Bound states in boson impurity models},\
  }}\href@noop {} {\  (\bibinfo {year} {2015}{\natexlab{b}})},\ \Eprint
  {http://arxiv.org/abs/1512.07238} {1512.07238} \BibitemShut {NoStop}%
\bibitem [{\citenamefont {Shi}\ and\ \citenamefont
  {Sun}(2009{\natexlab{b}})}]{Shi2009a}%
  \BibitemOpen
  \bibfield  {author} {\bibinfo {author} {\bibfnamefont {T.}~\bibnamefont
  {Shi}}\ and\ \bibinfo {author} {\bibfnamefont {C.~P.}\ \bibnamefont {Sun}},\
  }\bibfield  {title} {\emph {\bibinfo {title} {Two-photon scattering in one
  dimension by localized two-level system},\ }}\href@noop {} {\  (\bibinfo
  {year} {2009}{\natexlab{b}})},\ \Eprint {http://arxiv.org/abs/0907.2776}
  {0907.2776} \BibitemShut {NoStop}%
\bibitem [{\citenamefont {Lee}(1954)}]{Lee1954}%
  \BibitemOpen
  \bibfield  {author} {\bibinfo {author} {\bibfnamefont {T.~D.}\ \bibnamefont
  {Lee}},\ }\bibfield  {title} {\emph {\bibinfo {title} {Some special examples
  in renormalizable field theory},\ }}\href {\doibase 10.1103/PhysRev.95.1329}
  {\bibfield  {journal} {\bibinfo  {journal} {Phys. Rev.}\ }\textbf {\bibinfo
  {volume} {95}},\ \bibinfo {pages} {1329} (\bibinfo {year}
  {1954})}\BibitemShut {NoStop}%
\bibitem [{\citenamefont {K{\"a}ll{\'e}n}\ and\ \citenamefont
  {Pauli}(1955)}]{Kallen1955}%
  \BibitemOpen
  \bibfield  {author} {\bibinfo {author} {\bibfnamefont {G.}~\bibnamefont
  {K{\"a}ll{\'e}n}}\ and\ \bibinfo {author} {\bibfnamefont {W.}~\bibnamefont
  {Pauli}},\ }\bibfield  {title} {\emph {\bibinfo {title} {On the mathematical
  structure of {T. D. Lee's} model of a renormalizable field theory},\ }}\href
  {http://www.sdu.dk/media/bibpdf/Bind%2030-39%5CBind%5Cmfm-30-7.pdf}
  {\bibfield  {journal} {\bibinfo  {journal} {Dan. Mat. Fys. Medd.}\ }\textbf
  {\bibinfo {volume} {30}},\ \bibinfo {pages} {1} (\bibinfo {year}
  {1955})}\BibitemShut {NoStop}%
\bibitem [{\citenamefont {Glaser}\ and\ \citenamefont
  {K{\"a}ll{\'e}n}(1957)}]{Glaser1956–1957}%
  \BibitemOpen
  \bibfield  {author} {\bibinfo {author} {\bibfnamefont {V.}~\bibnamefont
  {Glaser}}\ and\ \bibinfo {author} {\bibfnamefont {G.}~\bibnamefont
  {K{\"a}ll{\'e}n}},\ }\bibfield  {title} {\emph {\bibinfo {title} {A model of
  an unstable particle},\ }}\href {\doibase
  http://dx.doi.org/10.1016/0029-5582(56)90116-X} {\bibfield  {journal}
  {\bibinfo  {journal} {Nuclear Physics}\ }\textbf {\bibinfo {volume} {2}},\
  \bibinfo {pages} {706} (\bibinfo {year} {1956--1957})}\BibitemShut {NoStop}%
\bibitem [{\citenamefont {Wick}(1955)}]{Wick1955}%
  \BibitemOpen
  \bibfield  {author} {\bibinfo {author} {\bibfnamefont {G.~C.}\ \bibnamefont
  {Wick}},\ }\bibfield  {title} {\emph {\bibinfo {title} {Introduction to some
  recent work in meson theory},\ }}\href {\doibase 10.1103/RevModPhys.27.339}
  {\bibfield  {journal} {\bibinfo  {journal} {Rev. Mod. Phys.}\ }\textbf
  {\bibinfo {volume} {27}},\ \bibinfo {pages} {339} (\bibinfo {year}
  {1955})}\BibitemShut {NoStop}%
\bibitem [{\citenamefont {Peskin}(2014)}]{Peskin2014}%
  \BibitemOpen
  \bibfield  {author} {\bibinfo {author} {\bibfnamefont {M.~E.}\ \bibnamefont
  {Peskin}},\ }\bibfield  {title} {\emph {\bibinfo {title} {{Ken Wilson}:
  Solving the strong interactions},\ }}\href {\doibase
  10.1007/s10955-014-1048-1} {\bibfield  {journal} {\bibinfo  {journal}
  {Journal of Statistical Physics}\ }\textbf {\bibinfo {volume} {157}},\
  \bibinfo {pages} {651} (\bibinfo {year} {2014})}\BibitemShut {NoStop}%
\bibitem [{\citenamefont {Schneider}\ \emph {et~al.}(2016)\citenamefont
  {Schneider}, \citenamefont {Sproll}, \citenamefont {Stawiarski},
  \citenamefont {Schmitteckert},\ and\ \citenamefont {Busch}}]{Schneider2016}%
  \BibitemOpen
  \bibfield  {author} {\bibinfo {author} {\bibfnamefont {M.~P.}\ \bibnamefont
  {Schneider}}, \bibinfo {author} {\bibfnamefont {T.}~\bibnamefont {Sproll}},
  \bibinfo {author} {\bibfnamefont {C.}~\bibnamefont {Stawiarski}}, \bibinfo
  {author} {\bibfnamefont {P.}~\bibnamefont {Schmitteckert}}, \ and\ \bibinfo
  {author} {\bibfnamefont {K.}~\bibnamefont {Busch}},\ }\bibfield  {title}
  {\emph {\bibinfo {title} {Green's-function formalism for waveguide {QED}
  applications},\ }}\href {\doibase 10.1103/PhysRevA.93.013828} {\bibfield
  {journal} {\bibinfo  {journal} {Phys. Rev. A}\ }\textbf {\bibinfo {volume}
  {93}},\ \bibinfo {pages} {013828} (\bibinfo {year} {2016})}\BibitemShut
  {NoStop}%
\bibitem [{\citenamefont {Shore}\ and\ \citenamefont
  {Knight}(1993)}]{Shore1993}%
  \BibitemOpen
  \bibfield  {author} {\bibinfo {author} {\bibfnamefont {B.~W.}\ \bibnamefont
  {Shore}}\ and\ \bibinfo {author} {\bibfnamefont {P.~L.}\ \bibnamefont
  {Knight}},\ }\bibfield  {title} {\emph {\bibinfo {title} {The
  {Jaynes-Cummings} model},\ }}\href {\doibase 10.1080/09500349314551321}
  {\bibfield  {journal} {\bibinfo  {journal} {Journal of Modern Optics}\
  }\textbf {\bibinfo {volume} {40}},\ \bibinfo {pages} {1195} (\bibinfo {year}
  {1993})}\BibitemShut {NoStop}%
\bibitem [{\citenamefont {Cohen-Tannoudji}\ \emph {et~al.}(1992)\citenamefont
  {Cohen-Tannoudji}, \citenamefont {Dupont-Roc},\ and\ \citenamefont
  {Grynberg}}]{Cohen-Tannoudji1992}%
  \BibitemOpen
  \bibfield  {author} {\bibinfo {author} {\bibfnamefont {C.}~\bibnamefont
  {Cohen-Tannoudji}}, \bibinfo {author} {\bibfnamefont {J.}~\bibnamefont
  {Dupont-Roc}}, \ and\ \bibinfo {author} {\bibfnamefont {G.}~\bibnamefont
  {Grynberg}},\ }\href@noop {} {\emph {\bibinfo {title} {Atom-photon
  interactions : basic processes and applications}}}\ (\bibinfo  {publisher}
  {Wiley},\ \bibinfo {address} {New York},\ \bibinfo {year} {1992})\BibitemShut
  {NoStop}%
\bibitem [{\citenamefont {Taylor}(2006)}]{Taylor2006}%
  \BibitemOpen
  \bibfield  {author} {\bibinfo {author} {\bibfnamefont {J.~R.}\ \bibnamefont
  {Taylor}},\ }\href@noop {} {\emph {\bibinfo {title} {Scattering theory: the
  quantum theory of nonrelativistic collisions}}}\ (\bibinfo  {publisher}
  {Dover},\ \bibinfo {year} {2006})\ Chap.\ \bibinfo {chapter}
  {16--20}\BibitemShut {NoStop}%
\bibitem [{\citenamefont {Maxon}\ and\ \citenamefont
  {Curtis}(1965)}]{Maxon1965}%
  \BibitemOpen
  \bibfield  {author} {\bibinfo {author} {\bibfnamefont {M.~S.}\ \bibnamefont
  {Maxon}}\ and\ \bibinfo {author} {\bibfnamefont {R.~B.}\ \bibnamefont
  {Curtis}},\ }\bibfield  {title} {\emph {\bibinfo {title}
  {{Lehmann-Symanzik-Zimmermann} formalism in the {Lee} model},\ }}\href
  {\doibase 10.1103/PhysRev.137.B996} {\bibfield  {journal} {\bibinfo
  {journal} {Phys. Rev.}\ }\textbf {\bibinfo {volume} {137}},\ \bibinfo {pages}
  {B996} (\bibinfo {year} {1965})}\BibitemShut {NoStop}%
\bibitem [{\citenamefont {Rephaeli}\ \emph {et~al.}(2010)\citenamefont
  {Rephaeli}, \citenamefont {Shen},\ and\ \citenamefont {Fan}}]{Rephaeli2010}%
  \BibitemOpen
  \bibfield  {author} {\bibinfo {author} {\bibfnamefont {E.}~\bibnamefont
  {Rephaeli}}, \bibinfo {author} {\bibfnamefont {J.-T.}\ \bibnamefont {Shen}},
  \ and\ \bibinfo {author} {\bibfnamefont {S.}~\bibnamefont {Fan}},\ }\bibfield
   {title} {\emph {\bibinfo {title} {Full inversion of a two-level atom with a
  single-photon pulse in one-dimensional geometries},\ }}\href {\doibase
  10.1103/PhysRevA.82.033804} {\bibfield  {journal} {\bibinfo  {journal} {Phys.
  Rev. A}\ }\textbf {\bibinfo {volume} {82}},\ \bibinfo {pages} {033804}
  (\bibinfo {year} {2010})}\BibitemShut {NoStop}%
\bibitem [{\citenamefont {Hewson}(1993)}]{Hewson1993}%
  \BibitemOpen
  \bibfield  {author} {\bibinfo {author} {\bibfnamefont {A.~C.}\ \bibnamefont
  {Hewson}},\ }\href@noop {} {\emph {\bibinfo {title} {The Kondo problem to
  heavy fermions}}}\ (\bibinfo  {publisher} {Cambridge Univ. Press},\ \bibinfo
  {address} {Cambridge},\ \bibinfo {year} {1993})\BibitemShut {NoStop}%
\bibitem [{\citenamefont {Maxon}(1966)}]{Maxon1966}%
  \BibitemOpen
  \bibfield  {author} {\bibinfo {author} {\bibfnamefont {M.~S.}\ \bibnamefont
  {Maxon}},\ }\bibfield  {title} {\emph {\bibinfo {title}
  {{Lehmann-Symanzik-Zimmermann} formalism in the {Lee} model
  ($v-\ensuremath{\theta}$ sector)},\ }}\href {\doibase
  10.1103/PhysRev.149.1273} {\bibfield  {journal} {\bibinfo  {journal} {Phys.
  Rev.}\ }\textbf {\bibinfo {volume} {149}},\ \bibinfo {pages} {1273} (\bibinfo
  {year} {1966})}\BibitemShut {NoStop}%
\bibitem [{\citenamefont {Newton}(1982)}]{Newton1982}%
  \BibitemOpen
  \bibfield  {author} {\bibinfo {author} {\bibfnamefont {R.~G.}\ \bibnamefont
  {Newton}},\ }\href@noop {} {\emph {\bibinfo {title} {Scattering Theory of
  Waves and Particles}}},\ \bibinfo {edition} {2nd}\ ed.\ (\bibinfo
  {publisher} {Springer-Verlag},\ \bibinfo {year} {1982})\ Chap.~\bibinfo
  {chapter} {16}\BibitemShut {NoStop}%
\bibitem [{\citenamefont {Lovelace}(1964)}]{Lovelace1964}%
  \BibitemOpen
  \bibfield  {author} {\bibinfo {author} {\bibfnamefont {C.}~\bibnamefont
  {Lovelace}},\ }\bibfield  {title} {\emph {\bibinfo {title} {Practical theory
  of three-particle states. i. {Nonrelativistic}},\ }}\href {\doibase
  10.1103/PhysRev.135.B1225} {\bibfield  {journal} {\bibinfo  {journal} {Phys.
  Rev.}\ }\textbf {\bibinfo {volume} {135}},\ \bibinfo {pages} {B1225}
  (\bibinfo {year} {1964})}\BibitemShut {NoStop}%
\bibitem [{\citenamefont {Alt}\ \emph {et~al.}(1967)\citenamefont {Alt},
  \citenamefont {Grassberger},\ and\ \citenamefont {Sandhas}}]{Alt1967}%
  \BibitemOpen
  \bibfield  {author} {\bibinfo {author} {\bibfnamefont {E.~O.}\ \bibnamefont
  {Alt}}, \bibinfo {author} {\bibfnamefont {P.}~\bibnamefont {Grassberger}}, \
  and\ \bibinfo {author} {\bibfnamefont {W.}~\bibnamefont {Sandhas}},\
  }\bibfield  {title} {\emph {\bibinfo {title} {Reduction of the three-particle
  collision problem to multi-channel two-particle {Lippmann-Schwinger}
  equations},\ }}\href {\doibase
  http://dx.doi.org/10.1016/0550-3213(67)90016-8} {\bibfield  {journal}
  {\bibinfo  {journal} {Nuclear Physics B}\ }\textbf {\bibinfo {volume} {2}},\
  \bibinfo {pages} {167 } (\bibinfo {year} {1967})}\BibitemShut {NoStop}%
\bibitem [{\citenamefont {Osborn}\ and\ \citenamefont
  {Kowalski}(1971)}]{Osborn1971}%
  \BibitemOpen
  \bibfield  {author} {\bibinfo {author} {\bibfnamefont {T.~A.}\ \bibnamefont
  {Osborn}}\ and\ \bibinfo {author} {\bibfnamefont {K.~L.}\ \bibnamefont
  {Kowalski}},\ }\bibfield  {title} {\emph {\bibinfo {title} {Optimal equations
  for three particle scattering},\ }}\href {\doibase
  http://dx.doi.org/10.1016/0003-4916(71)90128-X} {\bibfield  {journal}
  {\bibinfo  {journal} {Annals of Physics}\ }\textbf {\bibinfo {volume} {68}},\
  \bibinfo {pages} {361 } (\bibinfo {year} {1971})}\BibitemShut {NoStop}%
\bibitem [{\citenamefont {Ringel}\ and\ \citenamefont
  {Gritsev}(2013)}]{Ringel2013}%
  \BibitemOpen
  \bibfield  {author} {\bibinfo {author} {\bibfnamefont {M.}~\bibnamefont
  {Ringel}}\ and\ \bibinfo {author} {\bibfnamefont {V.}~\bibnamefont
  {Gritsev}},\ }\bibfield  {title} {\emph {\bibinfo {title} {Dynamical symmetry
  approach to path integrals of quantum spin systems},\ }}\href {\doibase
  10.1103/PhysRevA.88.062105} {\bibfield  {journal} {\bibinfo  {journal} {Phys.
  Rev. A}\ }\textbf {\bibinfo {volume} {88}},\ \bibinfo {pages} {062105}
  (\bibinfo {year} {2013})}\BibitemShut {NoStop}%
\bibitem [{\citenamefont {Muskhelishvili}(1958)}]{Muskhelishvili1958}%
  \BibitemOpen
  \bibfield  {author} {\bibinfo {author} {\bibfnamefont {N.~I.}\ \bibnamefont
  {Muskhelishvili}},\ }\href@noop {} {\emph {\bibinfo {title} {Singular
  integral equations : boundary problems of functions theory and their
  applications to mathematical physics}}}\ (\bibinfo  {publisher}
  {Wolters-Noordhoff},\ \bibinfo {address} {Groningen},\ \bibinfo {year}
  {1958})\BibitemShut {NoStop}%
\bibitem [{\citenamefont {Kirilov}\ and\ \citenamefont
  {Trott}(1994)}]{Kirilov1994}%
  \BibitemOpen
  \bibfield  {author} {\bibinfo {author} {\bibfnamefont {M.}~\bibnamefont
  {Kirilov}}\ and\ \bibinfo {author} {\bibfnamefont {M.}~\bibnamefont
  {Trott}},\ }\bibfield  {title} {\emph {\bibinfo {title} {K-space treatment of
  reflection and transmission at a potential step},\ }}\href {\doibase
  http://dx.doi.org/10.1119/1.17517} {\bibfield  {journal} {\bibinfo  {journal}
  {American Journal of Physics}\ }\textbf {\bibinfo {volume} {62}},\ \bibinfo
  {pages} {553} (\bibinfo {year} {1994})}\BibitemShut {NoStop}%
\bibitem [{\citenamefont {Aubert}\ and\ \citenamefont
  {Kornprobst}(2006)}]{Aubert2006}%
  \BibitemOpen
  \bibfield  {author} {\bibinfo {author} {\bibfnamefont {G.}~\bibnamefont
  {Aubert}}\ and\ \bibinfo {author} {\bibfnamefont {P.}~\bibnamefont
  {Kornprobst}},\ }\href@noop {} {\emph {\bibinfo {title} {Mathematical
  problems in image processing : partial differential equations and the
  calculus of variations}}}\ (\bibinfo  {publisher} {Springer},\ \bibinfo
  {address} {New York (NY)},\ \bibinfo {year} {2006})\ p.\ \bibinfo {pages}
  {285}\BibitemShut {NoStop}%
\bibitem [{\citenamefont {Evans}(2010)}]{Evans2010}%
  \BibitemOpen
  \bibfield  {author} {\bibinfo {author} {\bibfnamefont {L.~C.}\ \bibnamefont
  {Evans}},\ }\href@noop {} {\emph {\bibinfo {title} {Partial differential
  equations}}}\ (\bibinfo  {publisher} {American Mathematical Society},\
  \bibinfo {address} {Providence},\ \bibinfo {year} {2010})\BibitemShut
  {NoStop}%
\bibitem [{\citenamefont {Hartmann}\ \emph {et~al.}(2011)\citenamefont
  {Hartmann}, \citenamefont {Latorre},\ and\ \citenamefont
  {Ciccotti}}]{Hartmann2011}%
  \BibitemOpen
  \bibfield  {author} {\bibinfo {author} {\bibfnamefont {C.}~\bibnamefont
  {Hartmann}}, \bibinfo {author} {\bibfnamefont {J.~C.}\ \bibnamefont
  {Latorre}}, \ and\ \bibinfo {author} {\bibfnamefont {G.}~\bibnamefont
  {Ciccotti}},\ }\bibfield  {title} {\emph {\bibinfo {title} {On two possible
  definitions of the free energy for collective variables},\ }}\href {\doibase
  10.1140/epjst/e2011-01519-7} {\bibfield  {journal} {\bibinfo  {journal} {The
  European Physical Journal Special Topics}\ }\textbf {\bibinfo {volume}
  {200}},\ \bibinfo {pages} {73} (\bibinfo {year} {2011})}\BibitemShut
  {NoStop}%
\bibitem [{\citenamefont {Atkinson}\ and\ \citenamefont
  {Shampine}(2008)}]{Atkinson2008}%
  \BibitemOpen
  \bibfield  {author} {\bibinfo {author} {\bibfnamefont {K.~E.}\ \bibnamefont
  {Atkinson}}\ and\ \bibinfo {author} {\bibfnamefont {L.~F.}\ \bibnamefont
  {Shampine}},\ }\bibfield  {title} {\emph {\bibinfo {title} {Algorithm 876:
  Solving {Fredholm} integral equations of the second kind in {Matlab}},\
  }}\href {\doibase 10.1145/1377596.1377601} {\bibfield  {journal} {\bibinfo
  {journal} {ACM Trans. Math. Softw.}\ }\textbf {\bibinfo {volume} {34}},\
  \bibinfo {pages} {21:1} (\bibinfo {year} {2008})}\BibitemShut {NoStop}%
\bibitem [{\citenamefont {Amado}(1961)}]{Amado1961}%
  \BibitemOpen
  \bibfield  {author} {\bibinfo {author} {\bibfnamefont {R.~D.}\ \bibnamefont
  {Amado}},\ }\bibfield  {title} {\emph {\bibinfo {title}
  {$v-\ensuremath{\theta}$ collisions in the {Lee} model},\ }}\href {\doibase
  10.1103/PhysRev.122.696} {\bibfield  {journal} {\bibinfo  {journal} {Phys.
  Rev.}\ }\textbf {\bibinfo {volume} {122}},\ \bibinfo {pages} {696} (\bibinfo
  {year} {1961})}\BibitemShut {NoStop}%
\bibitem [{\citenamefont {Bolsterli}(1968)}]{Bolsterli1968}%
  \BibitemOpen
  \bibfield  {author} {\bibinfo {author} {\bibfnamefont {M.}~\bibnamefont
  {Bolsterli}},\ }\bibfield  {title} {\emph {\bibinfo {title} {Algebraic
  solution in the $v-\ensuremath{\theta}$ sector of the {Lee} model},\ }}\href
  {\doibase 10.1103/PhysRev.166.1760} {\bibfield  {journal} {\bibinfo
  {journal} {Phys. Rev.}\ }\textbf {\bibinfo {volume} {166}},\ \bibinfo {pages}
  {1760} (\bibinfo {year} {1968})}\BibitemShut {NoStop}%
\bibitem [{\citenamefont {Kazes}(1965)}]{Kazes1965}%
  \BibitemOpen
  \bibfield  {author} {\bibinfo {author} {\bibfnamefont {E.}~\bibnamefont
  {Kazes}},\ }\bibfield  {title} {\emph {\bibinfo {title} {Solution of an
  integral equation in $v-\theta$ scattering},\ }}\href {\doibase
  http://dx.doi.org/10.1063/1.1704722} {\bibfield  {journal} {\bibinfo
  {journal} {Journal of Mathematical Physics}\ }\textbf {\bibinfo {volume}
  {6}},\ \bibinfo {pages} {1772} (\bibinfo {year} {1965})}\BibitemShut
  {NoStop}%
\bibitem [{\citenamefont {Kenschaft}\ and\ \citenamefont
  {Amado}(1964)}]{Kenschaft1964}%
  \BibitemOpen
  \bibfield  {author} {\bibinfo {author} {\bibfnamefont {R.~P.}\ \bibnamefont
  {Kenschaft}}\ and\ \bibinfo {author} {\bibfnamefont {R.~D.}\ \bibnamefont
  {Amado}},\ }\bibfield  {title} {\emph {\bibinfo {title} {Solution of a
  singular integral equation from scattering theory},\ }}\href {\doibase
  http://dx.doi.org/10.1063/1.1704244} {\bibfield  {journal} {\bibinfo
  {journal} {Journal of Mathematical Physics}\ }\textbf {\bibinfo {volume}
  {5}},\ \bibinfo {pages} {1340} (\bibinfo {year} {1964})}\BibitemShut
  {NoStop}%
\bibitem [{\citenamefont {Pagnamenta}(1965)}]{Pagnamenta1965}%
  \BibitemOpen
  \bibfield  {author} {\bibinfo {author} {\bibfnamefont {A.}~\bibnamefont
  {Pagnamenta}},\ }\bibfield  {title} {\emph {\bibinfo {title} {Solution of the
  {K{\"a}ll\'{e}n-Pauli} equation},\ }}\href {\doibase
  http://dx.doi.org/10.1063/1.1704355} {\bibfield  {journal} {\bibinfo
  {journal} {Journal of Mathematical Physics}\ }\textbf {\bibinfo {volume}
  {6}},\ \bibinfo {pages} {955} (\bibinfo {year} {1965})}\BibitemShut {NoStop}%
\bibitem [{\citenamefont {Pagnamenta}(1966)}]{Pagnamenta1966}%
  \BibitemOpen
  \bibfield  {author} {\bibinfo {author} {\bibfnamefont {A.}~\bibnamefont
  {Pagnamenta}},\ }\bibfield  {title} {\emph {\bibinfo {title} {$v\theta$-bound
  state and uniqueness in the three-particle sector of the {Lee} model},\
  }}\href {\doibase http://dx.doi.org/10.1063/1.1704940} {\bibfield  {journal}
  {\bibinfo  {journal} {Journal of Mathematical Physics}\ }\textbf {\bibinfo
  {volume} {7}},\ \bibinfo {pages} {356} (\bibinfo {year} {1966})}\BibitemShut
  {NoStop}%
\bibitem [{\citenamefont {Sommerfield}(1965)}]{Sommerfield1965}%
  \BibitemOpen
  \bibfield  {author} {\bibinfo {author} {\bibfnamefont {C.~M.}\ \bibnamefont
  {Sommerfield}},\ }\bibfield  {title} {\emph {\bibinfo {title} {Solution of
  the integral equation for $v-\theta$ scattering in the {Lee} model},\ }}\href
  {\doibase http://dx.doi.org/10.1063/1.1704386} {\bibfield  {journal}
  {\bibinfo  {journal} {Journal of Mathematical Physics}\ }\textbf {\bibinfo
  {volume} {6}},\ \bibinfo {pages} {1170} (\bibinfo {year} {1965})}\BibitemShut
  {NoStop}%
\bibitem [{Sup()}]{Supplemental}%
  \BibitemOpen
  \href@noop {} {}\bibinfo {note} {See Supplemental Material at [URL will be
  inserted by publisher] for data and code required to generate all figures in
  the manuscript, the Mathematica notebook file used in evaluating integrals
  for pulse based scattering calculations, and a comparison of the results in
  this manuscript with those given in \cite{Shi2009a} for bound-to-bound
  scattering.}\BibitemShut {Stop}%
\bibitem [{\citenamefont {Bronzan}(1965)}]{Bronzan1965}%
  \BibitemOpen
  \bibfield  {author} {\bibinfo {author} {\bibfnamefont {J.~B.}\ \bibnamefont
  {Bronzan}},\ }\bibfield  {title} {\emph {\bibinfo {title} {Soluble model
  field theory with vertex function},\ }}\href {\doibase
  10.1103/PhysRev.139.B751} {\bibfield  {journal} {\bibinfo  {journal} {Phys.
  Rev.}\ }\textbf {\bibinfo {volume} {139}},\ \bibinfo {pages} {B751} (\bibinfo
  {year} {1965})}\BibitemShut {NoStop}%
\bibitem [{\citenamefont {Liu}\ and\ \citenamefont
  {Zimmerman}(1968)}]{Liu1968}%
  \BibitemOpen
  \bibfield  {author} {\bibinfo {author} {\bibfnamefont {T.}~\bibnamefont
  {Liu}}\ and\ \bibinfo {author} {\bibfnamefont {R.~L.}\ \bibnamefont
  {Zimmerman}},\ }\bibfield  {title} {\emph {\bibinfo {title} {Bound-state
  scattering in the {Lee} model},\ }}\href@noop {} {\bibfield  {journal}
  {\bibinfo  {journal} {Nuovo Cimento Suppl.}\ }\textbf {\bibinfo {volume}
  {6}},\ \bibinfo {pages} {1297} (\bibinfo {year} {1968})}\BibitemShut
  {NoStop}%
\bibitem [{\citenamefont {Liu}\ and\ \citenamefont
  {Zimmerman}(1970)}]{Liu1970}%
  \BibitemOpen
  \bibfield  {author} {\bibinfo {author} {\bibfnamefont {T.-H.}\ \bibnamefont
  {Liu}}\ and\ \bibinfo {author} {\bibfnamefont {R.~L.}\ \bibnamefont
  {Zimmerman}},\ }\bibfield  {title} {\emph {\bibinfo {title} {Scattering in
  general higher sectors of the {Lee} model},\ }}\href {\doibase
  http://dx.doi.org/10.1063/1.1665347} {\bibfield  {journal} {\bibinfo
  {journal} {Journal of Mathematical Physics}\ }\textbf {\bibinfo {volume}
  {11}},\ \bibinfo {pages} {1941} (\bibinfo {year} {1970})}\BibitemShut
  {NoStop}%
\bibitem [{\citenamefont {Hammer}\ and\ \citenamefont
  {Shrauner}(1988)}]{Hammer1988}%
  \BibitemOpen
  \bibfield  {author} {\bibinfo {author} {\bibfnamefont {C.~L.}\ \bibnamefont
  {Hammer}}\ and\ \bibinfo {author} {\bibfnamefont {J.~E.}\ \bibnamefont
  {Shrauner}},\ }\bibfield  {title} {\emph {\bibinfo {title} {Coherent-state
  path-integral {S-matrix} formalism applied to the {Lee} model},\ }}\href
  {\doibase http://dx.doi.org/10.1063/1.528091} {\bibfield  {journal} {\bibinfo
   {journal} {Journal of Mathematical Physics}\ }\textbf {\bibinfo {volume}
  {29}},\ \bibinfo {pages} {2507} (\bibinfo {year} {1988})}\BibitemShut
  {NoStop}%
\bibitem [{\citenamefont {Li}\ and\ \citenamefont {Song}(2015)}]{Li2015}%
  \BibitemOpen
  \bibfield  {author} {\bibinfo {author} {\bibfnamefont {C.}~\bibnamefont
  {Li}}\ and\ \bibinfo {author} {\bibfnamefont {Z.}~\bibnamefont {Song}},\
  }\bibfield  {title} {\emph {\bibinfo {title} {Polariton-photon transition in
  coupled-cavity {QED} system},\ }}\href@noop {} {\  (\bibinfo {year}
  {2015})},\ \Eprint {http://arxiv.org/abs/1503.00447} {1503.00447}
  \BibitemShut {NoStop}%
\bibitem [{\citenamefont {Hahn}(2005)}]{Hahn2005}%
  \BibitemOpen
  \bibfield  {author} {\bibinfo {author} {\bibfnamefont {T.}~\bibnamefont
  {Hahn}},\ }\bibfield  {title} {\emph {\bibinfo {title} {\textsc{cuba}--a
  library for multidimensional numerical integration},\ }}\href {\doibase
  http://dx.doi.org/10.1016/j.cpc.2005.01.010} {\bibfield  {journal} {\bibinfo
  {journal} {Computer Physics Communications}\ }\textbf {\bibinfo {volume}
  {168}},\ \bibinfo {pages} {78 } (\bibinfo {year} {2005})}\BibitemShut
  {NoStop}%
\bibitem [{\citenamefont {John}\ \emph {et~al.}(1997)\citenamefont {John},
  \citenamefont {Hasbun},\ and\ \citenamefont {Singh}}]{John1997}%
  \BibitemOpen
  \bibfield  {author} {\bibinfo {author} {\bibfnamefont {G.~C.}\ \bibnamefont
  {John}}, \bibinfo {author} {\bibfnamefont {J.~E.}\ \bibnamefont {Hasbun}}, \
  and\ \bibinfo {author} {\bibfnamefont {V.~A.}\ \bibnamefont {Singh}},\
  }\bibfield  {title} {\emph {\bibinfo {title} {Simple scheme for the numerical
  evaluation of nearly singular integrals},\ }}\href {\doibase
  10.1063/1.168605} {\bibfield  {journal} {\bibinfo  {journal} {Computers in
  Physics}\ }\textbf {\bibinfo {volume} {11}},\ \bibinfo {pages} {293}
  (\bibinfo {year} {1997})}\BibitemShut {NoStop}%
\bibitem [{\citenamefont {Richardson}(2004)}]{Richardson2004}%
  \BibitemOpen
  \bibfield  {author} {\bibinfo {author} {\bibfnamefont {S.}~\bibnamefont
  {Richardson}},\ }\bibfield  {title} {\emph {\bibinfo {title} {Integral
  equations},\ }}\href
  {http://www.mathematica-journal.com/issue/v9i2/IntegralEquations.html}
  {\bibfield  {journal} {\bibinfo  {journal} {The Mathematica Journal}\
  }\textbf {\bibinfo {volume} {9}},\ \bibinfo {pages} {460} (\bibinfo {year}
  {2004})}\BibitemShut {NoStop}%
\bibitem [{\citenamefont {Amado}(1963)}]{Amado1963}%
  \BibitemOpen
  \bibfield  {author} {\bibinfo {author} {\bibfnamefont {R.~D.}\ \bibnamefont
  {Amado}},\ }\bibfield  {title} {\emph {\bibinfo {title} {Soluble problems in
  the scattering from compound systems},\ }}\href {\doibase
  10.1103/PhysRev.132.485} {\bibfield  {journal} {\bibinfo  {journal} {Phys.
  Rev.}\ }\textbf {\bibinfo {volume} {132}},\ \bibinfo {pages} {485} (\bibinfo
  {year} {1963})}\BibitemShut {NoStop}%
\bibitem [{\citenamefont {Gra\ss{}}\ \emph {et~al.}(2015)\citenamefont
  {Gra\ss{}}, \citenamefont {Muschik}, \citenamefont {Celi}, \citenamefont
  {Chhajlany},\ and\ \citenamefont {Lewenstein}}]{Grass2015}%
  \BibitemOpen
  \bibfield  {author} {\bibinfo {author} {\bibfnamefont {T.}~\bibnamefont
  {Gra\ss{}}}, \bibinfo {author} {\bibfnamefont {C.}~\bibnamefont {Muschik}},
  \bibinfo {author} {\bibfnamefont {A.}~\bibnamefont {Celi}}, \bibinfo {author}
  {\bibfnamefont {R.~W.}\ \bibnamefont {Chhajlany}}, \ and\ \bibinfo {author}
  {\bibfnamefont {M.}~\bibnamefont {Lewenstein}},\ }\bibfield  {title} {\emph
  {\bibinfo {title} {Synthetic magnetic fluxes and topological order in
  one-dimensional spin systems},\ }}\href {\doibase 10.1103/PhysRevA.91.063612}
  {\bibfield  {journal} {\bibinfo  {journal} {Phys. Rev. A}\ }\textbf {\bibinfo
  {volume} {91}},\ \bibinfo {pages} {063612} (\bibinfo {year}
  {2015})}\BibitemShut {NoStop}%
\bibitem [{\citenamefont {Moeferdt}\ \emph {et~al.}(2013)\citenamefont
  {Moeferdt}, \citenamefont {Schmitteckert},\ and\ \citenamefont
  {Busch}}]{Moeferdt2013}%
  \BibitemOpen
  \bibfield  {author} {\bibinfo {author} {\bibfnamefont {M.}~\bibnamefont
  {Moeferdt}}, \bibinfo {author} {\bibfnamefont {P.}~\bibnamefont
  {Schmitteckert}}, \ and\ \bibinfo {author} {\bibfnamefont {K.}~\bibnamefont
  {Busch}},\ }\bibfield  {title} {\emph {\bibinfo {title} {Correlated photons
  in one-dimensional waveguides},\ }}\href {\doibase 10.1364/OL.38.003693}
  {\bibfield  {journal} {\bibinfo  {journal} {Opt. Lett.}\ }\textbf {\bibinfo
  {volume} {38}},\ \bibinfo {pages} {3693} (\bibinfo {year}
  {2013})}\BibitemShut {NoStop}%
\bibitem [{\citenamefont {Nysteen}\ \emph
  {et~al.}(2015{\natexlab{b}})\citenamefont {Nysteen}, \citenamefont
  {McCutcheon},\ and\ \citenamefont {M\o{}rk}}]{Nysteen2015b}%
  \BibitemOpen
  \bibfield  {author} {\bibinfo {author} {\bibfnamefont {A.}~\bibnamefont
  {Nysteen}}, \bibinfo {author} {\bibfnamefont {D.~P.~S.}\ \bibnamefont
  {McCutcheon}}, \ and\ \bibinfo {author} {\bibfnamefont {J.}~\bibnamefont
  {M\o{}rk}},\ }\bibfield  {title} {\emph {\bibinfo {title} {Strong
  nonlinearity-induced correlations for counterpropagating photons scattering
  on a two-level emitter},\ }}\href {\doibase 10.1103/PhysRevA.91.063823}
  {\bibfield  {journal} {\bibinfo  {journal} {Phys. Rev. A}\ }\textbf {\bibinfo
  {volume} {91}},\ \bibinfo {pages} {063823} (\bibinfo {year}
  {2015}{\natexlab{b}})}\BibitemShut {NoStop}%
\bibitem [{\citenamefont {Shen}\ and\ \citenamefont {Shen}(2015)}]{Shen2015a}%
  \BibitemOpen
  \bibfield  {author} {\bibinfo {author} {\bibfnamefont {Y.}~\bibnamefont
  {Shen}}\ and\ \bibinfo {author} {\bibfnamefont {J.-T.}\ \bibnamefont
  {Shen}},\ }\bibfield  {title} {\emph {\bibinfo {title} {Photonic-{Fock}-state
  scattering in a waveguide-{QED} system and their correlation functions},\
  }}\href {\doibase 10.1103/PhysRevA.92.033803} {\bibfield  {journal} {\bibinfo
   {journal} {Phys. Rev. A}\ }\textbf {\bibinfo {volume} {92}},\ \bibinfo
  {pages} {033803} (\bibinfo {year} {2015})}\BibitemShut {NoStop}%
\bibitem [{\citenamefont {Xu}\ and\ \citenamefont {Fan}(2015)}]{Xu2015}%
  \BibitemOpen
  \bibfield  {author} {\bibinfo {author} {\bibfnamefont {S.}~\bibnamefont
  {Xu}}\ and\ \bibinfo {author} {\bibfnamefont {S.}~\bibnamefont {Fan}},\
  }\bibfield  {title} {\emph {\bibinfo {title} {Input-output formalism for
  few-photon transport: A systematic treatment beyond two photons},\ }}\href
  {\doibase 10.1103/PhysRevA.91.043845} {\bibfield  {journal} {\bibinfo
  {journal} {Phys. Rev. A}\ }\textbf {\bibinfo {volume} {91}},\ \bibinfo
  {pages} {043845} (\bibinfo {year} {2015})}\BibitemShut {NoStop}%
\bibitem [{\citenamefont {Obi}\ and\ \citenamefont {Shen}(2015)}]{Obi2015}%
  \BibitemOpen
  \bibfield  {author} {\bibinfo {author} {\bibfnamefont {K.~C.}\ \bibnamefont
  {Obi}}\ and\ \bibinfo {author} {\bibfnamefont {J.-T.}\ \bibnamefont {Shen}},\
  }\bibfield  {title} {\emph {\bibinfo {title} {Perturbative and iterative
  methods for photon transport in one-dimensional waveguides},\ }}\href
  {\doibase http://dx.doi.org/10.1016/j.optcom.2015.01.022} {\bibfield
  {journal} {\bibinfo  {journal} {Optics Communications}\ }\textbf {\bibinfo
  {volume} {343}},\ \bibinfo {pages} {135 } (\bibinfo {year}
  {2015})}\BibitemShut {NoStop}%
\bibitem [{\citenamefont {Paulisch}\ \emph {et~al.}(2015)\citenamefont
  {Paulisch}, \citenamefont {Kimble},\ and\ \citenamefont
  {Gonzalez-Tudela}}]{Paulisch2015}%
  \BibitemOpen
  \bibfield  {author} {\bibinfo {author} {\bibfnamefont {V.}~\bibnamefont
  {Paulisch}}, \bibinfo {author} {\bibfnamefont {H.~J.}\ \bibnamefont
  {Kimble}}, \ and\ \bibinfo {author} {\bibfnamefont {A.}~\bibnamefont
  {Gonzalez-Tudela}},\ }\bibfield  {title} {\emph {\bibinfo {title} {Universal
  quantum computation in waveguide {QED} using decoherence free subspaces},\
  }}\href@noop {} {\  (\bibinfo {year} {2015})},\ \Eprint
  {http://arxiv.org/abs/1512.04803} {1512.04803} \BibitemShut {NoStop}%
\bibitem [{\citenamefont {Economou}(2006)}]{Economou2006}%
  \BibitemOpen
  \bibfield  {author} {\bibinfo {author} {\bibfnamefont {E.~N.}\ \bibnamefont
  {Economou}},\ }\href@noop {} {\emph {\bibinfo {title} {Green's functions in
  quantum physics}}}\ (\bibinfo  {publisher} {Springer},\ \bibinfo {address}
  {Berlin},\ \bibinfo {year} {2006})\BibitemShut {NoStop}%
\end{thebibliography}

\end{document}